\documentclass[fleqn,usenatbib]{mnras}
\usepackage{epsfig,amsmath,amssymb,bm,color,graphicx, natbib}
\usepackage{afterpage}
\usepackage{pdfpages}
\usepackage{caption, subcaption}
\def\araa{{\rm ARA\&A}}             
\def\jcap{{\rm J. Cosmology Astropart. Phys.}}
\def\apjl{\rm Astrophysical Journal Letters}
\def\mnras{{\rm MNRAS}}             
\def\aap{{\rm A\&A}}                

\def\aj{{\rm Astronomical Journal}}

\usepackage{booktabs} 
\usepackage{tabularx}

\newcommand{\lya}{Ly-$\alpha$}
\newcommand{\lyb}{Ly-$\beta$}
\newcommand{\HI}{H{\sc ~i}}
\newcommand{\HeII}{He{\sc ~ii}}

\newcommand{\CIV}{C{\sc ~iv}}
\newcommand{\OVI}{O{\sc ~vi}}
\newcommand{\SiIII}{Si{\sc ~iii}}

\newcommand{\angstrom}{\textup{\AA}}

\newcommand{\skm}{\mathrm{s\,km^{-1}}}
\newcommand{\kms}{\mathrm{km\,s^{-1}}}

\newcommand{\be}{\begin{equation}}
\newcommand{\ee}{\end{equation}}

\usepackage{color}

\title[\lyb\ forest power spectrum]{A measurement of the \lyb\ forest power spectrum and its cross with the \lya\ forest in X-Shooter XQ-100}

\author[Wilson et al.]{Bayu Wilson$^{1}$\thanks{E-mail: bwils033@ucr.edu}, 
Vid Ir\v{s}i\v{c}$^{1,2,3}$, Matthew McQuinn$^{1}$\\
$^{1}$University of Washington, Department of Astronomy, 3910 15th Ave
NE, WA 98195-1580 Seattle, USA\\
$^{2}$Kavli Institute for Cosmology, Department of Physics, University of Cambridge, Madingley Road, Cambridge CB3 0HA, UK\\
$^{3}$Cavendish Laboratory, University of Cambridge, 19 J. J. Thomson Ave., Cambridge CB3 0HE, UK
}

\pubyear{2021}

\begin{document}
\label{firstpage}
\maketitle

\begin{abstract}
The \lya\ forest is the large-scale structure probe for which we appear to have modeling control to the highest wavenumbers. This makes the \lya\ forest of great interest for constraining the warmness/fuzziness of dark matter and the timing of reionization processes. However, the standard statistic, the \lya\ forest power spectrum, is unable to strongly constrain the IGM temperature-density relation, and this inability further limits how well other high wavenumber-sensitive parameters can be constrained. With the aim of breaking these degeneracies, we measure the power spectrum of the \lyb\ forest and its cross correlation with the coeval \lya\ forest using the one hundred spectra of $z=3.5-4.5$ quasars in the VLT/X-Shooter XQ-100 Legacy Survey, motivated by the \lyb\ transition's smaller absorption cross section that makes it sensitive to somewhat higher densities relative to the \lya\ transition.  Our inferences from this measurement for the IGM temperature-density relation appear to latch consistently onto the recent tight lower-redshift \lya\ forest constraints.  The $z=3.4-4.7$ trends we find using the \lya--\lyb\ cross correlation show a flattening of the slope of the temperature-density relation with decreasing redshift.  This is the trend anticipated from ongoing \HeII\ reionization and there being sufficient time to reach the asymptotic temperature-density slope after hydrogen reionization completes. Furthermore, our measurements provide a consistency check on IGM models that explain the \lya\ forest, with the cross correlation being immune to systematics that are uncorrelated between the two forests, such as metal line contamination.
\end{abstract}


\section{Introduction}

The \lya\ forest has been used to constrain the Universe's initial conditions \citep{mcdonald00,zaldarriaga01,croft02,zaldarriaga03,seljak03,mcdonald03,viel04,viel04bis,viel04hrwmap,mcdonald05,mcdonald06,seljak06,slosar11,busca13,slosar13,palanque13,palanque15,bautista15,baur17,bautista17,bourboux17}, the timing of reionization processes \citep{schaye00,ricotti00,theuns00,mcdonald00,viel06,theuns02,bolton08,lidz10,bolton10,becker11,garzilli12,rudie12,lee14,boera14,bolton14, sanderbeck16, rorai17, hiss17, walther19, wu19}, and the warmness or fuzziness of the dark matter 
\citep{narayanan00,viel05,uros06,viel08,bird10,viel13WDM,baur15,yeche17,irsic17a, irsic17b,armengaud17,garzilli17,garzilli19,irsic20,rogers21}. Often the \lya\ forest is the standard bearer for the said constraints.  This superiority owes to having modeling control over its spectrum of fluctuations to higher wavenumbers than all other established large-scale structure probes \citep[e.g.][]{mcquinn15}. 
  
However, it is likely that better constraints can be extracted from intergalactic Lyman-series absorption.  Many studies have found that large degeneracies in the \lya\ forest's constraints on parameters (particularly those constrained best by the highest wavenumbers probed) when using the standard statistic, the power spectrum.  Namely, significant degeneracies exist between any two of the following: the particle mass in warm/fuzzy dark matter models, the gas temperature at mean density, and the trend in temperature with density \citep[e.g.][]{becker11, lidz10, irsic17a}. The latter two thermal parameters constrain reionization (as well as any other heating processes), and the degeneracies are so severe that the trend of temperature with density is essentially unconstrained by previous \lya\ power spectrum analyses.  While some degeneracies can be broken by measurements at multiple redshifts \citep{mcdonald05}, substantially improving constraints over existing ones likely either requires (1) combining \lya\ power spectrum measurements with that of other Lyman-series transitions such as \lyb\ \citep{dijkstra04} or (2) using \lya\ absorption statistics beyond just the power spectrum \citep{zaldarriaga, fangwhite, gaikwad20}.
  
 Here we present the first measurement of the \lyb\ forest auto power spectrum at the wavenumbers most sensitive to the gas temperature, as well as its cross power spectrum with the \lya\ forest. The cross power spectrum is estimated using a fourier transformation of \lya\ and \lyb\ absorption features. An intervening gas overdensity creates both a \lya\ and \lyb\ forest absorption feature but the \lyb\ forest would also contain lower redshift \lya\ absorption which biases the \lyb\ auto power spectrum. The cross power turns out to be our most constraining new diagnostic because the effective noise for the \lyb\ forest, set by the lower redshift \lya\ absorption, is higher than \lya\ forest. The \lyb\ transition has a smaller cross section for absorption than \lya, which makes it more sensitive to higher density gas for which the \lya\ absorption is more saturated \citep{dijkstra04, 2014JCAP...12..024I}. This sensitivity to higher densities breaks the aforementioned degeneracies, most obviously between the temperature at mean density ($T_0$) and its power-law trend with density (with index $\gamma-1$).  
 
 A precision measurement requires a large sample of \lyb\ forest spectra.  However, this goal is hindered by the shorter path length probed by each sightline relative to the \lya\ forest (with about a third as much of useful absorption), by foreground \lya\ forest absorption that contaminates the \lyb\ forest, and by the blueness of the \lyb\ transition that results in this forest being less likely to be captured than the \lya\ forest in existing QSO spectra. Perhaps as a result of these difficulties, there has been only one attempt to measure the \lyb\ power spectrum: \cite{irsic13} presented a measurement using 60,000 SDSS/BOSS quasar spectra. The low resolution of these spectra, with ${\cal R} \equiv \lambda/\Delta \lambda \approx 2000$, inhibit constraining thermal scales and breaking the associated degeneracies.  Here we measure the \lyb\ forest power spectrum (and its cross with the \lya\ forest) using the one hundred, ${\cal R} \sim 10^4$ spectra observed as part of the VLT/XSHOOTER XQ-100 survey \citep{lopez16}.  We also present a preliminary measurement using the ${\cal R} \sim 10^5$ VLT/UVES SQUAD DR1 sample \citep{2019MNRAS.482.3458M}.
 
 Additional motivations for measuring the \lyb\ forest power spectrum, and especially its cross with \lya, are (1) to test consistency with parameters derived with \lya\ forest (2) to further test the standard paradigm that cosmological simulations reproduce the low density intergalactic medium (built principally from comparing these simulations to \lya\ forest power spectrum measurements).  The \lya\ forest power spectrum at high wavenumbers can be contaminated by metal lines owing to their smaller thermal widths \citep[e.g.][]{lidz10} and there is some controversy over the severity of this effect \citep{day19}. The cross power spectrum is unbiased by contaminates that are not correlated between the two forests, including the majority of metals (with a small bias for select transitions that fall near Ly-$\alpha$/$\beta$) as well as instrumental noise. Thus, the cross power could serve as a check on parameters derived from the \lya\ forest alone.

 This paper is organized as follows.  Section~\ref{sec:analysis} describes the data analysis pipeline, as well as how mock spectra are created and used to test the pipeline.  Section~\ref{sec:measurement} describes how the power spectrum is estimated and how it is corrected for noise and resolution, concluding by presenting our power spectrum estimates.  We also compare our XQ-100 measurement with a preliminary \lyb\ measurement using archival VLT/UVES spectra.  Section~\ref{sec:interpretation} interprets our measurements in terms of the IGM thermal history. A series of appendices present details regarding our resolution, noise, DLAs, and metal corrections. 

\section{Data selection and mocks}
\label{sec:analysis}

\begin{figure}
\includegraphics[scale=0.38]{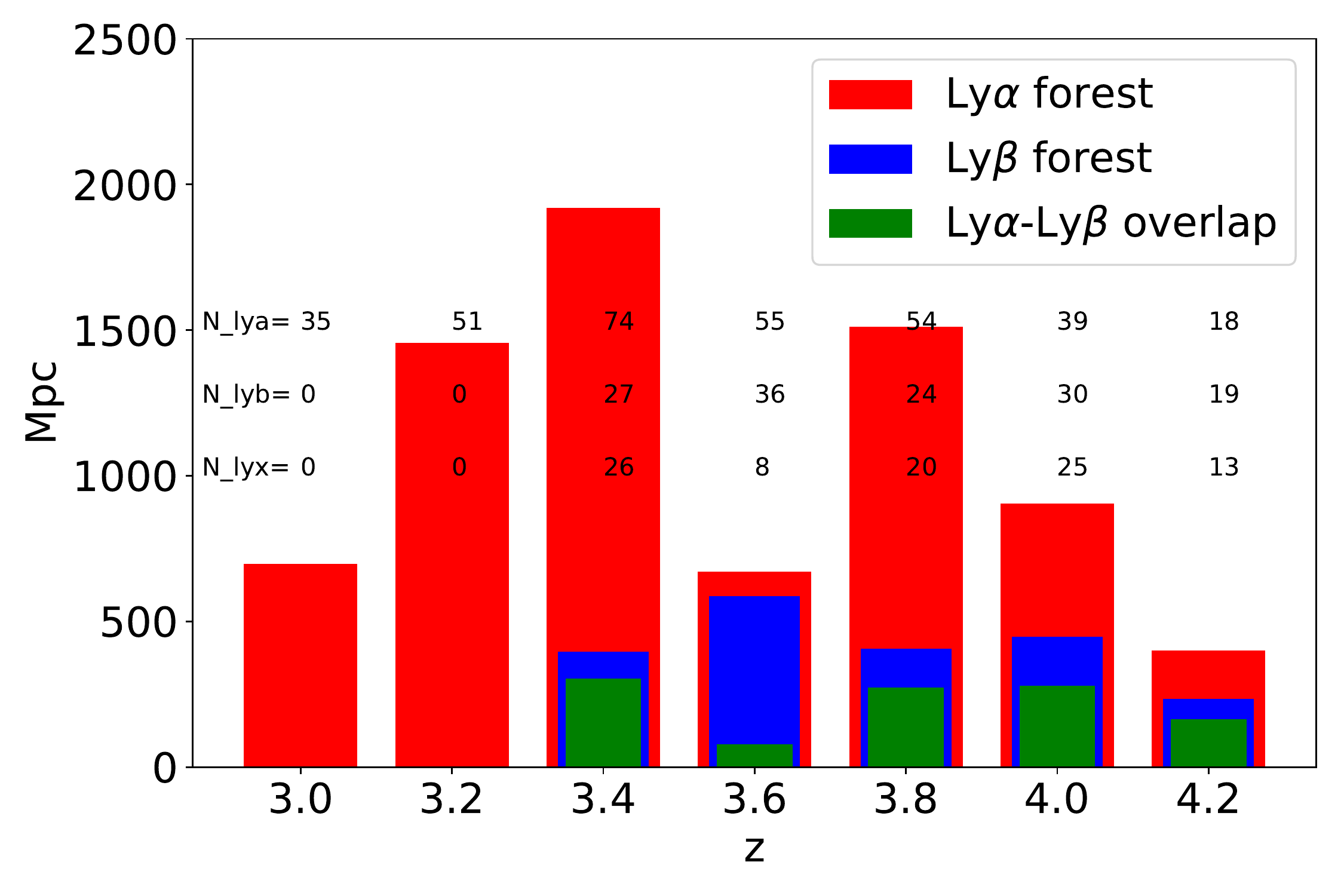}
\caption{The total comoving pathlength of our XQ-100 sample for the \lya\ forest, \lyb\ forest, and the intersection of these two paths (used to calculate the cross power), using the selection criteria outlined in \S~\ref{sec:analysis}.  The annotations give the total number of quasars that contribute in each redshift bin to the three pathlengths.  At the highest wavenumbers that we report power spectra estimates, these distances are marginally smaller as we use spectra for which the resolution correction is the most certain.
} \label{fig:Nz}
\end{figure}

Our primary measurement uses the XQ-100 Legacy Survey 
\citep{lopez16}, consisting of one hundred $3.4 < z < 4.7$ quasi-stellar object (QSO) spectra observed with the X-Shooter spectrograph on the
Very Large Telescope \citep{vernet11}.  The number of quasars  and path lengths of this sample are presented in Figure~\ref{fig:Nz}, quoting these quantities in the $\delta z=0.2$ redshift bins we use in our analysis.  

Each quasar was observed for 2-12 exposures with a 1.0'' for the UV arm or 0.9'' slit for the VIS arm.  The wavelengths for each exposure were calibrated off of known skylines (Appendix~\ref{ap:calibration}).  The individual exposures then used nearest grid point interpolation to average the exposures onto a $20~$km~s$^{-1}$ grid for the UVB arm and $11~$km~s$^{-1}$ for the visible arm.  In our mocks, discussed shortly, we replicate this binning to show that our measurements are not affected. This binning is chosen to be somewhat smaller than the full width half maximum (FWHM) spectral resolution of the X-shooter spectrograph for a fully illuminated slit of $56~$km~s$^{-1}$ for the UV arm and $34~$km~s$^{-1}$ for the VIS arm.  There is an overlap region at $5500-5600$\AA\ where both arms of the spectrograph process a significant fraction of the light, corresponding to Ly$\alpha$ at $z\approx 3.6$.  We exclude this region from our analysis, which results in significantly less \lya\ (and hence cross) data in the $z=3.6$ redshift bin (see Fig.~\ref{fig:Nz}). (This cut also results in this redshift bin always using data from the spectrograph's VIS arm.)

Our analysis uses quasar continuum estimates developed by the XQ-100 Legacy Survey team \citep{Berg2016}, which fitted a spline over several wavelength ranges within each spectrum.  We use these estimates to calculate the continuum-normalized flux in the Lyman forests. The same continuum estimate was used in the previous XQ-100 \lya\ forest analysis \citep{irsic17}.

Our \lya\ forest measurement uses the pixels within the  $1045 - 1185\, \angstrom$ QSO-frame wavelength range, a range chosen to omit absorption in the broad \lya\ and \lyb\ emission lines of the quasar where continuum fitting can be more challenging \citep[e.g.][]{mcdonald05}. Furthermore, this cut omits regions where the ionizing background is enhanced even modestly by the QSO owing to the proximity effect.  For similar reasons, our measurement of the \lyb\ forest uses the QSO-frame wavelength range $978 - 1014\, \angstrom$, following \citet{irsic13}.  This wavelength range for \lyb\ corresponds to using pixels somewhat closer to the quasar than in \lya\ (to an equivalent wavelength of 1202\AA\ in the \lya\ forest) as the \lyb\ line of the QSO is less broad than its \lya\ counterpart.

We mask regions around Damped \lya\ (DLA) systems
using the DLA sample of the XQ-100 survey team \citep{sanchez16}. Thirty percent of sightlines show a DLA, with the probability that a DLA falls in both \lya\ and \lyb\ reduced by the pathlength ratio (Fig.~\ref{fig:Nz}). We do not use data within $1/2$ the equivalent width of \lya\ and \lyb\ line centers of each DLA and additionally correct the mean flux outside of this range for the wings of the line.   Appendix~\ref{ap:metals} quantifies the effect of this masking on our measurements.

Our data analysis pipeline to estimate the mean flux and power spectra of the \lya\ and \lyb\ forests was tested with synthetic Lyman-forest data that were generated following the method for creating synthetic \lya\ forest mocks presented in more detail in \citet{irsic17a}, with the most significant difference here being the inclusion of \lyb\ forest absorption. In summary, we generate a realistic flux field with a QSO redshift distribution matching that of the  XQ-100 sample as well as approximating the XQ-100 pixel, resolution, and noise specifications.  Five thousand light-cone spectra are created using the simulation outputs spaced at $\Delta z = 0.1$.  We use the Sherwood simulation suite  of high resolution hydro-dynamical simulations, with $2\times2048^3$ particles in a 40~Mpc$/h$ box \citep{bolton16}. This simulation appears to be converged in its estimate for the \lya\ forest power spectrum at the redshifts of interest to better than $\sim$5\% \citep{bolton16,irsic17b}. The mean flux of our mocks is rescaled to match measurements. In addition to testing our pipeline, bootstraps of our mocks are used for calculating the covariance matrix.  

\section{Data analysis and measurement}
\label{sec:measurement}

This section describes both our mean flux and power spectrum measurements. The same analysis strategy was adopted for both real and synthetic data.

\subsection{Mean Flux}
\label{sec:meanF}

\begin{figure}
\begin{center}
\includegraphics[scale=0.31]{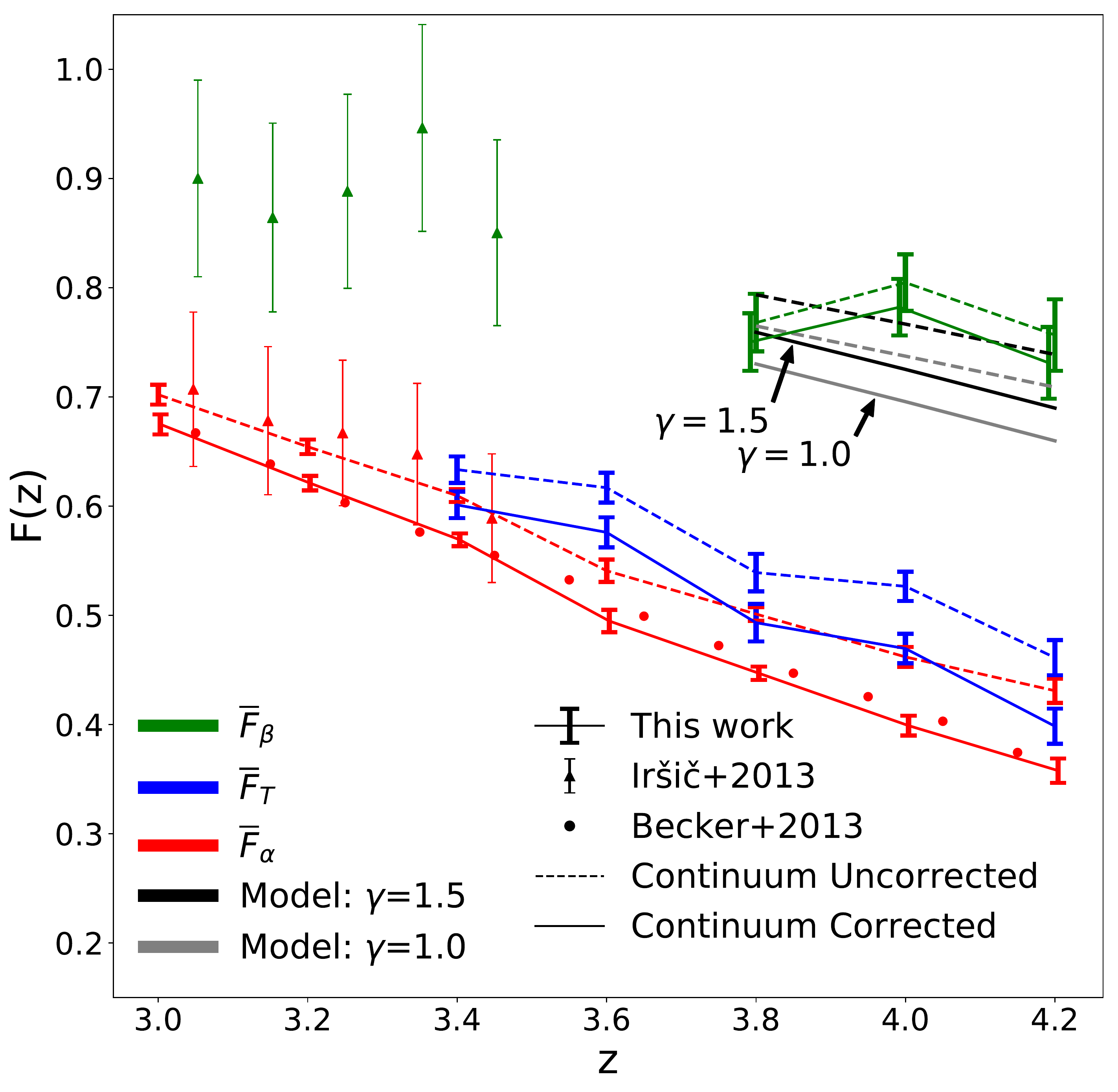}
\end{center}
\caption{Mean flux estimates for \lyb\ ($\bar F_\beta$; green), \lya +\lyb\ at redshift that corresponds to \lyb\ ($\bar F_{T}$; blue), and \lya\ ($\bar F_\alpha$; red).  The error bars that are linked by dashed lines are those estimated here using the XQ-100 data, and those linked with solid lines shows our measurement corrected for the human bias in continuum placement using the correction of \citet{faucher08}. For comparison, the triangles with errorbars are the $\bar F_\alpha$ and $\bar F_\beta$ estimates of \citet{irsic13} using SDSS/BOSS, and the circles are the $F_\alpha$ estimates of \citet{becker13} again on SDSS/BOSS data using a novel method for removing continuum (which results in minute statistical error bars).  The thin grey and black ``lines'' spanning $z=3.8-4.2$ show the predicted mean flux in \lyb\ ($\bar F_\beta$) using simulations that have respectively $\gamma = 1.5$ and $1.0$ and our measured $\bar F_\alpha$. The solid and dashed versions of these lines calibrate the simulations to our continuum corrected and uncorrected values of $\bar F_\alpha$: the fact that \citet{becker13} measurement falls near our continuum corrected estimates makes us favor our solid lines. } \label{fig:meanF_data}
\end{figure}

We average the continuum-normalized flux (i.e. the estimated transmission) in all pixels that fall into a redshift bin to obtain the mean transmission in \lya\ ($\bar F_\alpha$) and \lya + \lyb\  ($\bar F_T$).  Error bars are estimated by bootstrap resampling. Each sample constitutes pixels from a sightline's \lya\ forest spectrum within a redshift bin.  The mean transmission in \lyb\ ($\bar F_\beta$) in a redshift bin is estimated by dividing the value of $\bar F_\alpha$ for the foreground redshift that contributes \lya\ absorption from $\bar F_T$.  We only measure the foreground $\bar F_\alpha$ and, hence, can perform this subtraction for our three highest \lyb\ redshifts.  The errorbars that are connected with dashed lines in Fig.~\ref{fig:meanF_data} show our measurements of $\bar F_\alpha$, $\bar F_T$, and $\bar F_\beta$ (red, blue and green curves, respectively).

The most significant systematic in estimating the mean flux is the placement of the quasar continuum.  The error bars connected with the solid lines are our mean flux measurement corrected for the human bias from by-eye continuum fitting, using the correction factor in \citet{faucher08}.  Again red, blue and green solid curves respectively show our continuum-corrected estimates for $\bar F_\alpha$, $\bar F_T$, and $\bar F_\beta$.  The \cite{faucher08} correction, $\Delta C/C_{true}=1.58 \times  10^{-5} (1+z)^{5.63}$, was estimated by fitting the continua of mock spectra corrections to the mean flux in \lya. 
 While this correction was estimated on mocks that simulated the Keck/ESI spectra assuming FWHM=40~km~s$^{-1}$ and $S/N$=20, these specifications are not dissimilar to our X-Shooter spectra.  \citet{faucher08} further found that the corrections were similar if they considered higher-S/N and higher-resolution mock spectra reminiscent of Keck/HIRES.  However, the \citet{faucher08} estimates could overestimate the true continuum correction as they used low resolution $N$-body simulations to model the forest and a steep $T-\Delta$ relation with $\gamma =1.6$.  

Figure~\ref{fig:meanF_data} compares our measurements with those of \citet{irsic13} and \cite{becker13}.  Both of these measurements use different methodologies than our more traditional mean transmission measurement. The \citet{irsic13} measurements are done by fitting a parametric model to the measured power spectrum in a sample of thousands SDSS/BOSS quasars rather than directly measuring the mean transmission.  Also applied to the SDSS/BOSS sample, the \citet{becker13} measurement uses a different method still that estimates the mean transmission using stacked quasar spectra.  This measurement assumes that in-aggregate the stack's mean continuum shows little redshift evolution.  Both methods likely avoid continuum over-fitting issues that is a major systematic in our measurement.  The measurement of \citet{irsic13} have large errorbars and are consistent at the $1\sigma$ level with our $\bar F_\alpha$.  Their $\bar F_\beta$ measurements do not overlap in redshift with ours. The results by \cite{becker13} fall much closer to our continuum corrected estimate.\footnote{Another common correction made in mean flux measurements is for metal absorption contamination. We do not apply such a correction in our analysis. Metals result in a $9\%$ correction at $z=3$ and $5\%$ at $z=4$ using the metal correction estimates based on direct identification of \citet{2003ApJ...596..768S}, and a $6\%$ at $z=3$ and $2\%$ at $z=4$ using the statistical results of 
\citet{2004AJ....128.1058T}. For the principle aim of our analysis, a \lya\ and \lyb\ power spectrum measurement, the mean flux that is used to calculate the flux overdensity does not need to be corrected for metal absorption.  Metals would not affect our $\bar F_\beta$ estimates to the extent that the mean metal absorption does not differ between the coeval \lya\ and \lyb\ forests. 
} 
However, we note that the uncertainty in the continuum correction divides out in our power spectrum measurements (presented in the next section) and so is not a concern there.

These mean flux measurements alone have the potential to constrain the intergalactic temperature-density relation.  Figure~\ref{fig:meanF_data} investigates this possibility. The thin grey and black lines spanning $z=3.8-4.2$ show the predicted mean flux in \lyb\ ($\bar F_\beta$) using simulations that have respectively temperature-density relations with power-law slope $\gamma = 1.5$ and $1.0$ and our measured \lya\ mean flux (which falls approximately on a single line at these redshifts).  These values for $\gamma$ span most of the theoretically motivated range of $1-1.6$ \citep{hui97, mcquinnTrho}. The solid and dashed versions of these lines calibrate the simulations to the continuum corrected and uncorrected values of the \lya\ mean flux.\footnote{We note that simulations with different values for the temperature at the mean density, $T_0$, predict essentially the same $\bar F_\beta$ when calibrated to the same $\bar F_\alpha$.}  As the grey and solid models span the $1\sigma$ error bar of our measured $\bar F_\beta$ (the connected green points with errorbars), we conclude that our mean flux measurements alone are not strongly constraining of $\gamma$.  There is a slight preference to $\gamma=1.5$, especially if one takes the \citet{becker13} \lya\ mean flux measurement to indicate that our continuum corrected measurement is closer to the truth, as one should expect.  The value $\gamma=1.5$ is also closer to what we find in our power spectrum analysis (\S~\ref{sec:interpretation}).\footnote{The inferred  \lya\ mean flux from the much different power spectrum analysis is the same to a percent fractional level to the continuum corrected curve.  This analysis did use a prior centered on \citet{becker13} with a 5\% error.}

\subsection{Power Spectra}
\label{sec:Pk}
\label{sec:powerspectrum}
To estimate the power spectrum from the data, we follow the standard approach of Fourier transforming segments of our data that pass our cuts
\citep{croft99,croft02,kim04,viel04,Viel_2013}. A disadvantage of this approach is that the power spectrum estimate is not  weighted optimally to minimize errors. Additionally, if the analysis uses segments that are too short, the cutoffs at the edges result in spurious high-$k$ power.  The other approach that has been adopted is to use a quadratic estimator, which mitigates these effects \citep{mcdonald05}.  Quadratic estimators have predominantly been used only for SDSS data sets as they have difficulty converging for smaller data sets for which the power spectrum is less constrained.  Because our data set is small relative to SDSS, we adopt the direct Fourier transform approach. 

\begin{figure*}
\includegraphics[scale=0.45]{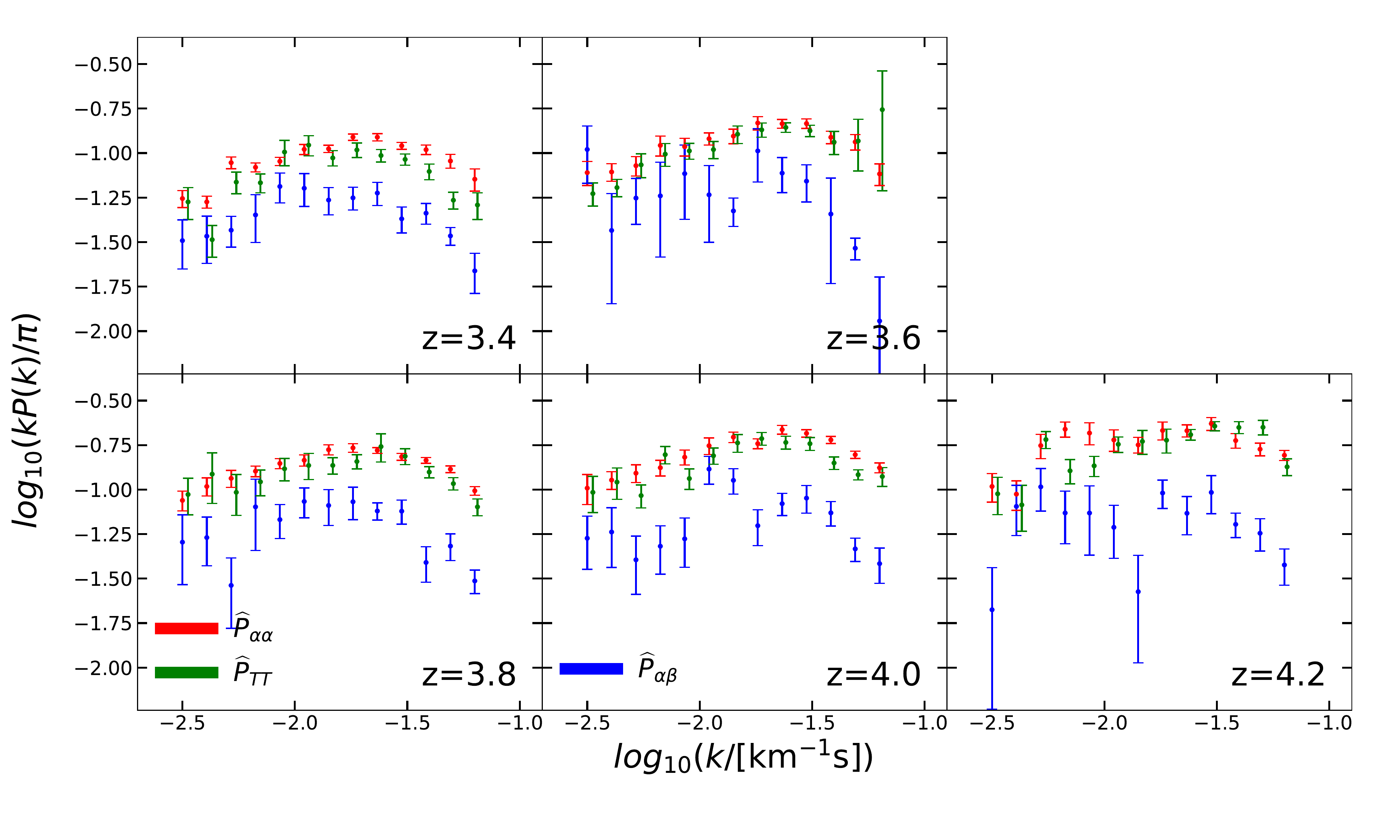}
\vspace{-.8cm}
\caption{The panels show our estimates for $\widehat{P}_{\alpha\alpha}$, $\widehat{P}_{TT}$, and $\widehat{P}_{\alpha T}$ ($=\widehat{P}_{\alpha \beta}$) in the five redshift bins.  Note that $\widehat{P}_{\alpha T}$ is equivalent to $\widehat{P}_{\alpha \beta}$ as the two Ly$\alpha$ forest segments are at such large distances that they essentially do not correlate.  The error bars do not include the allowance for 20\% uncertainty in the resolution parameter $\sigma_R$ and also the additional error for undersampling used in our final analysis (c.f.~\S~\ref{sec:Pk}).
 \label{fig:Pkest}}
\end{figure*}

\begin{figure*}
\includegraphics[scale=0.45]{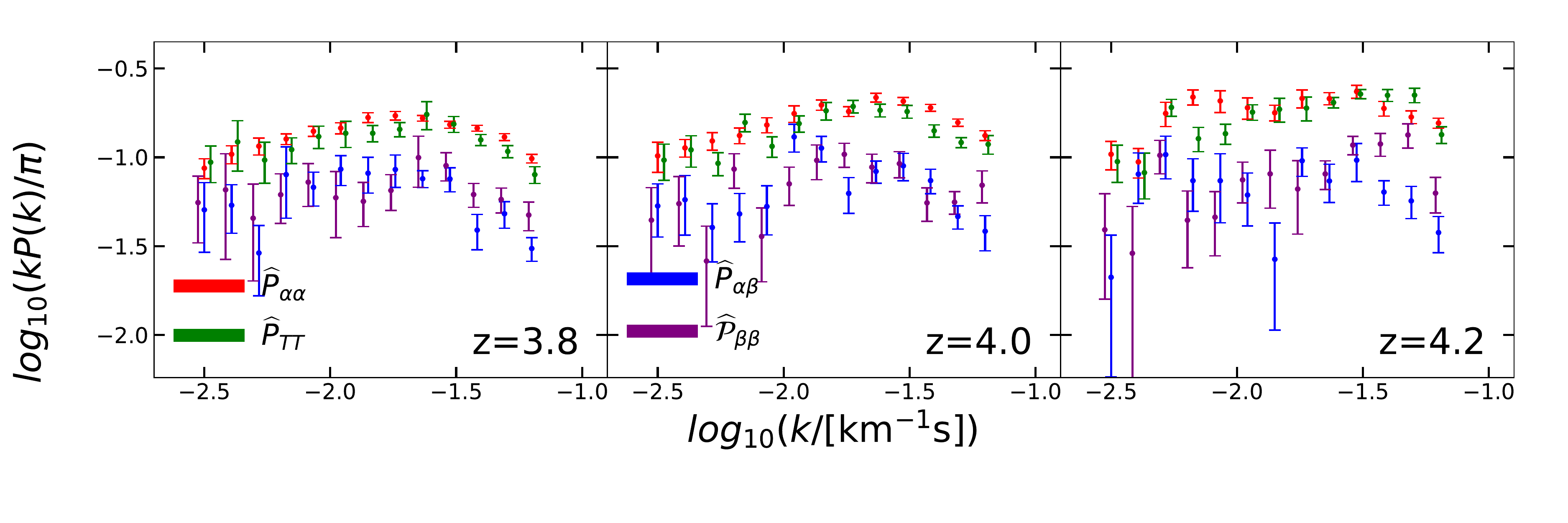}
\vspace{-.8cm}
\caption{The same as Fig.~\ref{fig:Pkest} but where we also show $\widehat{\cal P}_{\beta \beta} \equiv P_{TT}(z) - P_{\alpha\alpha}(z_f)$ where $z_f \equiv \lambda_\beta/\lambda_\alpha (1+ z)-1$. The three panels corresponds to the $z$ where $P_{\alpha\alpha}(z_f)$ is estimated so that this subtraction can be performed within our measurements.  Note that $\widehat{P}_{\beta \beta}$ is not exactly the \lyb\ auto-power spectrum as there is a convolution term that contributes at the $10\%$ level (c.f.~eqn.~\ref{eq:PTT})     
\label{fig:Pkestbeta}}
\end{figure*}

The flux power spectrum used in the analysis
has been calculated in five redshift bins, each with $\delta z = 0.2$, spanning $3.4<z<4.2$.  For each quasar spectra, we sort the data into redshift bins that corresponds to \lya\ and \lyb\ absorption in that redshift window.  Spectral segments are selected that fall within a given redshift bin are used for the power estimate in the bin. The flux in each segment is divided by our mean flux estimate, ${\bar F_X}$, to convert to an overdensity
\begin{equation}
\widehat{\delta}_{F_X}^s(\lambda, z_i)
  = \widehat{F}_X^s(\lambda)/{\bar F_X}(z_i) - 1, \label{eqn:deltaF}
\end{equation}
where $X = \{\alpha, T \}$ denotes which Lyman-series forest is measured and $s$ indexes the quasar spectrum.  Remember our convention that $T$ indicates the \lyb\ forest plus the foreground \lya\ that falls in the same spectral region.  The ${\bar F_X}(z_i)$ are our estimates for the mean flux presented in \ref{sec:meanF} (which are uncorrected for continuum bias and metal absorption as these then essentially cancel out in the computation of \ref{eqn:deltaF}).  Next, $\widehat{\delta}_{F_X}(\lambda, z_i)$, is Fourier transformed yielding $\widetilde{\delta}_{F_X}(k, z_i)$ and the auto and cross power spectrum is estimated as
\begin{eqnarray}
\widehat{P}_{XY}(k, z_j) &=& N_k^{-1} \sum_{\substack{|k_i - k| < \frac{\Delta k}{2} \\ s \in {\cal S}(k)}}  \Bigg\{\Big[\widetilde{\delta}^s_{F_X}(k_i, z_j) \widetilde{\delta}_{F_Y}^s(k_i, z_j)^*  L_s^{z_i}\nonumber \\ &-& P_{N, X}^s(k_i, z_j)\delta^{\rm K}_{X,Y} \Big]  W_{s,X}(k_i, z_j)^{-1}W_{s,Y}(k_i, z_j)^{-1}  \nonumber\\
 &-& \widehat{P}_{M, X}(k, z_j) \; \delta^{\rm K}_{X,Y}\Bigg\} ,
 \label{eqn:Pkest}
\end{eqnarray}
where the sum runs over all $N_k$ modes that fall in the band power in all segments that correspond to the desired redshift bin, ${\cal S}(k)$ is the sample of all quasar spectra where the effective resolution is reliably known to estimate the power at $k$, $L_s^{z_i}$ is the length of each spectral segment, $W_{s,X}$ are the kernels correcting for the effects of spectral resolution and pixel size, and $P_{N, X}^s$ ($\widehat{P}_{M, X}$) are the noise (estimated metal power).  The details of the resolution kernel and noise/metal power are described below. Since the noise between different Lyman series forests is uncorrelated, it only contributes to the auto power and, hence, the Kronecker delta function yields $\delta_{X,Y}^K = 1$ if ($X=Y$) and $0$ otherwise. Band power measurements are made in 13 logarithmic wavenumber bins, with bin centers spanning the range $-2.5\leq(k/$km$^{-1}$s $)\leq-1.2$. For cross power spectra ($X\neq Y$), the power spectra can be imaginary if translational invariance is broken, which can occur because of resonant metal contamination or because of imperfect wavelength calibration.  Appendix~\ref{ap:calibration} uses the value of the imaginary to test the wavelength calibration. We finally note that the minimum variance estimator would weight by the signal-to-noise squared, but since the statistics at all wavenumbers we report are limited by sample variance rather than detector noise, the above estimator should essentially be minimum variance. In what follows, we provide details regarding the treatment of noise, metal absorption, instrumental resolution, and wavelength calibration.\\

{\bf \noindent noise:} For the noise power spectrum $P_{N, X}^s$ in eqn.~(\ref{eqn:Pkest}), we assume it is white such that $P_{N,X}^s(k, z) = \sigma_s(z)^2 w_X(k, z)^2$, where  $\sigma_s(z)^2$ is the noise variance in each spectral segment for quasar $s$ in redshift bin $z$ and $w_X(k, z) \equiv  {\rm sinc}\left(k  \Delta v_X^z/2\right)$ owes to the boxcar spectral bins with velocity width $\Delta v_X^z$. The noise power $P_{N,X}^s$ is at least two orders of magnitude below the \lya\ power spectrum signal \citep{irsic17}, and there is no noise correction for our cross power measurement.   We have also tested a correction that accounts for spatial inhomogeneities in the noise and concluded that the associated correction would be negligible.  \\

{\bf \noindent metals:} Our measurements are at sufficiently high redshifts that the contamination from metal absorbers is a percent-level correction to the total power, a correction well below our quoted error bars.  Furthermore, the \lya\--\lyb\ cross power is unbiased by non-resonant metal absorption.  Nevertheless, for the auto power, we do correct for non-resonant metals using the standard procedure of using the absorption redward of the forest in our spectra to subtract their contribution as represented in eqn.~(\ref{eqn:Pkest}).  The Kronecker delta-function that multiplies the metal power $\widehat{P}_M(k_i, z_j)$ in that equation is only nonzero for the auto power spectra. For the power spectrum of the metals $\widehat{P}_M(k_i, z_j)$, we use the measurement of \citet{irsic17} with the same XQ-100 dataset from the red-side power spectrum, which decreases our auto power by a few percent (Appendix~\ref{ap:metals}).  We further find that the contamination from the resonantly enhanced metals, defined by that they fall near our \lya\ and \lyb\ (namely \OVI\ $\lambda,\lambda$1032, 1038\AA\ and \SiIII\ $1207$\AA), is at a similar level, and we do not correct for resonant metals.  See Appendix~\ref{ap:metals} for additional details.   \\  

{\bf \noindent resolution:} The correction for spectral resolution is  complex as slit spectrographs have resolutions that depend on the illumination of the slit and, hence, the seeing of individual observations.  Yet, for IGM thermal constraints, it is advantageous to use as high a wavenumber as possible, but high wavenumbers are also the most sensitive to uncertainties in the spectral resolution.  This issue has led to there being debate over the reliability of XQ-100 inferences from the \lya\ forest power spectrum, especially inferences from the lower resolution UV arm \citep{irsic17,yeche17,  walther19}.  Here we develop a better model for the seeing-dependent resolution of XQ-100.  Our approach suggests that some of the previous discrepancy in the quoted X-Shooter resolution may owe to the functional form that was assumed for the line spread function (Appendix~\ref{ap:resolution}). 

We model the line spread function $W_s(k_i, z_i)$ as a Gaussian multiplied by a sinc function:  
\be
W_{s, X} = e^{-\frac{1}{2}k^2 \sigma_R(X, s, z)^2} w_X(k,  z),
\label{eqn:WSX}
\ee
where $w_X(k, z) = {\rm sinc}\left(k  \Delta v_X^z/2\right)$ with $v_X^z = 11 ~(20)$km/s for wavelengths that fall on the visible (ultraviolet) arm. The Gaussian kernel that multiplies $w_X$ approximates the Fourier transform of the X-Shooter line spread function. 
In Appendix~\ref{ap:resolution}, we show that really the line spread function is better modeled as a Gaussian convolved with a tophat function.  We calibrate this model off of arc-lamp spectra.  However, we find that the Fourier transform of our model to yield $W_{s, X}$ is well approximated over the measured wavenumber range by a Gaussian with standard deviation $\sigma_R(X,s,z)$. 

Our model for $\sigma_R$ depends on the seeing conditions for each spectrum.  Among the exposures combined for a single quasar spectrum, the seeing can vary significantly (although for $80\%$ of the quasars the FWHM of seeing varies by $<0.2''$ over the exposures).  We estimate the minimum and maximum of  $\sigma_R$ using the minimum and maximum seeing of the exposures on an individual quasar.    We discard modes measured from spectra where using $\sigma_R^{\rm min}$ rather than $\sigma_R^{\rm max}$ to estimate the $W_{s,X}W_{s,Y}$ would lead to a 10\% difference in the estimated power, i.e. we discard modes for which
\begin{eqnarray}
 \left [ \sigma_{R}^{\rm max}(X, s, z)^2 + \sigma_{R}^{\rm max}(Y, s, z)^2 \right ]-&& \label{eqn:seeing} \\   \left [\sigma_{R}^{\rm min}(X, s, z)^2 + \sigma_{R}^{\rm min}(Y, s, z)^2 \right] &>& 0.1  \; k^{-2}.\nonumber
\end{eqnarray}
This selection criteria is combined with a 20\% allowance for uncertainty in the mean $\sigma_R$ when constraining thermal parameters in \S~\ref{sec:thermalhist}.\\

{\bf \noindent wavelength calibration:}
 Our \lyb\ power spectrum require wavelengths to be calibrated to the accuracy of $10$\, ($7$)\;km~s$^{-1}$ in order to make a 20 (10)\% error at the bin center of the maximum wavenumber we report, $k=10^{-1.2}$\;s~km$^{-1}$. 
  These errors would be approximately halved in our next-to-largest wavenumber bin.  These numbers hold both if the wavelength calibration is systematically offset or Gaussian random between sightlines with the accuracy quoted above being the standard deviation (Appendix~\ref{ap:calibration}).
 
  The wavelength calibration of our XQ-100 data set is done using skylines, first calibrated on a master integration and then adjusted for each exposure, and interpolating using the pipeline used by the XQ-100 team \citep{lopez16}. Tests show that the precision is likely better than 5km~s$^{-1}$, with the dominant error being interpolation and being more significant for the UV arm where there are fewer skylines (George Becker, private communication).  There can also be offsets owing to the positioning of the source within the slit for each arm; any offsets from center would result in a shift for that arm. The quoted VLT/X-Shooter precision of the alignment of the arms indicates that such offsets are likely controlled to a few km~s$^{-1}$ \footnote{\url{ https://www.eso.org/sci/facilities/paranal/instruments/xshooter/doc/XS_wlc_shift_150615.pdf }} and, if correct, such offsets would not be a significant systematic.\footnote{Another potential wavelength offset occurs owing to fine structure of the transitions shifts the wavelengths by $\Delta \lambda_\alpha=0.006$\AA\ ($1.5$~km~s$^{-1}$) and $\Delta \lambda_\beta=0.002$\AA\ ($0.6$ km~s$^{-1}$), which are insignificant. We use $\lambda_\alpha=1215.67$\AA\ and $\lambda_\beta=1025.72$\AA\ as our mean vacuum wavelengths.}

We can test that the wavelength calibrations likely meet the required calibration level by using the imaginary component of the \lya-\lyb\ cross power spectrum (Appendix~\ref{ap:calibration}).  Our measurement of this statistic limits any systematic offset to $\lesssim 10$~km~s$^{-1}$, with some evidence for an offset at this level at $z=3.8$ and $z=4.0$.  We have rerun our analysis presented in \S~\ref{sec:results} with a 5~km~s$^{-1}$ correction to offset this apparent shift, as the error is quadratic in this shift and so this splits the difference, and find our results for temperature only change at the $\sim 0.1 \sigma$ level.  Part of this insensitivity is because of our large allowance for resolution error also effectively increases the variance at the highest wavenumbers that are affected by such a shift.  The imaginary component of the cross power is less sensitive to a positive and negative offsets that are random between each spectra, with our measurements of the imaginary power suggesting that the random offsets have RMS of $<15$~km~s$^{-1}$.

Seven of the XQ-100 quasars were observed without their atmospheric dispersion corrector, which corrects for differential atmospheric refraction.  A lack of correction could result in larger offsets, particular for objects observed with large zenith angles.  We have done the analysis with and without these quasars included and find negligible differences.\\

\noindent {\bf masks:} After masking to account for our restframe wavelength cuts and DLA contamination, the minimum contiguous number of pixels in a segment that we still use to make a power spectrum measurement is 100 pixels.  A minimum allowed segment reduces the extra power at high wavenumbers that owes to the discontinuities at the end of each segment.  We wrote a lognormal mocks code to understand what biases result from this cut (and from the nonperiodicity of the segment at the edge), finding that even if all of our segments are at the 100 pixel threshold, the biases are negligible:  $\lesssim 3$\% for the UVB arm and double this bias for the VIS where $100$ pixels corresponds to a shorter pathlength. 
In our data, only $\sim 10\%$ of segments tend to fall within a factor of two of this minimum pixel threshold and so these percentiles significantly overestimate the effect.\footnote{This bias can can be substantial at higher wavenumbers that are used in high-resolution data sets.}\\

\begin{figure*}
\includegraphics[scale=0.45]{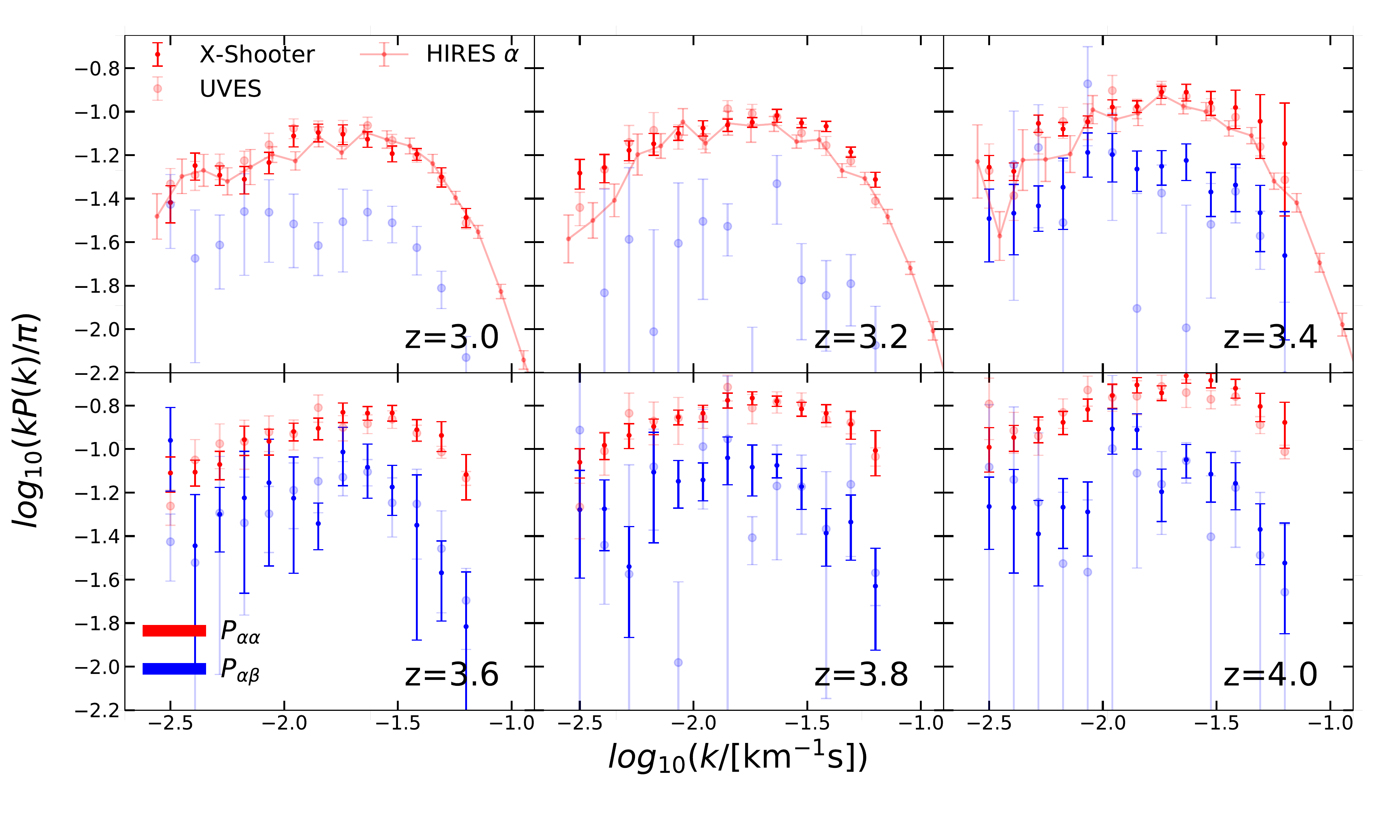}
\vspace{-.8cm}
\caption{A comparison of our XQ-100 power spectrum measurements with the Keck/HIRES $P_{\alpha\alpha}$ measurement of \citet{walther18} at $z=3.0, 3.2$ and $3.4$ and our measurement of $P_{\alpha\alpha}$ and $P_{\alpha\beta}$ using VLT/UVES archival data in all the redshift bins \citep{irsicinprep}. The systematic error for resolution has been added to the X-Shooter measurements, except at $z=3.0$ and $z=3.2$ $P_{\alpha \alpha}$, which are not used in our thermal parameter analysis.
\label{fig:Pkcomparison}}
\end{figure*}

{\bf \noindent covariance matrix:} We estimate the measurement covariance matrix by first bootstrap sampling our mocks to measure the covariance cross correlation coefficient matrix
\begin{equation}
    \widehat{C}_{ij} = \left \langle \left( \widehat{P_i} - \langle {\widehat{P}_i} \rangle  \right) \left( \widehat{P}_j - {\langle \widehat{P}_j \rangle} \right) \right\rangle,
\end{equation}
where $i,~j$ enumerate both the types of power spectra, $P_{\alpha \alpha}, P_{TT}, P_{\alpha T}$, the redshift bins, and the $k$-bin.  We use the $3000$ mock spectra to estimate the full covariance, denoting this as $\widehat{C}_{ij}^{\rm M}$, although we also will use the diagonals of the covariance matrix measured with XQ-100, which we denote as $\widehat{C}_{ii}^{\rm XS}$.  We further set correlations between different redshift bins to zero as these should be nearly zero. Because our mocks use the same skewers through different times in the simulations, there are spurious correlations that this step suppresses.  
We evaluate the correlation coefficient from the mock sample
\begin{equation}
\widehat{r}_{ij}^M \equiv \frac{\widehat{C}_{ij}^M}{\sqrt{\widehat{C}_{ii}^M \widehat{C}_{jj}^M}},
\end{equation}
which carries the information about the off-diagonal structure in the covariance matrix. As the last step we rescale the correlation coefficient from the mocks by the diagonal elements of $\widehat{C}_{ii}^{\rm XS}$, as measured on the XQ-100 data.  This rescaling allows us to potentially capture additional variance that is not in our mocks (such as from e.g. continuum fitting errors or large-scale modes not in our simulations) and also to likely take out some of the model dependence of the mock covariance matrix. Since $\widehat{C}_{ii}^{\rm XS}$ is measured on far fewer sightlines than the mock covariance matrix, there is roughly 10\% scatter around the mean relation from the mocks (re-scaled to the same path-length). In the subsequent MCMC analysis we have tested that replacing the $\widehat{C}_{ii}^{\rm XS}$ with the mean values from the re-scaled mock sample does not impact the conclusions of this paper. Due to re-sampling from the fixed pool of sightlines the bootstrap method can underestimate the variance \citep{rollinde13,viel13,irsic17}. This effect is more severe for the redshift bins with shorter pathlength. To correct for that effect, we multiply the full covariance matrix by a factor of 
\begin{equation}
\widehat{C}_{ij}^2 = [\widehat{C}_{ij}^{\rm XS}]^2 \left( 1 + 0.3\sqrt{\frac{\Delta x_{\rm max}}{\Delta x(z_i)}} \right) \left( 1 + 0.3\sqrt{\frac{\Delta x_{\rm max}}{\Delta x(z_j)}} \right),  
\label{eq:perr}
\end{equation}
where $\Delta x(z_{i,j})$ denotes the pathlength for the type of power spectrum ($P_{\alpha\alpha}$,$P_{\alpha T}$,$P_{\rm TT}$), index ${i,j}$ corresponds to the redshift bin, and $\Delta x_{\rm max}$ is the maximum pathlength in the sample of our redshift bins and spectra. The pathlengths used in this rescaling correspond to the values in Fig.~\ref{fig:Nz}, with $\Delta x_{\rm max}$ corresponding to the pathlength of \lya\ forest at $z=3.4$. This procedure effectively boosts the covariance matrix by an average factor of $1.44$ ($1.9$) for the covariance corresponding to $P_{\alpha\alpha}$ ($P_{\alpha T}$), with larger boosts for smaller pathlengths. This corresponds to the boost in the the power spectrum errors by roughly $20$\% ($40$\%) on average for the $P_{\alpha\alpha}$ ($P_{\alpha T}$).  The amplitude of $0.3$ is chosen to reproduce the typical value of $0.44$ in the XQ-100 \lya\ analysis of \citet{irsic13} and the functional form for this correction is motivated by a sampling argument.\footnote{The boost aims to account for that our estimate for the covariance matrix diagonals should have an error that scales as the inverse of the number of samples, which we assume is proportional to the path length $\Delta x(z)$.  Eqn.~\ref{eq:perr} is adding back this typical error so that we are unlikely to substantially underestimate the covariance in any redshift bin.}   We note that our results are not significantly changed if we instead use a constant $(1+ 0.44)^2$ enhancement over all redshifts as in \citet{irsic13} rather than the terms in parentheses in eqn.~(\ref{eq:perr}).

In the MCMC analysis we also include a systematic error budget owing primarily to uncertainty in the resolution measurement. The systematic error is modeled as uncorrelated -- contributing in quadrature only to diagonal elements of the covariance matrix.\footnote{Generally one would expect that the resolution uncertainty induces correlated systematic uncertainty across the sightlines, as the uncertainty is fixed per wavelength calibration and thus the same for all sightlines. This is further complicated in the presence of seeing corrections that could add sightline-to-sightline variations. While these effects are accounted for in the mean measurement, a simplistic model is sufficient for the covariance matrix that we use in the MCMC analysis of this paper.} In this simple model the resolution uncertainty in UVB ($\Delta \sigma^{\rm UVB}_R/\sigma^{\rm UVB}_R$) and VIS ($\Delta \sigma^{\rm VIS}_R/\sigma^{\rm VIS}_R$) arms are propagated to the power spectrum measurement between X and Y fields as
\begin{equation}
    \left(\frac{\sigma_{P_{\rm XY}}(k,z_j)}{\widehat{P}_{\rm XY}(k,z_j)}\right)^2 = \sum_{W\in\{X,Y\}} \left.\frac{\Delta \sigma_R}{\sigma_R}\right|_{z_j^W} \left( k\, \sigma_R(z_j^W) \right)^2,
\end{equation}
with the sum accounting for the contributions of both fields $X$ and $Y$. In the case of auto-power ($X=Y$) the two terms in the sum are identical. Each of the parts in the sum depends on the {\it observed wavelength} range for that transition, as the distinction between UVB and VIS arms is in the frame of the spectrograph. Thus each of the arms covers a range of redshifts of absorptions that depends on the transition. For $W = \alpha$ the redshifts are already expressed for the \lya\ transition, and so $z_j^W = z_j$. For the case of $W = T$ the redshifts have to be shifted by the ratio of the wavelengths of the \lyb\ and \lya\ transitions, and thus $1+z_j^W = (1+z_j)\lambda_{\beta}/\lambda_{\alpha}$. We then imposed the condition that if $z_j^W < 3.6$ the contribution is coming from the UVB arm. This is exact, as in our measurement of the $z=3.6$ bin we have only include data from the VIS arm (\S~\ref{sec:analysis}).  For our analysis, we take $\Delta \sigma^{\rm UVB}_R/\sigma^{\rm UVB}_R =  \Delta \sigma^{\rm VIS}_R/\sigma^{\rm VIS}_R = 0.2$, as motivated earlier.

\subsection{Measurement}
Figure~\ref{fig:Pkest} shows the resulting power spectrum estimates,  with the diagonal errors computed from bootstrapping the data in the five redshift bins in which the \lyb\ forest can be measured.  
  For our three redshift bins with $z\geq3.8$, we are able  to isolate the \lyb\ auto power spectrum as we are able to estimate and then subtract the foreground \lya\ power from $\widehat{P}_{TT}$.  These estimates, which are the purple errors in Figure~\ref{fig:Pkestbeta}, are not exactly the \lyb\ auto spectrum $\widehat{P}_{\beta \beta}$ as $ {P}_{T T}$ is not just a sum of ${P}_{\alpha\alpha}$ and ${P}_{\beta \beta}$ but rather given by
\begin{equation}
 {P}_{T T}  = {P}_{\alpha\alpha}(k,z_f) +\overbrace{{P}_{\beta \beta}(k, z) + \int \frac{dk'}{2\pi} {P}_{\beta \beta}(k',z) {P}_{\alpha\alpha}(k-k',z_f)}^{{\cal P}_{\beta \beta}},\label{eq:PTT}
\end{equation}
where $z_f \equiv \lambda_\beta/\lambda_\alpha (1+z) -1$.  Thus, we denote what we measure when we subtract ${P}_{\alpha\alpha}$ as ${\cal P}_{\beta \beta}$, which is the quantity indicated by the overbrace in the previous formula.  
A rough estimate for the size of $P_{\beta\beta}$-contaminating convolution term is $\sigma_\alpha^2 {P}_{\beta \beta}(k, z)$ where $\sigma_\alpha^2\sim k P_{\alpha \alpha}(k)/\pi \sim 0.1$ approximates the variance in the \lya\ forest over our surveyed redshifts.  (We have done full calculations that verify this estimate.)  This additional contribution needs to be modeled for accurate inference from ${\cal P}_{\beta \beta}$, although we suspect generally it will be more useful to forward model $P_{T T}$.  As our measurement of $P_{\alpha\beta}$ is more constraining, we do not use ${\cal P}_{\beta \beta}$ nor $P_{T T}$ in our analysis to estimate the thermal properties of the IGM (\S~\ref{sec:interpretation}).  Note that ${P}_{\alpha T}$ is equivalent to ${P}_{\alpha \beta}$ as the two Ly$\alpha$ forest segments are at such large distances that they essentially do not correlate, and so we subsequently refer to this cross power as ${P}_{\alpha \beta}$.

Figure~\ref{fig:Pkcomparison} compares our measurement to the Keck/HIRES $P_{\alpha\alpha}$ measurement of \citet{walther18} using the KODIAQ reductions \citep{2015AJ....150..111O} and a preliminary measurement of $P_{\alpha\alpha}$, $P_{\alpha\beta}$ ${P}_{TT}$ using the SQUAD DR1 reductions of VLT/UVES archival data \citep{2019MNRAS.482.3458M}.  Both the HIRES and UVES datasets are at significantly higher resolution (with $\langle {\cal R} \rangle \sim 50,000$) than our XQ-100 data, which makes the resolution correction essentially negligible over the wavenumbers we report for XQ-100 data. \footnote{The purpose of including the preliminary VLT/UVES measurement in this paper is to evaluate the spectral resolution correction for the XQ-100 data. The joint analysis between VLT/UVES and XQ-100 data is left for a follow-up paper.}. We first concentrate on the comparison with the \citet{walther18} Keck/HIRES measurement of $P_{\alpha\alpha}$.  The top panels show the power at $z=3.0, 3.2$ and $3.4$.  The former two redshifts are lower than the redshifts where we measure Ly$\beta$ power, but they overlap with this Keck analysis. Our Ly$\alpha$ power spectrum measurement agrees better with this Keck/HIRES measurement in the first two redshift bins, but overshoots in the $z=3.4$ bin at the highest wavenumbers.  A similar overshoot has been seen before when comparing with the XQ-100 measurement of \citet{irsic17}, with \citet{walther18} contending that it owed to \citet{irsic17} using a value of $\sigma_R$ that is too small. While our resolution correction is different than \citet{irsic17}, the results are not dissimilar.  In our analysis (and the error bars shown here), we allow for a $20\%$ uncertainty in the resolution parameter $\sigma_R$ that mitigates this discrepancy such that the overshoot is now within the error bar.  Note that this error is particularly significant for the lower resolution UV arm that is used for the $z=3.4$ $P_{\alpha \alpha}$ measurement.

Next, let us compare with our preliminary measurement of the VLT/UVES SQUAD data set.  This was analyzed using a slightly adapted pipeline as that used here for XQ-100.  We further chose a minimum continuum-to-noise of $20$ in a $2.5$km/s pixel at $5500$~\AA.  This resulted in our UVES analysis considering 62 $z>3$ quasars, about half of the full SQUAD sample at these redshifts.  In future work, we intend to use lower quality spectra to improve the S/N, and more thoroughly investigate systematics that crop up especially in the lower quality spectra \citep{irsicinprep}.  However, one can see that our UVES measurement is less precise at higher redshifts for estimating the cross power; we do not expect this property to change.  Rather, the VLT/UVES data set excels at lower redshifts. This figure also presents $z=3$ and $z=3.2$ cross power estimates with VLT/UVES, lower redshifts than where this can be measured with X-Shooter.  By and large, the VLT/UVES measurements appear consistent with X-Shooter measurements.  We take this agreement as further support for our new X-Shooter resolution correction.

\section{Thermal history estimation}
\label{sec:thermalhist}\label{sec:interpretation}

While the temperature at the mean gas density has been routinely measured, more previous measurements of $P_{\alpha\alpha}$ were not sensitive enough to draw conclusions on the slope of the temperature density relation $\gamma-1 \equiv d\ln{T}/d\ln{\rho}$. This stems from the fact that $P_{\alpha\alpha}$ is sensitive to a fairly narrow range of gas densities, and effectively traces the temperature at one characteristic gas overdensity \citep{becker11}. To the extent that it does not probe a range of overdensities, the measurement of the temperature is not sensitive to the trend with density  \citep{lidz10, becker11,2011MNRAS.415..977M,irsic18}. To alleviate this problem, several suggestions have been put forward including: (a) other statistics than the two point function of the \lya\ forest \citep{dijkstra04,bolton14,hiss17,telikova19,gaikwad20} or (b) bluer Lyman series transitions \citep{dijkstra04,irsic13,irsic14b,boera16}. We follow the latter, presenting the first measurements of the IGM thermal history from the \lyb\ forest power spectra and its cross with \lya.

\begin{figure*}
\includegraphics[scale=0.55]{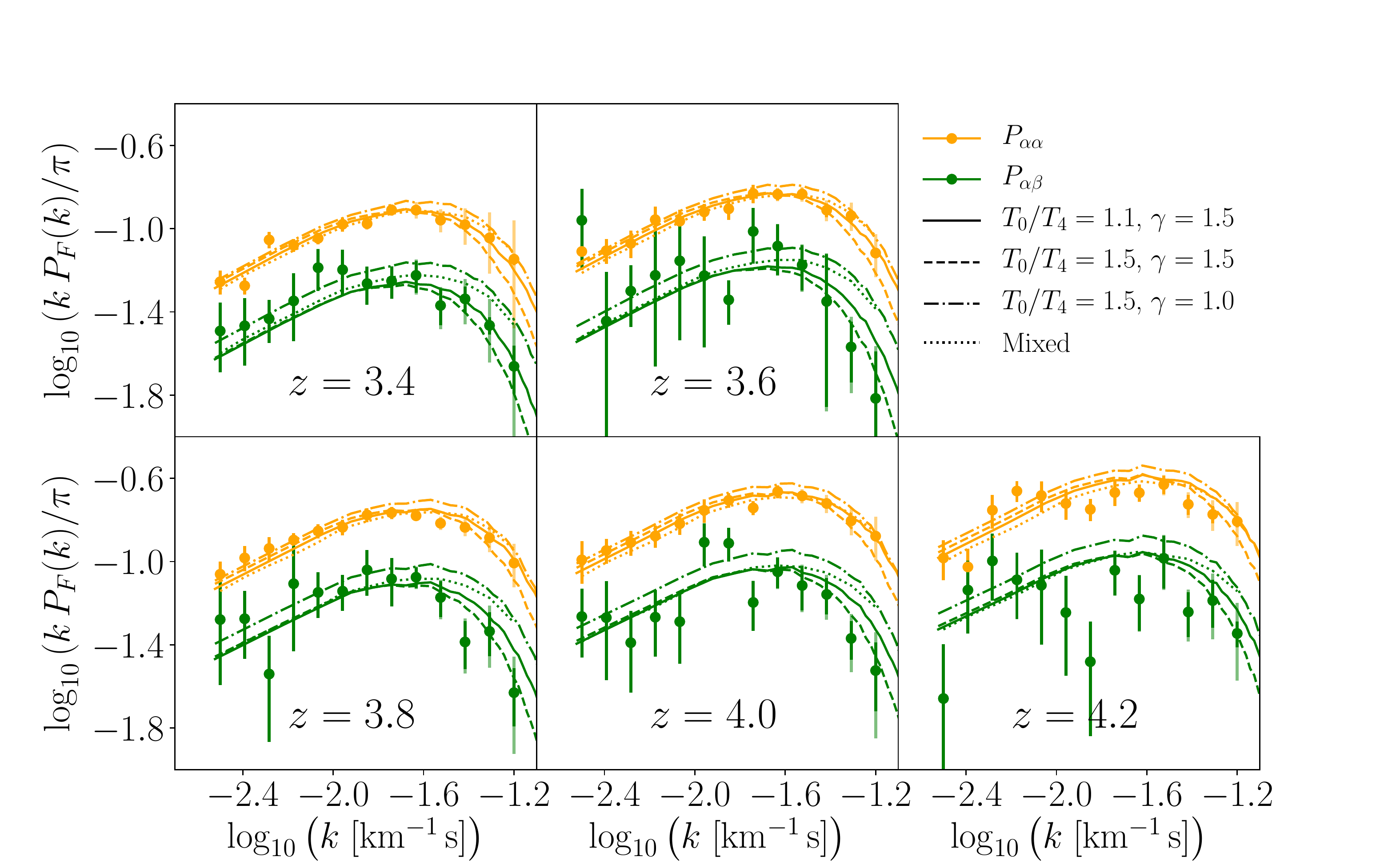}
\caption{Model predictions (and not best fit models!) for several temperature-density relations (curves) alongside our XQ-100 measurement (points with errorbars).  Three of the models show power-law temperature-density relations specified by $T_0/T_4,\gamma$, where $T_4 \equiv 10^4$K.  In addition, a `Mixed' model where the IGM is half filled with gas at $(T_0/T_4,\gamma) = (0.7, 1.5)$ and half with $(T_0/T_4,\gamma) = (1.5, 1.0)$, which exaggerates the features anticipated for midway through \HeII\ reionization. That the differences between the various models are larger in $P_{\alpha \beta}$ compared to $P_{\alpha \alpha}$ illustrates how our measurement of $P_{\alpha \beta}$ has the potential to break degeneracies in the thermal model.
}\label{fig:pk_models}
\end{figure*}

\subsection{Simulations}

In our Bayesian analysis, we sample the parameter space $\left \{ {\bar F}(z_i),T_0(z_i),\gamma(z_i),z_{\rm rei},\sigma_8,n_{\rm eff} \right \}$, where $z_i$ goes over the observed redshift bins $(3.4,3.6,3.8,4.0,4.2)$, using Monte Carlo Markov Chain (MCMC) sampler based on the Metropolis-Hastings algorithm (e.g. \citealt{irsic17b}).  All parameters are computed self consistently in the simulation (and not in post processing), so effects like heating from structure formation shocks are captured. To compare the models to the measurement, we compute mock \lya\ forest absorption spectra from a suite of hydro-dynamical simulations and, then, compute  from each simulation their 1D flux power spectra, namely $P_{\alpha\alpha}$, $P_{\alpha\beta}$ and $P_{\beta\beta}$.

The grid of simulations used to cover the parameter space is described in detail in \citet{irsic17b}. The simulations were run using the Gadget-2 cosmology N-body+Smooth particle hydrodynamics code with $768^3$ dark matter particles and $768^3$ gas particles in a $20\mathrm{Mpc/h}$ comoving box.  Additional simulations were run using $2\times 1024^3$ particles in the same box size to calibrate a correction for the mass resolution. The resulting corrected power spectra are converged in both resolution and boxsize to within 5\% over the redshift and wavenumber range ($k < 0.1\skm$) considered in our measurements \citep{irsic17b}.

Fig.~\ref{fig:pk_models} showcases the power spectra computed from a sample of our simulations (curves), alongside our flux power spectra measurements (points with errorbars).  We caution however of the direct comparison here of models and data, as the models are not best fit and such parameters like mean flux have not been chosen carefully. The simulations span a large range of the IGM mean temperatures and temperature-density relations (see the $T_0$ and $\gamma$ labels), and the flux power spectra are sensitive to the differences between the model at highest wavenumbers ($k > 0.02\;\skm$). Moreover, the differences between the models are larger for the cross power spectra, $P_{\alpha \beta}$ compared to the \lya\ auto power spectrum $P_{\alpha \alpha}$. Therefore, the addition of \lyb\ power spectra measurements is likely to increase the sensitivity to the IGM parameters. We explicitly show this to be the case in the next section. 

Also shown in Fig.~\ref{fig:pk_models} is a ``Mixed'' case,  which combines two simulations so the power spectrum is computed so that half the skewers are through a simulation with $(T_0,\gamma) = (7000\;\mathrm{K}, 1.5)$ and the other half from a simulation with $T_0,\gamma) =(15000\;\mathrm{K}, 1.0)$.  This mixed case idealizes the situation that may be expected halfway through (inhomogeneous) \HeII\ reionization \citep{2011MNRAS.415..977M}.  
Qualitatively, the flattening does not appear consistent with our measurements of the cross power, but the mixed model likely exaggerates the actual picture and, again, this is not a best fit model.  Our best-fit models presented in \S~\ref{sec:results} are shifted somewhat down relative to the data relative to the models here.

\subsection{Results}
\label{sec:results}

Here we describe the IGM thermal parameter analysis that results from the MCMC analysis. The analysis fits for all the redshifts together, with three IGM parameters per redshift bin (${\bar F}$,$T_0$,$\gamma$) and three global parameters that are not changing with redshift: the redshift of instantaneous reionization $z_{\rm rei}$ in the simulations (which is a standard parameter in such analyses adopted as a proxy for the amount of Jeans smoothing of the gas by sound waves) and the two cosmological parameters $\sigma_8$ and $n_{\rm eff}$ (where the later non-standard parameter is the slope of the matter power spectrum at $k = 1 \;\mathrm{h/Mpc}$).  For all the results in this section we fix the cosmological parameters by imposing a tight Gaussian prior around the values as measured by Planck+2018 ($\sigma_8 = 0.811 \pm 0.006$ and $n_{\rm eff} = -2.30 \pm 0.005$). While these two cosmological parameters are marginalized over in our analysis, including them makes little difference in the final results. For the redshift of reionization, we adopt a flat prior over $6<z< 14$.  We generally find that this parameter is not well constrained, which is expected as our measurements are well after the end of reionization. 

Furthermore we have used Gaussian priors on the mean transmission in the \lya\ forest at each redshift centered at the measurement of \citet{becker13} with 5\% standard deviation.  This 5\% standard deviation is motivated by the differences we find when correcting and not correcting for continuum and larger than the differences between \citet{becker11} and our mean flux measurement (\S~\ref{sec:meanF}).  We note that the mean transmission in the \lyb\ forest is then a prediction of the simulations once calibrating to the \lya\ mean flux and so does not need to be modeled.  Similarly we have used a Gaussian prior on the IGM temperature around the results of \citet{becker11} and taking the reported $T_0$ values for $\gamma=1.3$ as the mean with standard deviation of $1500$ K as our fiducial value. As these are relatively tight priors based on prior measurements of $T_0$, we have also explored the effects of weakening the $T_0$ priors by increasing the standard deviation to $3000$ K (T3) and $6000$ K (T6), finding that our measurement of $\gamma$ is robust to these priors.

As described in Sec.~\ref{sec:powerspectrum} we have added a systematic error budget to the covariance matrix. The nominal values that we choose based on our estimated spectral resolution for the median seeing were $\sigma_R^{\rm UVB} = 18.3\;\mathrm{\kms}$ ($\Delta \sigma_R^{\rm UVB} = 3.7\;\mathrm{\kms}$) for the UVB arm, and $\sigma_R^{\rm VIS} = 12.0\;\mathrm{\kms}$ ($\Delta \sigma_R^{\rm VIS} = 2.40\;\mathrm{\kms}$) for the VIS arm. This corresponds to 20\% uncertainty in $\sigma_R$ in each of the spectral arms. This is perhaps on the conservative side, and we have tested that reducing the resolution error bar by half does not have a large impact on our conclusions. 

\begin{figure*}
\centering
\begin{subfigure}{.5\textwidth}
    \centering
    \includegraphics[scale=0.35]{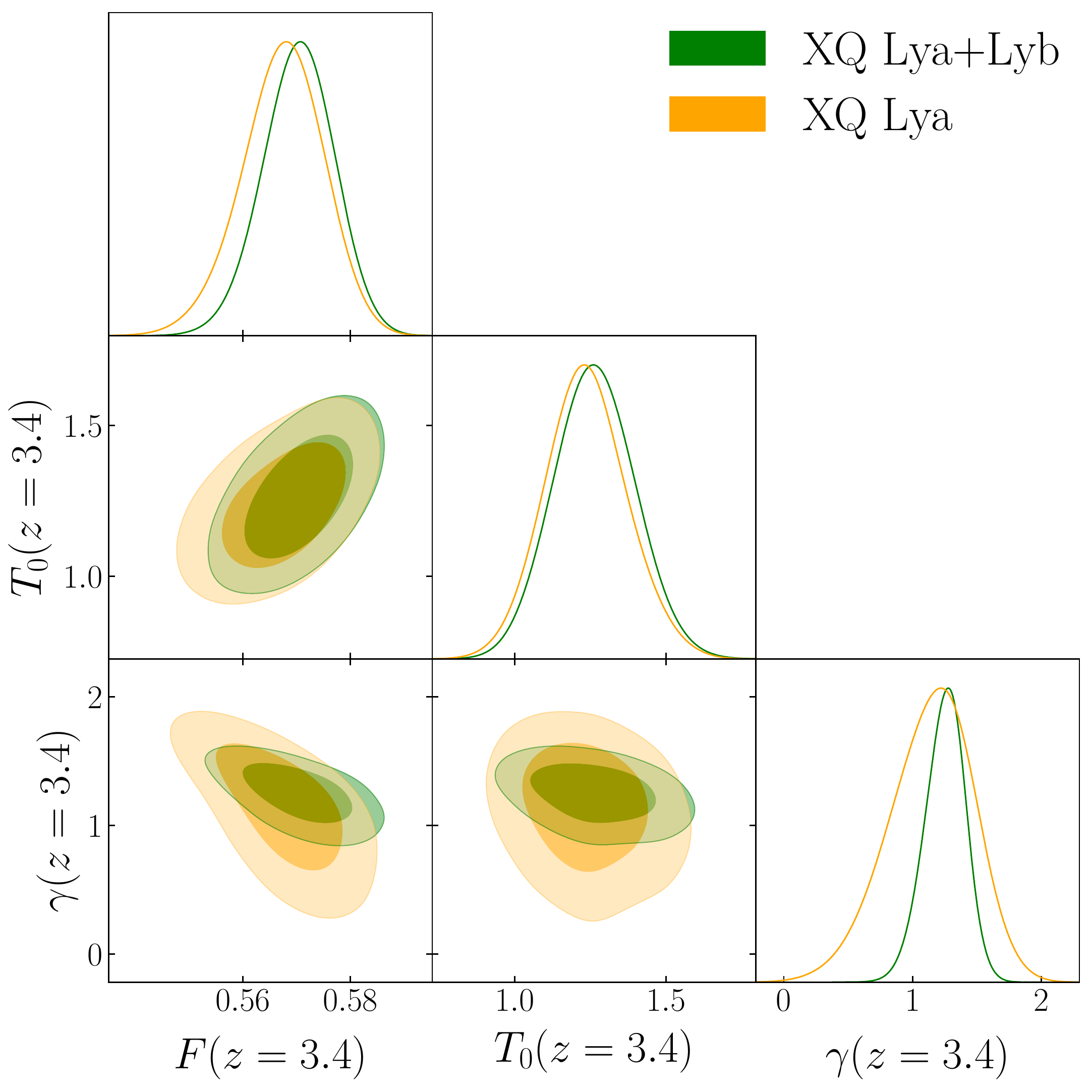}
    \caption{}
\end{subfigure}%
\begin{subfigure}{.5\textwidth}
    \centering
    \includegraphics[scale=0.43]{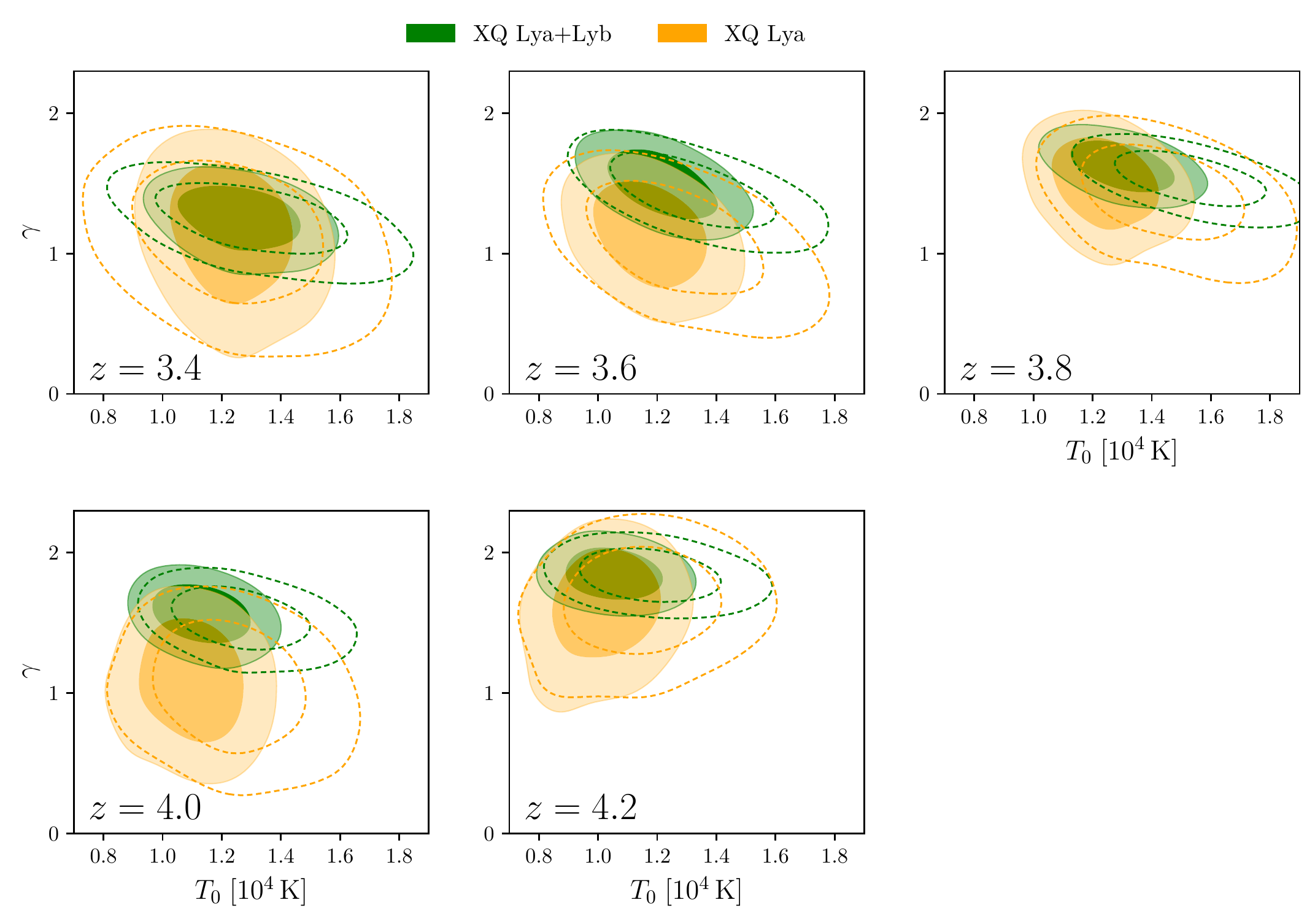}
    \caption{}
\end{subfigure}
\caption{{\it Left:} Contour plot for the thermal parameters at $z=3.4$ using our $\widehat{P}_{\alpha \alpha}$ estimates alone (orange) and these plus our $\widehat{P}_{\alpha \beta}$ estimates (green).  The constraint on $\gamma$ is especially improved by the inclusion of the cross power spectrum. {\it Right:} Contour plot in the $T_0-\gamma$ plane for all the measured redshifts. The colour scheme is the same as in the left panel, and $T_0$ is again in units of $10^4$K. The full contours are for the same analysis as in the left panel, while the dashed contours are showing the results when weaker $T_0$ priors were used in the analysis (T3; using $\Delta T_0 = 3,000\;\mathrm{K}$ as the standard deviation in our prior). The constraints on $T_0$ are largely set by the $T_0$ prior, but constraints on $\gamma$ are not driven by any $\gamma$ prior nor are they shaped by which $T_0$ prior is used.}
\label{fig:posterior}
\end{figure*}

We now present the results of this analysis.
As different redshifts are nearly independent, the degeneracy structure between the IGM parameters is similar in all the redshift bins.  To illustrate this structure, we first consider one redshift. The left panel of Fig.~\ref{fig:posterior} shows 2D posterior distributions for the IGM parameter at $z=3.4$, with the orange contours our \lya--only analysis and with the green contours showing our combined \lya\ and \lyb\ analysis (i.e. that uses also $P_{\alpha \beta}$ in addition to $P_{\alpha \alpha}$). As anticipated, adding the \lyb\ measurement helps principally to constrain $\gamma$, the power-law parameter of the temperature-density relation of the IGM. Adding the \lyb\ forest information does not particularly help in constraints on the IGM temperature nor the mean transmission, rather it appears to multiply our \lya--only posterior by a function $p(\gamma)$, thus acting to constrain only $\gamma$. This behavior is present in every redshift bin. This is illustrated in the right panel of Fig.~\ref{fig:posterior} that shows 2d posterior distributions in the plane of $T_0-\gamma$ for every redshift in our measurement. The solid contours in the panel are for our fiducial prior choice on $T_0$, whereas the empty, dashed contours are for a weaker prior choice on $T_0$, where the standard deviation on the parameter was $3,000$ K (T3). The primary effect of the weaker temperature prior is to broaden the posterior distribution in the temperature direction, without significantly affecting the width of the posterior in $\gamma$ direction.

\begin{figure*}
\centering
\begin{subfigure}{.5\textwidth}
    \centering
    \includegraphics[scale=0.3]{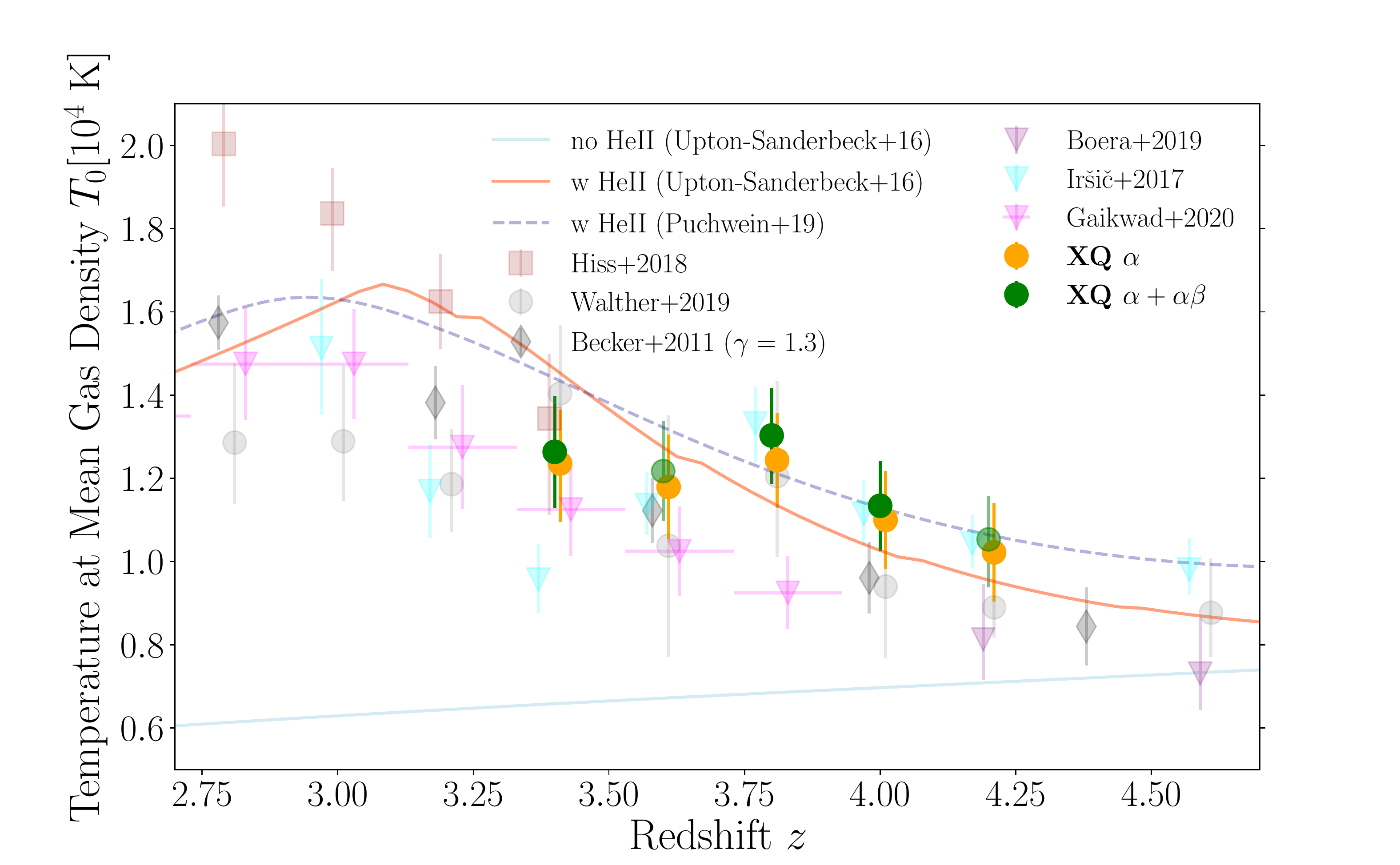}
    \caption{}
\end{subfigure}%
\begin{subfigure}{.5\textwidth}
    \centering
    \includegraphics[scale=0.3]{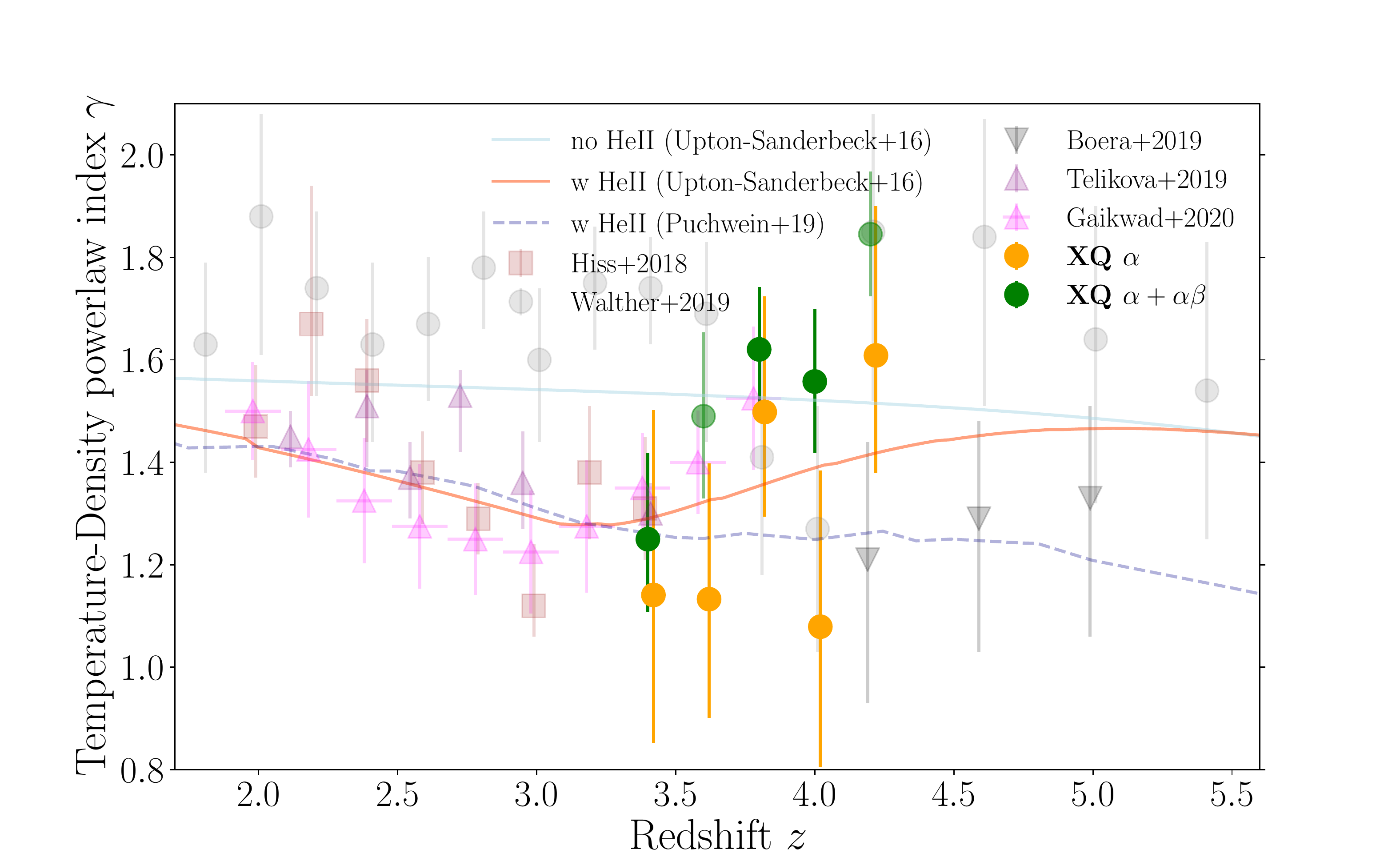}
    \caption{}
\end{subfigure}
\vskip\baselineskip
\begin{subfigure}{.5\textwidth}
    \centering
    \includegraphics[scale=0.3]{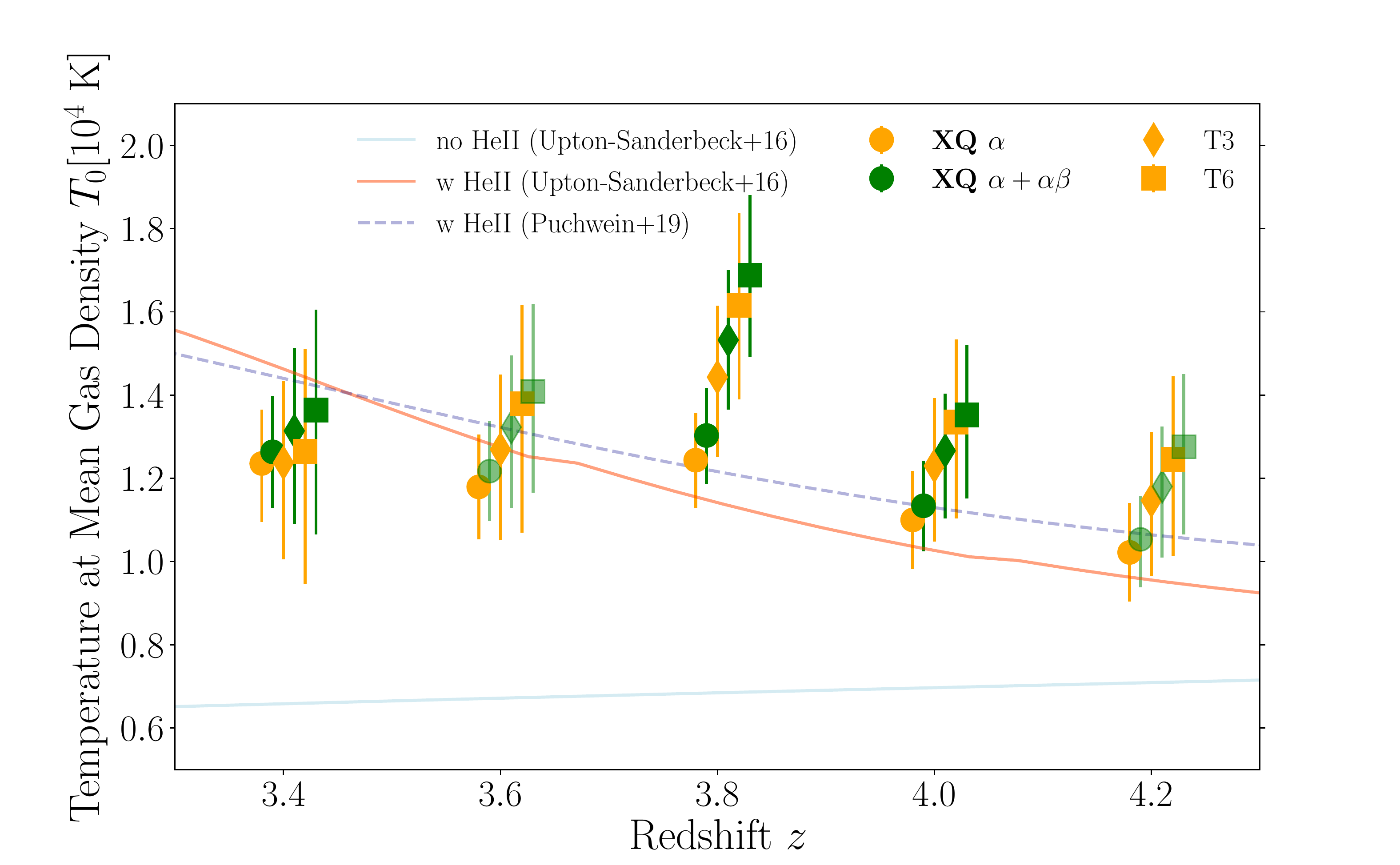}
    \caption{}
\end{subfigure}%
\begin{subfigure}{.5\textwidth}
    \centering
    \includegraphics[scale=0.3]{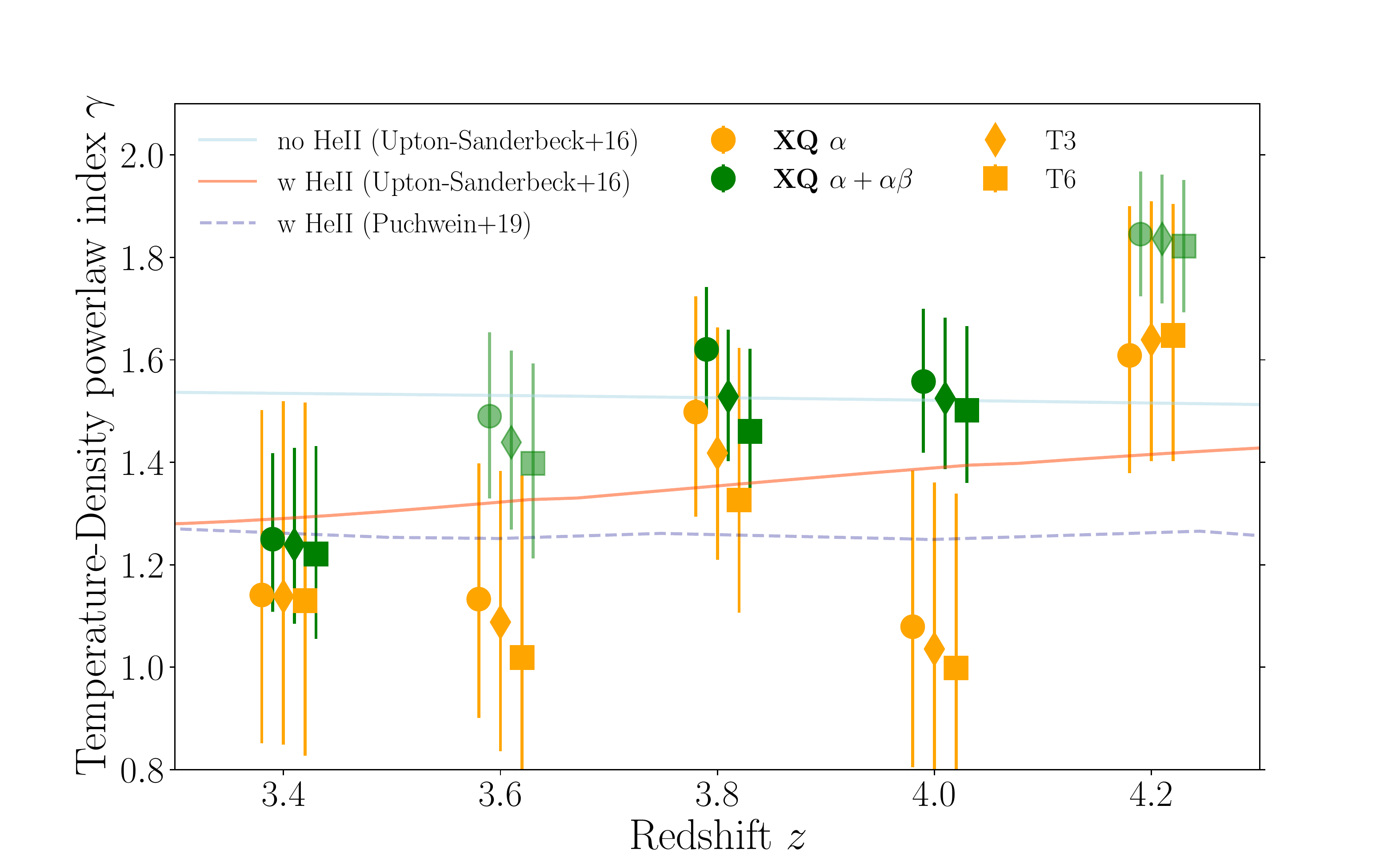}
    \caption{}
\end{subfigure}
\caption{Constraints on the temperature at mean density ($T_0$) and the power-law index of the temperature-density relation ($\gamma-1$) of our work (green and orange) as well as other recent measurements.  The solid curves are the models of \citet{upton} with and and without \HeII\ reionization, and the dashed curves are models of \citet{puchwein19} with \HeII\ reionization. {\it Top:} Our measurements using a fiducial $T_0$ prior plotted against other measurements in the literature. We have indicated that our $z=3.6$ measurement of the cross and both measurements at $z=4.2$ used a much smaller pathlength than our other measurements by making them transparent. {\it Bottom:} Our measurements for different choices of $T_0$ prior, plotted only within the redshift range of our observations. The $T_0$ priors differ only in the value of the standard deviation: the fiducal assumes $1500$ K (circles); T3 assumes $3000$ K (diamonds); and T6 assumes $6000$ K (squares). The posterior on $\gamma(z_i)$ does not change significantly when relaxing the priors on $T_0$.
}
\label{fig:zbin_bestfit}
\end{figure*}

The recovered redshift evolution of the IGM temperature and the power-law index of the temperature-density relation are shown in Fig.~\ref{fig:zbin_bestfit}, highlighting the difference between the \lya--only (orange) and \lya + \lya\ $\times$ \lyb\ analysis (green). The top panels compare our measurements to other measurements in the literature, using the fiducial $T_0$ prior that is centered around the \citet{becker11} measurement. The addition of the cross power spectrum does not add much information to the measurement of the temperature, with the estimated errors on the best-fit $T_0$ improving by less than $\sim 5$\% between the \lya--only and joint \lya\ and \lyb\ analysis. The additional information in the \lyb\ forest is primarily in the $\gamma$ evolution. The improvement in the uncertainty on the recovered $\gamma$ parameters is, on average, a factor of two in joint \lya\ and \lyb\ analysis compared to the \lya--only analysis. The best-fit values for the temperature-density relation in the combined \lya\ analysis for the $1500$K $T_0$ prior case are $\gamma = (1.25_{-0.14}^{+0.17},1.49_{-0.16}^{+0.16},1.62_{-0.12}^{+0.12},1.56_{-0.14}^{+0.14},1.85_{-0.12}^{+0.12})$, with the values corresponding to the measured redshift bins $(3.4,3.6,3.8,4.0,4.2)$. Note that the central values reduce by $0.1-0.2$ at $z=3.6$ and $z=3.8$ when we reduce this strong prior on $T_0$.   The highest redshift bin at $z=4.2$ is inconsistent at $2\sigma$ with the theoretical upper limit of $1.6$ for models that assume the only source of heat is photoheating from a uniform background \citealt{hui03, mcquinn15}. As this is also one of the two redshift bins where we have a very short pathlength (and small number of quasars contributing; c.f.~Fig.~\ref{fig:Nz}), we are suspicious of our relatively small errorbar of $\pm0.12$ here.

The bottom panels of Fig.~\ref{fig:zbin_bestfit} highlight the robustness of the recovered $\gamma$ posteriors against the imposed priors on $T_0$. From the left bottom panel it is clear that the $T_0$ measurements from XQ-100 sample are heavily influenced by the choice of the $T_0$ prior, where our fiducial choice is a very tight prior with $1500$K standard deviation around \citet{becker11}. Increasing the standard deviation on the Gaussian priors by a factor of $2\times$ ($4\times$) increases the uncertainty on $T_0$ parameters by $1.5\times$ ($2.2\times$). The benefit of the joint \lya\ and \lyb\ analysis on the $T_0$ measurement is slightly more pronounced for weaker priors, with the improvement on the estimated temperature uncertainty being 9\% (15\%) for the T3 and T6 prior choices respectively. The XQ-100 power spectrum measurements do not competitively constrain $T_0$ relative to determinations from higher resolution data, and adding the \lyb\ forest is only of minor help.

Our measurement of $\gamma$ is very stable with respect to the choice of $T_0$ prior and, when including the cross power, is competitive with previous constraints. 
The bottom right panel of Fig.~\ref{fig:zbin_bestfit} shows that the recovered $\gamma$ values change very little between the different prior choices. This is true for both the \lya--only and joint \lya\ and \lyb\ analyses, with the uncertainty in $\gamma$ increasing by 4\% (18\%) and 6\% (19\%) respectively, when considering the T3 (T6) prior choice.  In addition, the errors are significantly reduced when including the cross. The reason why the cross power is able to competitively constrain $\gamma$ is illustrated in Fig.~\ref{fig:pk_models}.  The effect of different possible $\gamma$ have more effect at high $k$ than allowed variations in $T_0$.  The effect on the power of changing $\gamma$ by $0.5$, a change that is inline with current scatter in $\gamma$ values, is shown by the dot dashed and dashed curves in this figure.  This effect on the cross power is much larger than that of changing $T_0$ by $4000$K but fixing $\gamma$ (compare the solid and dashed curves), a change in $T_0$ that is significantly larger than the reported errors on this parameter in the literature.

\begin{figure*}
\includegraphics[scale=0.45]{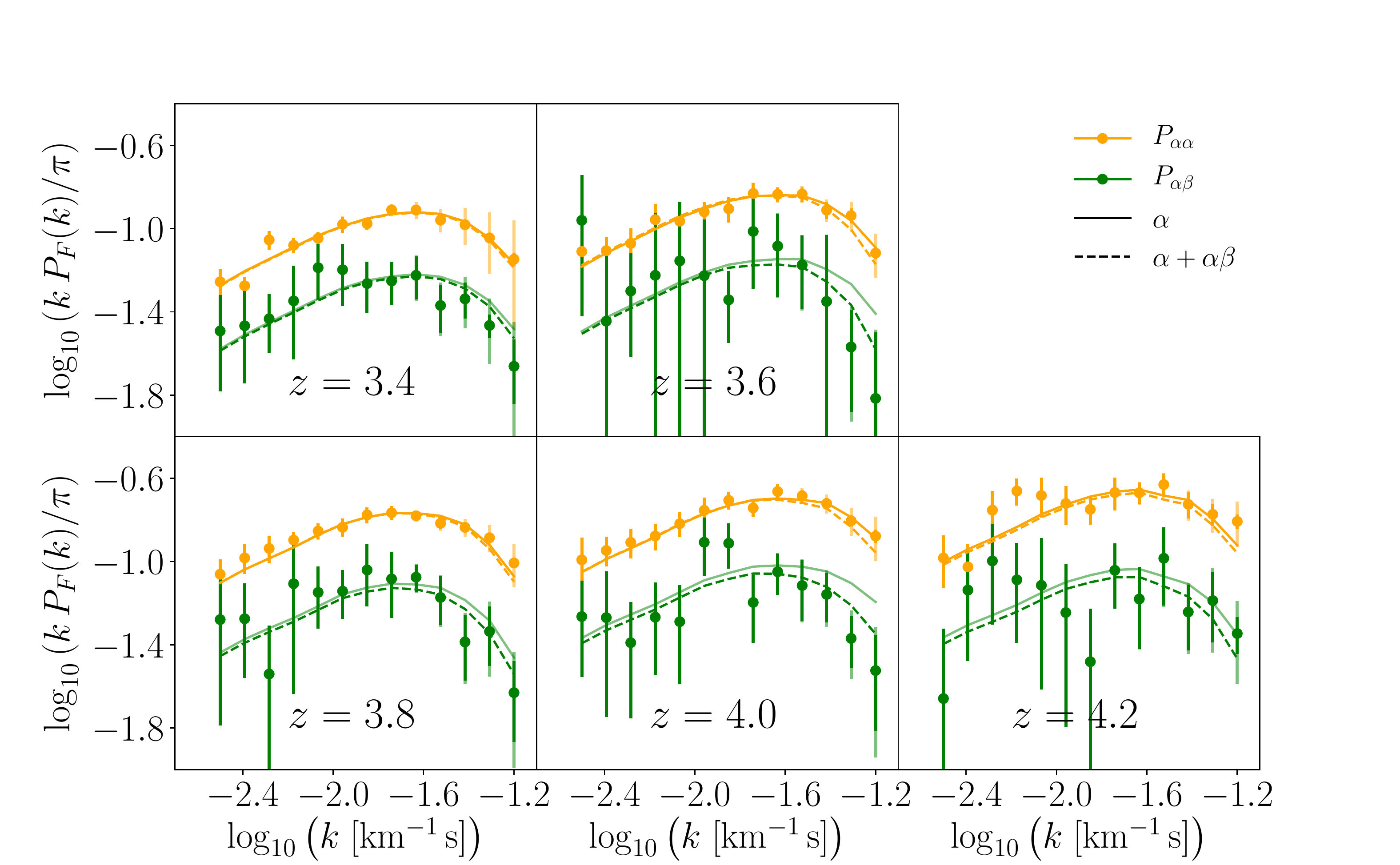}
\caption{ The best-fit models of the analysis using $P_{\alpha\alpha}$ only (solid orange) and of the analysis combining $P_{\alpha\alpha}$ with the cross $P_{\alpha\beta}$ (dashed). We also show the predicted $P_{\alpha\beta}$ using the best-fit parameters of the \lya\ only analysis (solid green). The errorbars shown are from statistical errors, with added systematic error budget of 20\% in the resolution parameter $\sigma_R$. The recovered goodness-of-fit as indicated by $\chi^2$ (including systematic error budget) is $48.7$ for $47$ degrees of freedom, and $115.5$ for $112$ degrees of freedom for \lya--only and joint $P_{\alpha\alpha}$ and $P_{\alpha\beta}$ analysis respectively. The $T_0$ prior choice was the fiducial prior with $1500$ K for the value of standard deviation. 
\label{fig:pk_bestfit}}
\end{figure*}

Fig.~\ref{fig:pk_bestfit} shows the best-fit model to only the \lya\ flux power spectrum (solid orange curve) as well as the fit that includes the cross power with Ly$\beta$ (dashed orange and green curves, respectively).  The errorbars show the measurements. The best-fit model yields $\chi^2 = 48.7$ for $47$ degrees of freedom.  For the analysis that adds the cross, $\chi^2 = 115.5$ for $112$ degrees of freedom. The final $\chi^2=115.5$ only reduces by 0.5 (2.4) for the weaker T3 (T6) priors on $T_0$, again indicating that our measurement is not very sensitive to $T_0$.

While the $\chi^2$ values indicate a good fit, one odd aspect of our result is that the combined analysis shifts $\gamma$ in \emph{all} of our redshift bins by $0.5-1.5\sigma$ relative to the \lya\ analysis.  While not of high statistical significance, this trend may suggest a systematic effect.  
We have performed several tests to assess the robustness of our competitive constraints on $\gamma(z)$. As discussed in more detail in Appendixes \ref{ap:resolution} and \ref{ap:calibration}, resolution and wavelength calibration are possible sources of systematic budget at smaller scales, considerations that led to our error budget of 20\% in $\sigma_R$. Comparison with high resolution HIRES/UVES datasets the resolution differences are largest at $z=3.4$ at the level of 15\% in $\sigma_R$, which results in a factor of two smaller errors at our maximum wavenumber. If this lower resolution uncertainty is assumed in the analysis the results change on average by less than $\sim 0.2\sigma$ with the main effect in $T_0$ values, and there are only two redshift bins where the shift is $> 0.2\sigma$ in both the combined and \lya-only analysis (and the maximum change in the estimated parameter value is always $< 0.7\sigma$). On average the changes in the parameters are smaller for the combined analysis. Similarly, when including a systematic shift in the wavelength calibration between the UVB and VIS arms of the order of $5~\kms$ to minimize an imaginary component of the cross power that may indicate a systematic wavelength miscalibration (see Appendix \ref{ap:resolution}), the main change is for $\gamma(z=3.4)$, where the value changes by $1\sigma$, a redshift where we do not expect any significant miscalibration since the cross is done using one arm of the spectrograph. We found that other redshifts have much smaller shifts of $\sim 0.1\sigma$. 

The last test we report concerns measuring the covariance matrix on a small statistical sample. 
We have re-run our analysis but \emph{only} using the mocks to estimate the covariance matrix rather than rescaling the diagonal by bootstrapping the data. To do this, we rescaled the covariance matrix to match the observed pathlengths in both the Lyman series forests. This also yields similar results, with the largest changes observed in $\gamma(z=3.8)$ and $\gamma(z=4.0)$ at below a $1\sigma$ shift.

Returning to Figure~\ref{fig:pk_bestfit}, where the solid curves show the \lya\ auto and cross power from the best-fit model that only uses the $P_{\alpha\alpha}$ measurement, whereas the dashed curves show the best fit model that also includes our $P_{\alpha\beta}$ measurement.  Even though it was not used in the fit, the green solid curve shows the cross power in this model.  For the redshifts where the cross has the largest effect on shifting the best-fit model posterior, namely $z=3.6$ and $z=4.0$, this figure illustrates that the added constraining-power from using the cross is coming from high $k$.  (The improvement in error relative to \lya\ analysis alone in the other redshift owes to same cross-power sensitivity.) It is the lowness of the power at high $k$ that is driving our high $\gamma$ values. Thus, we should be concerned with systematics whose affect is at the higher wavenumbers: To bias towards higher $\gamma$, one needs to find systematics that result in lower cross power particular relative to \lya.\footnote{Another possibility is that it adds additional variance in the cross that affects our results and that by chance this variance affects our measurements in the several redshift bins in the same direction. This variance should be picked up in the bootstrap error estimates and should be reflected in the errors.  Let us take the case of variance from continuum errors since this might be singled out owing to the perception (originating we think from higher redshift studies) that the continuum may be harder to fit there. At our redshifts, it is not obviously harder to fit the continuum in \lyb\ region than \lya\ since the \lyb\ region is actually less absorbed at a fixed redshift (Fig.~\ref{fig:meanF_data}).}  Since the cross power is immune to systematics that do not appear in both forests, it may be easier to put such a systematic in the \lya\ auto power.  More auto-power could be due to residual metal contamination, structure in the spectrograph's noise, or residual continuum fitting.  However, our XQ-100 \lya\ power normalization agrees with others' analysis of Keck data and our analysis of VLT/UVES (Fig.~\ref{fig:Pkcomparison}), and we note that all of these systematic effects would be outside the mainstream understanding of how these systematics affect the \lya\ forest power spectrum.  The cross power could be affected by resolution uncertainty and wavelength calibration.  While these calibrations are discussed and tested in the appendix, we note that \emph{if anything} our resolution correction appears to be too large when comparing with the Keck/HIRES and our VLT/UVES high-resolution measurements of the \lya\ auto power (See $z=3.2$ and $z=3.4$ panels in Fig.~\ref{fig:Pkcomparison} where our measurement falls slightly above these at high $k$, although at $z=3.0$ the agreement appears greater.  The $z=3-3.4$ \lya\ measurements are relevant since they use the lower resolution UV arm that is used for all of our \lyb\ measurements.). Also, our preliminary VLT/UVES \lyb--\lya\ cross power spectrum appears to show similar high wavenumber behavior: There is no evidence for a systematic underestimate at high wavenumbers.

The shift in $\gamma$ to higher values that owes to the cross could also indicate some missing ingredient in our IGM models.
  For example, the standard cosmological hydrodynamical simulations, like those employed here, do not model the complexities of how reionization processes heat the gas.  During the \HeII\ reionization -- which the important process the redshift range of $z=3-4$ -- segments of sightlines should pierce colder IGM while others should pierce recently reionized \HeII\ regions that are hotter.  We suspect that adding this inhomogeneity will not lead to large improvements in the quality of the best fit as \citet{2011MNRAS.415..977M} found that the \lya\ power spectrum in realistic models of \HeII\ reionization could be well described by a single power-law temperature-density model (despite their simulations of \HeII\ reionization showing significant temperature dispersion). However, they did not investigate the impact on a joint measurement like ours, where maybe the same model cannot fit both. We note that our simple {\it mixed model} for \HeII\ reionzation in  Fig.~\ref{fig:pk_models} does show a rather large effect, although likely in the direction that would result in \lyb\ appearing colder.  Further investigation is merited to see if such missing physics could result in the systematic shift between the two analyses. 
  
  Some other source of heating/disruption that is more important in the somewhat higher density regions that are probed by the \lyb\ forest could also drive the posterior shift between our \lya\ analysis and the combined analysis with the cross.  Perhaps the most obvious source there would be cosmic rays from galaxies as these can easily reach the low density gas probed by our forests.  However, as cosmic rays are more likely to be a source of pressure than heating, pressure alone may not be enough since much of the power spectrum cutoff at high $k$ owes more to thermal broadening than pressure smoothing.  To throw a final effect out there, feedback from galactic and quasar winds could be more disruptive to the \lyb\ forest than \lya\ since it probes gas somewhat closer to galaxies. A potentially difficulty with all these possibilities is that the densities that \lyb\ is sensitive to are likely less than a factor of two larger than those for \lya. 

\subsection{Interpretation in context of standard model for IGM heating}

Figure~\ref{fig:zbin_bestfit} also compares our measurement to other measurements \citep{lidz10,becker11,rorai17,irsic17a,hiss17,walther19,boera19,telikova19,gaikwad20}. As just mentioned, when we relax the standard deviation of our prior on $T_0$ that is centered on previous measurements, our data mildly prefers larger temperatures than the most constraining of these measurements.  Also shown are two models.  The solid red curve are the fiducial \HeII\ reionization model in \citet{sanderbeck16} assuming parameters to achieve a minimum temperature at $z\sim 5$ to match the low temperature values at high redshifts. Thus, it is the case that the low values of many of the previous measurements are difficult to explain with thermal models, and our mildly higher temperatures appear more consistent with thermal history models.  The model of \citet{puchwein19} favors slightly higher temperatures (dashed blue curve).  Our and other temperature measurements are more consistent with the models that include \HeII\ reionization than the one that does not (solid blue curve).

More interesting is our constraints on $\gamma$.   The right panels in Figure \ref{fig:zbin_bestfit} show select prior measurements of $\gamma$.  The error bars are generally large and there is not clear agreement among these measurements.  The reason for large error bars is that Ly$\alpha$ alone is sensitive to a narrow range of densities and so does not provide much of the lever arm needed to constrain this parameter \citep{lidz10,becker11}. \citealt{gaikwad20} is the first to precisely measure $\gamma$ at multiple redshifts, aided by using several \lya\ forest statistics rather than just the power spectrum alone.\footnote{Also of note is a a constraining measurement at $z=2.4$ of $\gamma = 1.54\pm 0.11$ of \citet{2014MNRAS.438.2499B}, exploiting that this redshift has the best \lya\ forest data.}  Our measurement suggests that $\gamma$ flattens with decreasing redshift from $\gamma \sim 1.6$ at $z=4.0$ to $\gamma \sim 1.2$ at $z=3.4$, confirming the trend found at lower redshift by \citealt{gaikwad20}. This flattening is consistent with expectations from the heating from \HeII\ reionization by quasars found in simulations \citep{mcquinn09} and models that attempt to replicate the inhomogeneous heating seen in simulations like \citet[red solid line]{sanderbeck16}. 

It is interesting to ask how the $\gamma$ curve changes for different models.  The $T_0$ measurements suggest a relatively brief \HeII\ reionization over $z\sim 3-4.5$, which is followed by the  \citet{sanderbeck16} model.  If \HeII\ reionizaiton is more extended, $\gamma$ would evolve more slowly and be somewhat steeper.  The \citet{sanderbeck16} model assumes an instantaneous hydrogen reionization at $z=8$ that imparts a flat temperature-density relation at this redshift ($\gamma =1$).  If most of the hydrogen is reionized as late as possible, at $z\approx6$, the $\gamma$ values are pushed down at high redshifts, with the $z=6$ reionization model in \citet{mcquinn09} suggesting that they can have a flat value of $\gamma\approx 1.3$ over our redshift range (similar, but somewhat higher, than the 1 zone \citealt{puchwein19} model shown in Fig.~\ref{fig:zbin_bestfit}).  Thus, our measurement may tentatively constrain hydrogen reionization, favoring significant ionization at $z> 6$ \citep{raskutti12,kulkarni19,keating20}.  Our high values of $\gamma$ at $z\sim 4$ are in mild tension with the measurements of \citet{boera19} and more consistent with \citet{walther19}.

\section{Conclusion}
We have presented a measurement of the \lyb\ forest auto power spectrum as well as the \lya --\lyb\ cross power spectrum using one hundred $\lambda/\Delta \lambda\sim 10^4$, $3.4<z<4.2$ quasar spectra from the XQ-100 Legacy Survey. Previously these statistics have only been investigated using low-S/N, low resolution SDSS spectra \citep{irsic13}, where the low resolution inhibited much of the IGM temperature science that is our primary motivation.  A secondary motivation is that the cross power spectrum is essentially immune to the metal line contamination, which substantially complicates \lya\ forest temperature analyses \citep{lidz10, day19}, or any systematic that is uncorrelated between the two spectral regions.
  
The \lyb\ cross power helps break a strong degeneracy between the mean flux, the temperature at mean density and the power-law index of the temperature-density relation ($\gamma-1$) that is present in \lya\ forest analyses, allowing potentially a much better constraint on $\gamma$.  Our analysis demonstrated this advantage, reducing the error bar on $\gamma$ considerably compared to our analysis that use the \lya\ auto-power alone.  Our measurements suggest that $\gamma$ flattens with decreasing redshift from $\gamma \sim 1.6$ at $z=4.0$ to $\gamma \sim 1.2$ at $z=3.4$, confirming the trend found at lower redshift by \citealt{gaikwad20}.  
(A steep value at $z \sim 4$ also seems consistent with our mean flux analysis in \S~\ref{sec:meanF}.)
This flattening in the temperature-density relation is the expected trend from \HeII\ reionization, and relatively steep values with $\gamma \sim 1.5$ are anticipated before \HeII\ reionization and well after hydrogen reionization.  The steep values may be inconsistent with very late reionization in which the bulk of the IGM is reionized at $z\approx 6$.    

The value of $\gamma$ increased across \emph{all} redshift bins at the $0.5-1.5\sigma$ when using the cross power compared to the analysis that uses the \lya\ auto-power alone. Although this trend is not of high statistical significance, it could indicate some additional contribution to the power or some effect that makes \lyb\ appear relatively hotter. Our extensive investigation of the XQ-100 resolution and wavelength calibration (and comparison to Keck/HIRES and VLT/UVES high resolution measurements) suggests that these are not at play.  Any systematic error seems more likely to affect the \lya\ auto power since the cross is relatively immune to systematics. However, this would indicate contamination from e.g. metal lines that is well beyond what most \lya\ power spectrum studies have concluded. This trend could also indicate missing IGM physics in our models that makes more overdense regions appear hotter or more pressurized.

For our \lyb\ measurements, we had to generalize and expand methods adopted for \lya\ to \lyb.  In particular, this included understanding metal contamination of our \lyb\ statistics (including especially the resonant contamination of \OVI\ $\lambda,\lambda 1032,1038$\AA) and how to remove foreground contamination of the \lyb\ forest (where we found that removal in the power spectrum is imperfect with just subtraction).  We also had to generalize the previous XQ-100 \lya\ analysis to deal with a $4\times$ larger covariance matrix and for our MCMC code to search a larger parameter space.  In addition, wavelength calibration becomes a more pressing concern when disparate lines are being used to measure correlations on $10\;\kms$ separations.  We used the imaginary part of the cross power spectrum as a diagnostic of such systematic miscalibration.

This work shows that the \lyb\ forest can be valuable tool for IGM thermal evolution studies at these redshifts.  We put significant work into understanding the resolution of the X-Shooter spectrograph, calibrating a physical seeing-dependent model for the line spread function on arc-lamp lines.  Yet, our final thermal parameter constraints were largely limited by resolution uncertainties.   Our competitive constraints marginalize over a generous allowance for this uncertainty.  Future \lya--\lyb\ cross-power measurements with higher resolution data sets such as SQUAD \citep{2019MNRAS.482.3458M} and KODIAQ \citep{2015AJ....150..111O} have the potential to set our tightest constraints on $\gamma$.  We presented a preliminary analysis of the high resolution VLT/UVES SQUAD DR1 data over the overlapping redshift interval that showed broad agreement.  This data set will be even more powerful at lower redshifts than presented here owing to the redshift distribution of the quasars in this sample.  We plan to perform a full \lyb\ analysis on the SQUAD spectra in future work \citep{irsicinprep}.\\

\section{Acknowledgements}

We would like to especially thank George Becker for discussion of X-Shooter and possible systematics, Guido Cupani for help with X-Shooter arc lamp spectra, Bob Carswell for many long discussions on the spectra resolution, Matteo Viel for discussion and help with running the simulations, and John O'Meara for a long discussion on spectral reductions. We would also like to thank Phoebe Upton-Sanderback and Ewald Puchwein for providing the theoretical models to which we compare our results. VI is supported by the Kavli foundation.  This work is supported by NSF grant AST-1514734 and NASA grant NNX17AH68G.

Based on observations collected at the European Organisation for Astronomical Research in the Southern Hemisphere under ESO programme 189.A-0424.  This work made use of the DiRAC High Performance  Computing  System  (HPCS)  and  the  COSMOS shared memory service at the University of Cambridge. These are operated on behalf of the STFC DiRAC HPC facility. This  equipment  is  funded  by  BIS  National  E-infrastructure  capital  grant  ST/J005673/1  and  STFC grants ST/H008586/1, ST/K00333X/1.

\section{Data Availability}

The spectroscopic data used in this article were obtained through VLT/XSHOOTER XQ-100 program and are publicly available in the form of ESO Phase 3 material (\url{http://archive.eso.org/wdb/wdb/adp/phase3_main/form}), as described in \citep{lopez16}. The mean flux and power spectrum measurements used in this paper may be accessed in the Github repository (\url{https://github.com/bayu-wilson/lyb_pk/tree/main/output}).

\bibliographystyle{mnras}
\interlinepenalty=10000
\bibliography{main}

\begin{thebibliography}{}
\makeatletter
\relax
\def\mn@urlcharsother{\let\do\@makeother \do\$\do\&\do\#\do\^\do\_\do\%\do\~}
\def\mn@doi{\begingroup\mn@urlcharsother \@ifnextchar [ {\mn@doi@}
  {\mn@doi@[]}}
\def\mn@doi@[#1]#2{\def\@tempa{#1}\ifx\@tempa\@empty \href
  {http://dx.doi.org/#2} {doi:#2}\else \href {http://dx.doi.org/#2} {#1}\fi
  \endgroup}
\def\mn@eprint#1#2{\mn@eprint@#1:#2::\@nil}
\def\mn@eprint@arXiv#1{\href {http://arxiv.org/abs/#1} {{\tt arXiv:#1}}}
\def\mn@eprint@dblp#1{\href {http://dblp.uni-trier.de/rec/bibtex/#1.xml}
  {dblp:#1}}
\def\mn@eprint@#1:#2:#3:#4\@nil{\def\@tempa {#1}\def\@tempb {#2}\def\@tempc
  {#3}\ifx \@tempc \@empty \let \@tempc \@tempb \let \@tempb \@tempa \fi \ifx
  \@tempb \@empty \def\@tempb {arXiv}\fi \@ifundefined
  {mn@eprint@\@tempb}{\@tempb:\@tempc}{\expandafter \expandafter \csname
  mn@eprint@\@tempb\endcsname \expandafter{\@tempc}}}

\bibitem[\protect\citeauthoryear{{Aguirre}, {Schaye}, {Kim}, {Theuns}, {Rauch}
  \& {Sargent}}{{Aguirre} et~al.}{2004}]{2004ApJ...602...38A}
{Aguirre} A.,  {Schaye} J.,  {Kim} T.-S.,  {Theuns} T.,  {Rauch} M.,
  {Sargent} W. L.~W.,  2004, \mn@doi [\apj] {10.1086/380961}, \href
  {https://ui.adsabs.harvard.edu/abs/2004ApJ...602...38A} {602, 38}

\bibitem[\protect\citeauthoryear{{Aguirre}, {Dow-Hygelund}, {Schaye}  \&
  {Theuns}}{{Aguirre} et~al.}{2008}]{2008ApJ...689..851A}
{Aguirre} A.,  {Dow-Hygelund} C.,  {Schaye} J.,   {Theuns} T.,  2008, \mn@doi
  [\apj] {10.1086/592554}, \href
  {https://ui.adsabs.harvard.edu/abs/2008ApJ...689..851A} {689, 851}

\bibitem[\protect\citeauthoryear{{Armengaud}, {Palanque-Delabrouille},
  {Y{\`e}che}, {Marsh}  \& {Baur}}{{Armengaud} et~al.}{2017}]{armengaud17}
{Armengaud} E.,  {Palanque-Delabrouille} N.,  {Y{\`e}che} C.,  {Marsh} D.
  J.~E.,   {Baur} J.,  2017, \mn@doi [\mnras] {10.1093/mnras/stx1870}, \href
  {https://ui.adsabs.harvard.edu/abs/2017MNRAS.471.4606A} {471, 4606}

\bibitem[\protect\citeauthoryear{{Baur}, {Palanque-Delabrouille}, {Y{\`e}che},
  {Magneville}  \& {Viel}}{{Baur} et~al.}{2016}]{baur15}
{Baur} J.,  {Palanque-Delabrouille} N.,  {Y{\`e}che} C.,  {Magneville} C.,
  {Viel} M.,  2016, \mn@doi [\jcap] {10.1088/1475-7516/2016/08/012}, \href
  {http://adsabs.harvard.edu/abs/2016JCAP...08..012B} {8, 012}

\bibitem[\protect\citeauthoryear{{Baur}, {Palanque-Delabrouille}, {Y{\`e}che},
  {Boyarsky}, {Ruchayskiy}, {Armengaud}  \& {Lesgourgues}}{{Baur}
  et~al.}{2017}]{baur17}
{Baur} J.,  {Palanque-Delabrouille} N.,  {Y{\`e}che} C.,  {Boyarsky} A.,
  {Ruchayskiy} O.,  {Armengaud} {\'E}.,   {Lesgourgues} J.,  2017, \mn@doi
  [\jcap] {10.1088/1475-7516/2017/12/013}, \href
  {https://ui.adsabs.harvard.edu/abs/2017JCAP...12..013B} {2017, 013}

\bibitem[\protect\citeauthoryear{{Bautista} et~al.,}{{Bautista}
  et~al.}{2015}]{bautista15}
{Bautista} J.~E.,  et~al., 2015, \mn@doi [\jcap]
  {10.1088/1475-7516/2015/05/060}, \href
  {http://adsabs.harvard.edu/abs/2015JCAP...05..060B} {5, 060}

\bibitem[\protect\citeauthoryear{{Bautista} et~al.,}{{Bautista}
  et~al.}{2017}]{bautista17}
{Bautista} J.~E.,  et~al., 2017, \mn@doi [\aap] {10.1051/0004-6361/201730533},
  \href {http://adsabs.harvard.edu/abs/2017A%26A...603A..12B} {603, A12}

\bibitem[\protect\citeauthoryear{{Becker}, {Bolton}, {Haehnelt}  \&
  {Sargent}}{{Becker} et~al.}{2011}]{becker11}
{Becker} G.~D.,  {Bolton} J.~S.,  {Haehnelt} M.~G.,   {Sargent} W.~L.~W.,
  2011, \mn@doi [\mnras] {10.1111/j.1365-2966.2010.17507.x}, \href
  {http://adsabs.harvard.edu/abs/2011MNRAS.410.1096B} {410, 1096}

\bibitem[\protect\citeauthoryear{{Becker}, {Hewett}, {Worseck}  \&
  {Prochaska}}{{Becker} et~al.}{2013}]{becker13}
{Becker} G.~D.,  {Hewett} P.~C.,  {Worseck} G.,   {Prochaska} J.~X.,  2013,
  \mn@doi [\mnras] {10.1093/mnras/stt031}, \href
  {https://ui.adsabs.harvard.edu/abs/2013MNRAS.430.2067B} {430, 2067}

\bibitem[\protect\citeauthoryear{{Berg} et~al.,}{{Berg}
  et~al.}{2016}]{Berg2016}
{Berg} T.~A.~M.,  et~al., 2016, \mn@doi [\mnras] {10.1093/mnras/stw2232}, \href
  {https://ui.adsabs.harvard.edu/abs/2016MNRAS.463.3021B} {463, 3021}

\bibitem[\protect\citeauthoryear{{Bird}, {Peiris}, {Viel}  \& {Verde}}{{Bird}
  et~al.}{2011}]{bird10}
{Bird} S.,  {Peiris} H.~V.,  {Viel} M.,   {Verde} L.,  2011, \mn@doi [\mnras]
  {10.1111/j.1365-2966.2011.18245.x}, \href
  {http://adsabs.harvard.edu/abs/2011MNRAS.413.1717B} {413, 1717}

\bibitem[\protect\citeauthoryear{{Boera}, {Murphy}, {Becker}  \&
  {Bolton}}{{Boera} et~al.}{2014}]{boera14}
{Boera} E.,  {Murphy} M.~T.,  {Becker} G.~D.,   {Bolton} J.~S.,  2014, \mn@doi
  [\mnras] {10.1093/mnras/stu660}, \href
  {http://adsabs.harvard.edu/abs/2014MNRAS.441.1916B} {441, 1916}

\bibitem[\protect\citeauthoryear{{Boera}, {Murphy}, {Becker}  \&
  {Bolton}}{{Boera} et~al.}{2016}]{boera16}
{Boera} E.,  {Murphy} M.~T.,  {Becker} G.~D.,   {Bolton} J.~S.,  2016, \mn@doi
  [\mnras] {10.1093/mnrasl/slv172}, \href
  {https://ui.adsabs.harvard.edu/abs/2016MNRAS.456L..79B} {456, L79}

\bibitem[\protect\citeauthoryear{{Boera}, {Becker}, {Bolton}  \&
  {Nasir}}{{Boera} et~al.}{2019}]{boera19}
{Boera} E.,  {Becker} G.~D.,  {Bolton} J.~S.,   {Nasir} F.,  2019, \mn@doi
  [\apj] {10.3847/1538-4357/aafee4}, \href
  {https://ui.adsabs.harvard.edu/abs/2019ApJ...872..101B} {872, 101}

\bibitem[\protect\citeauthoryear{{Bolton}, {Viel}, {Kim}, {Haehnelt}  \&
  {Carswell}}{{Bolton} et~al.}{2008}]{bolton08}
{Bolton} J.~S.,  {Viel} M.,  {Kim} T.-S.,  {Haehnelt} M.~G.,   {Carswell}
  R.~F.,  2008, \mn@doi [\mnras] {10.1111/j.1365-2966.2008.13114.x}, \href
  {http://adsabs.harvard.edu/abs/2008MNRAS.386.1131B} {386, 1131}

\bibitem[\protect\citeauthoryear{{Bolton}, {Becker}, {Wyithe}, {Haehnelt}  \&
  {Sargent}}{{Bolton} et~al.}{2010}]{bolton10}
{Bolton} J.~S.,  {Becker} G.~D.,  {Wyithe} J.~S.~B.,  {Haehnelt} M.~G.,
  {Sargent} W.~L.~W.,  2010, \mn@doi [\mnras]
  {10.1111/j.1365-2966.2010.16701.x}, \href
  {http://adsabs.harvard.edu/abs/2010MNRAS.406..612B} {406, 612}

\bibitem[\protect\citeauthoryear{{Bolton}, {Becker}, {Haehnelt}  \&
  {Viel}}{{Bolton} et~al.}{2014a}]{bolton14}
{Bolton} J.~S.,  {Becker} G.~D.,  {Haehnelt} M.~G.,   {Viel} M.,  2014a,
  \mn@doi [\mnras] {10.1093/mnras/stt2374}, \href
  {http://adsabs.harvard.edu/abs/2014MNRAS.438.2499B} {438, 2499}

\bibitem[\protect\citeauthoryear{{Bolton}, {Becker}, {Haehnelt}  \&
  {Viel}}{{Bolton} et~al.}{2014b}]{2014MNRAS.438.2499B}
{Bolton} J.~S.,  {Becker} G.~D.,  {Haehnelt} M.~G.,   {Viel} M.,  2014b,
  \mn@doi [\mnras] {10.1093/mnras/stt2374}, \href
  {https://ui.adsabs.harvard.edu/abs/2014MNRAS.438.2499B} {438, 2499}

\bibitem[\protect\citeauthoryear{{Bolton}, {Puchwein}, {Sijacki}, {Haehnelt},
  {Kim}, {Meiksin}, {Regan}  \& {Viel}}{{Bolton} et~al.}{2017}]{bolton16}
{Bolton} J.~S.,  {Puchwein} E.,  {Sijacki} D.,  {Haehnelt} M.~G.,  {Kim} T.-S.,
   {Meiksin} A.,  {Regan} J.~A.,   {Viel} M.,  2017, \mn@doi [\mnras]
  {10.1093/mnras/stw2397}, \href
  {https://ui.adsabs.harvard.edu/abs/2017MNRAS.464..897B} {464, 897}

\bibitem[\protect\citeauthoryear{{Busca} et~al.,}{{Busca}
  et~al.}{2013}]{busca13}
{Busca} N.~G.,  et~al., 2013, \mn@doi [\aap] {10.1051/0004-6361/201220724},
  \href {http://adsabs.harvard.edu/abs/2013A%26A...552A..96B} {552, A96}

\bibitem[\protect\citeauthoryear{{Croft}, {Weinberg}, {Pettini}, {Hernquist}
  \& {Katz}}{{Croft} et~al.}{1999}]{croft99}
{Croft} R.~A.~C.,  {Weinberg} D.~H.,  {Pettini} M.,  {Hernquist} L.,   {Katz}
  N.,  1999, \mn@doi [\apj] {10.1086/307438}, \href
  {http://adsabs.harvard.edu/abs/1999ApJ...520....1C} {520, 1}

\bibitem[\protect\citeauthoryear{{Croft}, {Weinberg}, {Bolte}, {Burles},
  {Hernquist}, {Katz}, {Kirkman}  \& {Tytler}}{{Croft} et~al.}{2002}]{croft02}
{Croft} R.~A.~C.,  {Weinberg} D.~H.,  {Bolte} M.,  {Burles} S.,  {Hernquist}
  L.,  {Katz} N.,  {Kirkman} D.,   {Tytler} D.,  2002, \mn@doi [\apj]
  {10.1086/344099}, \href {http://adsabs.harvard.edu/abs/2002ApJ...581...20C}
  {581, 20}

\bibitem[\protect\citeauthoryear{Day, Tytler  \& Kambalur}{Day
  et~al.}{2019}]{day19}
Day A.,  Tytler D.,   Kambalur B.,  2019, \mn@doi [Monthly Notices of the Royal
  Astronomical Society] {10.1093/mnras/stz2214}, 489, 2536

\bibitem[\protect\citeauthoryear{{Dijkstra}, {Lidz}  \& {Hui}}{{Dijkstra}
  et~al.}{2004}]{dijkstra04}
{Dijkstra} M.,  {Lidz} A.,   {Hui} L.,  2004, \mn@doi [\apj] {10.1086/382199},
  \href {http://adsabs.harvard.edu/abs/2004ApJ...605....7D} {605, 7}

\bibitem[\protect\citeauthoryear{{Fang} \& {White}}{{Fang} \&
  {White}}{2004}]{fangwhite}
{Fang} T.,  {White} M.,  2004, \mn@doi [\apjl] {10.1086/420965}, \href
  {http://adsabs.harvard.edu/abs/2004ApJ...606L...9F} {606, L9}

\bibitem[\protect\citeauthoryear{{Faucher-Gigu{\`e}re}, {Prochaska}, {Lidz},
  {Hernquist}  \& {Zaldarriaga}}{{Faucher-Gigu{\`e}re}
  et~al.}{2008}]{faucher08}
{Faucher-Gigu{\`e}re} C.-A.,  {Prochaska} J.~X.,  {Lidz} A.,  {Hernquist} L.,
  {Zaldarriaga} M.,  2008, \mn@doi [\apj] {10.1086/588648}, \href
  {https://ui.adsabs.harvard.edu/abs/2008ApJ...681..831F} {681, 831}

\bibitem[\protect\citeauthoryear{{Gaikwad}, {Srianand}, {Haehnelt}  \&
  {Choudhury}}{{Gaikwad} et~al.}{2020}]{gaikwad20}
{Gaikwad} P.,  {Srianand} R.,  {Haehnelt} M.~G.,   {Choudhury} T.~R.,  2020,
  arXiv e-prints, \href {https://ui.adsabs.harvard.edu/abs/2020arXiv200900016G}
  {p. arXiv:2009.00016}

\bibitem[\protect\citeauthoryear{{Garzilli}, {Bolton}, {Kim}, {Leach}  \&
  {Viel}}{{Garzilli} et~al.}{2012}]{garzilli12}
{Garzilli} A.,  {Bolton} J.~S.,  {Kim} T.-S.,  {Leach} S.,   {Viel} M.,  2012,
  \mn@doi [\mnras] {10.1111/j.1365-2966.2012.21223.x}, \href
  {http://adsabs.harvard.edu/abs/2012MNRAS.424.1723G} {424, 1723}

\bibitem[\protect\citeauthoryear{{Garzilli}, {Ruchayskiy}, {Magalich}  \&
  {Boyarsky}}{{Garzilli} et~al.}{2019a}]{garzilli19}
{Garzilli} A.,  {Ruchayskiy} O.,  {Magalich} A.,   {Boyarsky} A.,  2019a, arXiv
  e-prints, \href {https://ui.adsabs.harvard.edu/abs/2019arXiv191209397G} {p.
  arXiv:1912.09397}

\bibitem[\protect\citeauthoryear{{Garzilli}, {Magalich}, {Theuns}, {Frenk},
  {Weniger}, {Ruchayskiy}  \& {Boyarsky}}{{Garzilli}
  et~al.}{2019b}]{garzilli17}
{Garzilli} A.,  {Magalich} A.,  {Theuns} T.,  {Frenk} C.~S.,  {Weniger} C.,
  {Ruchayskiy} O.,   {Boyarsky} A.,  2019b, \mn@doi [\mnras]
  {10.1093/mnras/stz2188}, \href
  {https://ui.adsabs.harvard.edu/abs/2019MNRAS.489.3456G} {489, 3456}

\bibitem[\protect\citeauthoryear{{Hiss}, {Walther}, {Hennawi}, {O{\~n}orbe},
  {O'Meara}, {Rorai}  \& {Luki{\'c}}}{{Hiss} et~al.}{2018}]{hiss17}
{Hiss} H.,  {Walther} M.,  {Hennawi} J.~F.,  {O{\~n}orbe} J.,  {O'Meara} J.~M.,
   {Rorai} A.,   {Luki{\'c}} Z.,  2018, \mn@doi [\apj]
  {10.3847/1538-4357/aada86}, \href
  {https://ui.adsabs.harvard.edu/abs/2018ApJ...865...42H} {865, 42}

\bibitem[\protect\citeauthoryear{{Hui} \& {Gnedin}}{{Hui} \&
  {Gnedin}}{1997}]{hui97}
{Hui} L.,  {Gnedin} N.~Y.,  1997, \mnras, \href
  {http://adsabs.harvard.edu/abs/1997MNRAS.292...27H} {292, 27}

\bibitem[\protect\citeauthoryear{{Hui} \& {Haiman}}{{Hui} \&
  {Haiman}}{2003}]{hui03}
{Hui} L.,  {Haiman} Z.,  2003, \mn@doi [\apj] {10.1086/377229}, \href
  {http://adsabs.harvard.edu/abs/2003ApJ...596....9H} {596, 9}

\bibitem[\protect\citeauthoryear{{Ir{\v s}i{\v c}} et~al.,}{{Ir{\v s}i{\v c}}
  et~al.}{2013}]{irsic13}
{Ir{\v s}i{\v c}} V.,  et~al., 2013, \mn@doi [\jcap]
  {10.1088/1475-7516/2013/09/016}, \href
  {http://adsabs.harvard.edu/abs/2013JCAP...09..016I} {9, 16}

\bibitem[\protect\citeauthoryear{{Ir{\v s}i{\v c}} et~al.,}{{Ir{\v s}i{\v c}}
  et~al.}{2017a}]{irsic17b}
{Ir{\v s}i{\v c}} V.,  et~al., 2017a, \mn@doi [\prd]
  {10.1103/PhysRevD.96.023522}, \href
  {http://adsabs.harvard.edu/abs/2017PhRvD..96b3522I} {96, 023522}

\bibitem[\protect\citeauthoryear{{Ir{\v s}i{\v c}}, {Viel}, {Haehnelt},
  {Bolton}  \& {Becker}}{{Ir{\v s}i{\v c}} et~al.}{2017b}]{irsic17a}
{Ir{\v s}i{\v c}} V.,  {Viel} M.,  {Haehnelt} M.~G.,  {Bolton} J.~S.,
  {Becker} G.~D.,  2017b, \mn@doi [Physical Review Letters]
  {10.1103/PhysRevLett.119.031302}, \href
  {http://adsabs.harvard.edu/abs/2017PhRvL.119c1302I} {119, 031302}

\bibitem[\protect\citeauthoryear{{Ir{\v s}i{\v c}}, {Wilson}  \&
  {McQuinn}}{{Ir{\v s}i{\v c}} et~al.}{2021}]{irsicinprep}
{Ir{\v s}i{\v c}} V.,  {Wilson} B.,   {McQuinn} M.,  2021, in prep

\bibitem[\protect\citeauthoryear{{Ir{\v{s}}i{\v{c}}} \&
  {McQuinn}}{{Ir{\v{s}}i{\v{c}}} \& {McQuinn}}{2018}]{irsic18}
{Ir{\v{s}}i{\v{c}}} V.,  {McQuinn} M.,  2018, \mn@doi [\jcap]
  {10.1088/1475-7516/2018/04/026}, \href
  {https://ui.adsabs.harvard.edu/abs/2018JCAP...04..026I} {2018, 026}

\bibitem[\protect\citeauthoryear{{Ir{\v{s}}i{\v{c}}} \&
  {Viel}}{{Ir{\v{s}}i{\v{c}}} \& {Viel}}{2014a}]{2014JCAP...12..024I}
{Ir{\v{s}}i{\v{c}}} V.,  {Viel} M.,  2014a, \mn@doi [\jcap]
  {10.1088/1475-7516/2014/12/024}, \href
  {https://ui.adsabs.harvard.edu/abs/2014JCAP...12..024I} {2014, 024}

\bibitem[\protect\citeauthoryear{{Ir{\v{s}}i{\v{c}}} \&
  {Viel}}{{Ir{\v{s}}i{\v{c}}} \& {Viel}}{2014b}]{irsic14b}
{Ir{\v{s}}i{\v{c}}} V.,  {Viel} M.,  2014b, \mn@doi [\jcap]
  {10.1088/1475-7516/2014/12/024}, \href
  {https://ui.adsabs.harvard.edu/abs/2014JCAP...12..024I} {2014, 024}

\bibitem[\protect\citeauthoryear{{Ir{\v{s}}i{\v{c}}}
  et~al.,}{{Ir{\v{s}}i{\v{c}}} et~al.}{2017}]{irsic17}
{Ir{\v{s}}i{\v{c}}} V.,  et~al., 2017, \mn@doi [\mnras]
  {10.1093/mnras/stw3372}, \href
  {https://ui.adsabs.harvard.edu/abs/2017MNRAS.466.4332I} {466, 4332}

\bibitem[\protect\citeauthoryear{{Ir{\v{s}}i{\v{c}}}, {Xiao}  \&
  {McQuinn}}{{Ir{\v{s}}i{\v{c}}} et~al.}{2020}]{irsic20}
{Ir{\v{s}}i{\v{c}}} V.,  {Xiao} H.,   {McQuinn} M.,  2020, \mn@doi [\prd]
  {10.1103/PhysRevD.101.123518}, \href
  {https://ui.adsabs.harvard.edu/abs/2020PhRvD.101l3518I} {101, 123518}

\bibitem[\protect\citeauthoryear{{Keating}, {Weinberger}, {Kulkarni},
  {Haehnelt}, {Chardin}  \& {Aubert}}{{Keating} et~al.}{2020}]{keating20}
{Keating} L.~C.,  {Weinberger} L.~H.,  {Kulkarni} G.,  {Haehnelt} M.~G.,
  {Chardin} J.,   {Aubert} D.,  2020, \mn@doi [\mnras] {10.1093/mnras/stz3083},
  \href {https://ui.adsabs.harvard.edu/abs/2020MNRAS.491.1736K} {491, 1736}

\bibitem[\protect\citeauthoryear{{Kim}, {Viel}, {Haehnelt}, {Carswell}  \&
  {Cristiani}}{{Kim} et~al.}{2004}]{kim04}
{Kim} T.-S.,  {Viel} M.,  {Haehnelt} M.~G.,  {Carswell} R.~F.,   {Cristiani}
  S.,  2004, \mn@doi [\mnras] {10.1111/j.1365-2966.2004.07221.x}, \href
  {http://adsabs.harvard.edu/abs/2004MNRAS.347..355K} {347, 355}

\bibitem[\protect\citeauthoryear{{Kulkarni}, {Keating}, {Haehnelt}, {Bosman},
  {Puchwein}, {Chardin}  \& {Aubert}}{{Kulkarni} et~al.}{2019}]{kulkarni19}
{Kulkarni} G.,  {Keating} L.~C.,  {Haehnelt} M.~G.,  {Bosman} S. E.~I.,
  {Puchwein} E.,  {Chardin} J.,   {Aubert} D.,  2019, \mn@doi [\mnras]
  {10.1093/mnrasl/slz025}, \href
  {https://ui.adsabs.harvard.edu/abs/2019MNRAS.485L..24K} {485, L24}

\bibitem[\protect\citeauthoryear{{Lee} et~al.,}{{Lee} et~al.}{2015}]{lee14}
{Lee} K.-G.,  et~al., 2015, \mn@doi [\apj] {10.1088/0004-637X/799/2/196}, \href
  {https://ui.adsabs.harvard.edu/abs/2015ApJ...799..196L} {799, 196}

\bibitem[\protect\citeauthoryear{{Lidz}, {Faucher-Gigu{\`e}re}, {Dall'Aglio},
  {McQuinn}, {Fechner}, {Zaldarriaga}, {Hernquist}  \& {Dutta}}{{Lidz}
  et~al.}{2010}]{lidz10}
{Lidz} A.,  {Faucher-Gigu{\`e}re} C.-A.,  {Dall'Aglio} A.,  {McQuinn} M.,
  {Fechner} C.,  {Zaldarriaga} M.,  {Hernquist} L.,   {Dutta} S.,  2010,
  \mn@doi [\apj] {10.1088/0004-637X/718/1/199}, \href
  {http://adsabs.harvard.edu/abs/2010ApJ...718..199L} {718, 199}

\bibitem[\protect\citeauthoryear{{L{\'o}pez} et~al.,}{{L{\'o}pez}
  et~al.}{2016}]{lopez16}
{L{\'o}pez} S.,  et~al., 2016, \mn@doi [\aap] {10.1051/0004-6361/201628161},
  \href {http://adsabs.harvard.edu/abs/2016A%26A...594A..91L} {594, A91}

\bibitem[\protect\citeauthoryear{{McDonald}}{{McDonald}}{2003}]{mcdonald03}
{McDonald} P.,  2003, \mn@doi [\apj] {10.1086/345945}, \href
  {http://adsabs.harvard.edu/abs/2003ApJ...585...34M} {585, 34}

\bibitem[\protect\citeauthoryear{{McDonald}, {Miralda-Escud{\'e}}, {Rauch},
  {Sargent}, {Barlow}, {Cen}  \& {Ostriker}}{{McDonald}
  et~al.}{2000}]{mcdonald00}
{McDonald} P.,  {Miralda-Escud{\'e}} J.,  {Rauch} M.,  {Sargent} W.~L.~W.,
  {Barlow} T.~A.,  {Cen} R.,   {Ostriker} J.~P.,  2000, \mn@doi [\apj]
  {10.1086/317079}, \href {http://adsabs.harvard.edu/abs/2000ApJ...543....1M}
  {543, 1}

\bibitem[\protect\citeauthoryear{{McDonald} et~al.,}{{McDonald}
  et~al.}{2005}]{mcdonald05}
{McDonald} P.,  et~al., 2005, \mn@doi [\apj] {10.1086/497563}, \href
  {http://adsabs.harvard.edu/abs/2005ApJ...635..761M} {635, 761}

\bibitem[\protect\citeauthoryear{{McDonald} et~al.,}{{McDonald}
  et~al.}{2006}]{mcdonald06}
{McDonald} P.,  et~al., 2006, \mn@doi [\apjs] {10.1086/444361}, \href
  {http://adsabs.harvard.edu/abs/2006ApJS..163...80M} {163, 80}

\bibitem[\protect\citeauthoryear{{McQuinn}}{{McQuinn}}{2016}]{mcquinn15}
{McQuinn} M.,  2016, \mn@doi [\araa] {10.1146/annurev-astro-082214-122355},
  \href {https://ui.adsabs.harvard.edu/abs/2016ARA&A..54..313M} {54, 313}

\bibitem[\protect\citeauthoryear{{McQuinn} \& {Upton Sanderbeck}}{{McQuinn} \&
  {Upton Sanderbeck}}{2016}]{mcquinnTrho}
{McQuinn} M.,  {Upton Sanderbeck} P.~R.,  2016, \mn@doi [\mnras]
  {10.1093/mnras/stv2675}, \href
  {https://ui.adsabs.harvard.edu/abs/2016MNRAS.456...47M} {456, 47}

\bibitem[\protect\citeauthoryear{{McQuinn}, {Lidz}, {Zaldarriaga}, {Hernquist},
  {Hopkins}, {Dutta}  \& {Faucher-Gigu{\`e}re}}{{McQuinn}
  et~al.}{2009}]{mcquinn09}
{McQuinn} M.,  {Lidz} A.,  {Zaldarriaga} M.,  {Hernquist} L.,  {Hopkins} P.~F.,
   {Dutta} S.,   {Faucher-Gigu{\`e}re} C.-A.,  2009, \mn@doi [\apj]
  {10.1088/0004-637X/694/2/842}, \href
  {http://adsabs.harvard.edu/abs/2009ApJ...694..842M} {694, 842}

\bibitem[\protect\citeauthoryear{{McQuinn}, {Hernquist}, {Lidz}  \&
  {Zaldarriaga}}{{McQuinn} et~al.}{2011}]{2011MNRAS.415..977M}
{McQuinn} M.,  {Hernquist} L.,  {Lidz} A.,   {Zaldarriaga} M.,  2011, \mn@doi
  [\mnras] {10.1111/j.1365-2966.2011.18788.x}, \href
  {https://ui.adsabs.harvard.edu/abs/2011MNRAS.415..977M} {415, 977}

\bibitem[\protect\citeauthoryear{{Murphy}, {Kacprzak}, {Savorgnan}  \&
  {Carswell}}{{Murphy} et~al.}{2019}]{2019MNRAS.482.3458M}
{Murphy} M.~T.,  {Kacprzak} G.~G.,  {Savorgnan} G. A.~D.,   {Carswell} R.~F.,
  2019, \mn@doi [\mnras] {10.1093/mnras/sty2834}, \href
  {https://ui.adsabs.harvard.edu/abs/2019MNRAS.482.3458M} {482, 3458}

\bibitem[\protect\citeauthoryear{{Narayanan}, {Spergel}, {Dav{\'e}}  \&
  {Ma}}{{Narayanan} et~al.}{2000}]{narayanan00}
{Narayanan} V.~K.,  {Spergel} D.~N.,  {Dav{\'e}} R.,   {Ma} C.-P.,  2000,
  \mn@doi [\apjl] {10.1086/317269}, \href
  {http://adsabs.harvard.edu/abs/2000ApJ...543L.103N} {543, L103}

\bibitem[\protect\citeauthoryear{{O'Meara} et~al.,}{{O'Meara}
  et~al.}{2015}]{2015AJ....150..111O}
{O'Meara} J.~M.,  et~al., 2015, \mn@doi [\aj] {10.1088/0004-6256/150/4/111},
  \href {https://ui.adsabs.harvard.edu/abs/2015AJ....150..111O} {150, 111}

\bibitem[\protect\citeauthoryear{{Palanque-Delabrouille}
  et~al.,}{{Palanque-Delabrouille} et~al.}{2013}]{palanque13}
{Palanque-Delabrouille} N.,  et~al., 2013, \mn@doi [\aap]
  {10.1051/0004-6361/201322130}, \href
  {http://adsabs.harvard.edu/abs/2013A%26A...559A..85P} {559, A85}

\bibitem[\protect\citeauthoryear{{Palanque-Delabrouille}
  et~al.,}{{Palanque-Delabrouille} et~al.}{2015}]{palanque15}
{Palanque-Delabrouille} N.,  et~al., 2015, \mn@doi [\jcap]
  {10.1088/1475-7516/2015/11/011}, \href
  {http://adsabs.harvard.edu/abs/2015JCAP...11..011P} {11, 011}

\bibitem[\protect\citeauthoryear{{Puchwein}, {Haardt}, {Haehnelt}  \&
  {Madau}}{{Puchwein} et~al.}{2019}]{puchwein19}
{Puchwein} E.,  {Haardt} F.,  {Haehnelt} M.~G.,   {Madau} P.,  2019, \mn@doi
  [\mnras] {10.1093/mnras/stz222}, \href
  {https://ui.adsabs.harvard.edu/abs/2019MNRAS.485...47P} {485, 47}

\bibitem[\protect\citeauthoryear{{Raskutti}, {Bolton}, {Wyithe}  \&
  {Becker}}{{Raskutti} et~al.}{2012}]{raskutti12}
{Raskutti} S.,  {Bolton} J.~S.,  {Wyithe} J.~S.~B.,   {Becker} G.~D.,  2012,
  \mn@doi [\mnras] {10.1111/j.1365-2966.2011.20401.x}, \href
  {http://adsabs.harvard.edu/abs/2012MNRAS.421.1969R} {421, 1969}

\bibitem[\protect\citeauthoryear{{Ricotti}, {Gnedin}  \& {Shull}}{{Ricotti}
  et~al.}{2000}]{ricotti00}
{Ricotti} M.,  {Gnedin} N.~Y.,   {Shull} J.~M.,  2000, \mn@doi [\apj]
  {10.1086/308733}, \href {http://adsabs.harvard.edu/abs/2000ApJ...534...41R}
  {534, 41}

\bibitem[\protect\citeauthoryear{{Rogers} \& {Peiris}}{{Rogers} \&
  {Peiris}}{2021}]{rogers21}
{Rogers} K.~K.,  {Peiris} H.~V.,  2021, \mn@doi [\prl]
  {10.1103/PhysRevLett.126.071302}, \href
  {https://ui.adsabs.harvard.edu/abs/2021PhRvL.126g1302R} {126, 071302}

\bibitem[\protect\citeauthoryear{{Rollinde}, {Theuns}, {Schaye}, {P{\^a}ris}
  \& {Petitjean}}{{Rollinde} et~al.}{2013}]{rollinde13}
{Rollinde} E.,  {Theuns} T.,  {Schaye} J.,  {P{\^a}ris} I.,   {Petitjean} P.,
  2013, \mn@doi [\mnras] {10.1093/mnras/sts057}, \href
  {https://ui.adsabs.harvard.edu/abs/2013MNRAS.428..540R} {428, 540}

\bibitem[\protect\citeauthoryear{{Rorai} et~al.,}{{Rorai}
  et~al.}{2017}]{rorai17}
{Rorai} A.,  et~al., 2017, \mn@doi [Science] {10.1126/science.aaf9346}, \href
  {http://adsabs.harvard.edu/abs/2017Sci...356..418R} {356, 418}

\bibitem[\protect\citeauthoryear{{Rudie}, {Steidel}  \& {Pettini}}{{Rudie}
  et~al.}{2012}]{rudie12}
{Rudie} G.~C.,  {Steidel} C.~C.,   {Pettini} M.,  2012, \mn@doi [\apjl]
  {10.1088/2041-8205/757/2/L30}, \href
  {http://adsabs.harvard.edu/abs/2012ApJ...757L..30R} {757, L30}

\bibitem[\protect\citeauthoryear{{S{\'a}nchez-Ram{\'{\i}}rez}
  et~al.,}{{S{\'a}nchez-Ram{\'{\i}}rez} et~al.}{2016}]{sanchez16}
{S{\'a}nchez-Ram{\'{\i}}rez} R.,  et~al., 2016, \mn@doi [\mnras]
  {10.1093/mnras/stv2732}, \href
  {http://adsabs.harvard.edu/abs/2016MNRAS.456.4488S} {456, 4488}

\bibitem[\protect\citeauthoryear{{Schaye}, {Theuns}, {Rauch}, {Efstathiou}  \&
  {Sargent}}{{Schaye} et~al.}{2000}]{schaye00}
{Schaye} J.,  {Theuns} T.,  {Rauch} M.,  {Efstathiou} G.,   {Sargent} W.~L.~W.,
   2000, \mn@doi [\mnras] {10.1046/j.1365-8711.2000.03815.x}, \href
  {http://adsabs.harvard.edu/abs/2000MNRAS.318..817S} {318, 817}

\bibitem[\protect\citeauthoryear{{Schaye}, {Aguirre}, {Kim}, {Theuns}, {Rauch}
  \& {Sargent}}{{Schaye} et~al.}{2003}]{2003ApJ...596..768S}
{Schaye} J.,  {Aguirre} A.,  {Kim} T.-S.,  {Theuns} T.,  {Rauch} M.,
  {Sargent} W.~L.~W.,  2003, \mn@doi [\apj] {10.1086/378044}, \href
  {http://adsabs.harvard.edu/abs/2003ApJ...596..768S} {596, 768}

\bibitem[\protect\citeauthoryear{{Seljak}, {McDonald}  \& {Makarov}}{{Seljak}
  et~al.}{2003}]{seljak03}
{Seljak} U.,  {McDonald} P.,   {Makarov} A.,  2003, \mn@doi [\mnras]
  {10.1046/j.1365-8711.2003.06809.x}, \href
  {http://adsabs.harvard.edu/abs/2003MNRAS.342L..79S} {342, L79}

\bibitem[\protect\citeauthoryear{{Seljak}, {Slosar}  \& {McDonald}}{{Seljak}
  et~al.}{2006a}]{seljak06}
{Seljak} U.,  {Slosar} A.,   {McDonald} P.,  2006a, \mn@doi [\jcap]
  {10.1088/1475-7516/2006/10/014}, \href
  {http://adsabs.harvard.edu/abs/2006JCAP...10..014S} {10, 14}

\bibitem[\protect\citeauthoryear{{Seljak}, {Makarov}, {McDonald}  \&
  {Trac}}{{Seljak} et~al.}{2006b}]{uros06}
{Seljak} U.,  {Makarov} A.,  {McDonald} P.,   {Trac} H.,  2006b, \mn@doi
  [Physical Review Letters] {10.1103/PhysRevLett.97.191303}, \href
  {http://adsabs.harvard.edu/abs/2006PhRvL..97s1303S} {97, 191303}

\bibitem[\protect\citeauthoryear{{Slosar} et~al.,}{{Slosar}
  et~al.}{2011}]{slosar11}
{Slosar} A.,  et~al., 2011, \mn@doi [\jcap] {10.1088/1475-7516/2011/09/001},
  \href {http://adsabs.harvard.edu/abs/2011JCAP...09..001S} {9, 1}

\bibitem[\protect\citeauthoryear{{Slosar} et~al.,}{{Slosar}
  et~al.}{2013}]{slosar13}
{Slosar} A.,  et~al., 2013, \mn@doi [\jcap] {10.1088/1475-7516/2013/04/026},
  \href {http://adsabs.harvard.edu/abs/2013JCAP...04..026S} {4, 26}

\bibitem[\protect\citeauthoryear{{Telikova}, {Balashev}  \&
  {Shternin}}{{Telikova} et~al.}{2018}]{telikova19}
{Telikova} K.~N.,  {Balashev} S.~A.,   {Shternin} P.~S.,  2018, arXiv e-prints,
  \href {https://ui.adsabs.harvard.edu/abs/2018arXiv180601319T} {p.
  arXiv:1806.01319}

\bibitem[\protect\citeauthoryear{{Theuns} \& {Zaroubi}}{{Theuns} \&
  {Zaroubi}}{2000}]{theuns00}
{Theuns} T.,  {Zaroubi} S.,  2000, \mn@doi [\mnras]
  {10.1046/j.1365-8711.2000.03729.x}, \href
  {http://adsabs.harvard.edu/abs/2000MNRAS.317..989T} {317, 989}

\bibitem[\protect\citeauthoryear{{Theuns}, {Zaroubi}, {Kim}, {Tzanavaris}  \&
  {Carswell}}{{Theuns} et~al.}{2002}]{theuns02}
{Theuns} T.,  {Zaroubi} S.,  {Kim} T.-S.,  {Tzanavaris} P.,   {Carswell} R.~F.,
   2002, \mn@doi [\mnras] {10.1046/j.1365-8711.2002.05316.x}, \href
  {http://adsabs.harvard.edu/abs/2002MNRAS.332..367T} {332, 367}

\bibitem[\protect\citeauthoryear{{Tytler}, {O'Meara}, {Suzuki}, {Kirkman},
  {Lubin}  \& {Orin}}{{Tytler} et~al.}{2004}]{2004AJ....128.1058T}
{Tytler} D.,  {O'Meara} J.~M.,  {Suzuki} N.,  {Kirkman} D.,  {Lubin} D.,
  {Orin} A.,  2004, \mn@doi [\aj] {10.1086/423293}, \href
  {https://ui.adsabs.harvard.edu/abs/2004AJ....128.1058T} {128, 1058}

\bibitem[\protect\citeauthoryear{{Upton Sanderbeck}, {D'Aloisio}  \&
  {McQuinn}}{{Upton Sanderbeck} et~al.}{2016a}]{upton}
{Upton Sanderbeck} P.~R.,  {D'Aloisio} A.,   {McQuinn} M.~J.,  2016a, \mn@doi
  [\mnras] {10.1093/mnras/stw1117}, \href
  {http://adsabs.harvard.edu/abs/2016MNRAS.460.1885U} {460, 1885}

\bibitem[\protect\citeauthoryear{{Upton Sanderbeck}, {D'Aloisio}  \&
  {McQuinn}}{{Upton Sanderbeck} et~al.}{2016b}]{sanderbeck16}
{Upton Sanderbeck} P.~R.,  {D'Aloisio} A.,   {McQuinn} M.~J.,  2016b, \mn@doi
  [\mnras] {10.1093/mnras/stw1117}, \href
  {https://ui.adsabs.harvard.edu/abs/2016MNRAS.460.1885U} {460, 1885}

\bibitem[\protect\citeauthoryear{{Vernet} et~al.,}{{Vernet}
  et~al.}{2011}]{vernet11}
{Vernet} J.,  et~al., 2011, \mn@doi [\aap] {10.1051/0004-6361/201117752}, \href
  {http://adsabs.harvard.edu/abs/2011A%26A...536A.105V} {536, A105}

\bibitem[\protect\citeauthoryear{{Viel} \& {Haehnelt}}{{Viel} \&
  {Haehnelt}}{2006}]{viel06}
{Viel} M.,  {Haehnelt} M.~G.,  2006, \mn@doi [\mnras]
  {10.1111/j.1365-2966.2005.09703.x}, \href
  {http://adsabs.harvard.edu/abs/2006MNRAS.365..231V} {365, 231}

\bibitem[\protect\citeauthoryear{{Viel}, {Matarrese}, {Heavens}, {Haehnelt},
  {Kim}, {Springel}  \& {Hernquist}}{{Viel} et~al.}{2004a}]{viel04bis}
{Viel} M.,  {Matarrese} S.,  {Heavens} A.,  {Haehnelt} M.~G.,  {Kim} T.-S.,
  {Springel} V.,   {Hernquist} L.,  2004a, \mn@doi [\mnras]
  {10.1111/j.1365-2966.2004.07404.x}, \href
  {http://adsabs.harvard.edu/abs/2004MNRAS.347L..26V} {347, L26}

\bibitem[\protect\citeauthoryear{{Viel}, {Haehnelt}  \& {Springel}}{{Viel}
  et~al.}{2004b}]{viel04}
{Viel} M.,  {Haehnelt} M.~G.,   {Springel} V.,  2004b, \mn@doi [\mnras]
  {10.1111/j.1365-2966.2004.08224.x}, \href
  {http://adsabs.harvard.edu/abs/2004MNRAS.354..684V} {354, 684}

\bibitem[\protect\citeauthoryear{{Viel}, {Weller}  \& {Haehnelt}}{{Viel}
  et~al.}{2004c}]{viel04hrwmap}
{Viel} M.,  {Weller} J.,   {Haehnelt} M.~G.,  2004c, \mn@doi [\mnras]
  {10.1111/j.1365-2966.2004.08498.x}, \href
  {http://adsabs.harvard.edu/abs/2004MNRAS.355L..23V} {355, L23}

\bibitem[\protect\citeauthoryear{{Viel}, {Lesgourgues}, {Haehnelt}, {Matarrese}
   \& {Riotto}}{{Viel} et~al.}{2005}]{viel05}
{Viel} M.,  {Lesgourgues} J.,  {Haehnelt} M.~G.,  {Matarrese} S.,   {Riotto}
  A.,  2005, \mn@doi [\prd] {10.1103/PhysRevD.71.063534}, \href
  {http://adsabs.harvard.edu/abs/2005PhRvD..71f3534V} {71, 063534}

\bibitem[\protect\citeauthoryear{{Viel}, {Becker}, {Bolton}, {Haehnelt},
  {Rauch}  \& {Sargent}}{{Viel} et~al.}{2008}]{viel08}
{Viel} M.,  {Becker} G.~D.,  {Bolton} J.~S.,  {Haehnelt} M.~G.,  {Rauch} M.,
  {Sargent} W.~L.~W.,  2008, \mn@doi [Physical Review Letters]
  {10.1103/PhysRevLett.100.041304}, \href
  {http://adsabs.harvard.edu/abs/2008PhRvL.100d1304V} {100, 041304}

\bibitem[\protect\citeauthoryear{Viel, Becker, Bolton  \& Haehnelt}{Viel
  et~al.}{2013a}]{Viel_2013}
Viel M.,  Becker G.~D.,  Bolton J.~S.,   Haehnelt M.~G.,  2013a, \mn@doi
  [Physical Review D] {10.1103/physrevd.88.043502}, 88

\bibitem[\protect\citeauthoryear{{Viel}, {Becker}, {Bolton}  \&
  {Haehnelt}}{{Viel} et~al.}{2013b}]{viel13WDM}
{Viel} M.,  {Becker} G.~D.,  {Bolton} J.~S.,   {Haehnelt} M.~G.,  2013b,
  \mn@doi [\prd] {10.1103/PhysRevD.88.043502}, \href
  {http://adsabs.harvard.edu/abs/2013PhRvD..88d3502V} {88, 043502}

\bibitem[\protect\citeauthoryear{{Viel}, {Schaye}  \& {Booth}}{{Viel}
  et~al.}{2013c}]{viel13}
{Viel} M.,  {Schaye} J.,   {Booth} C.~M.,  2013c, \mn@doi [\mnras]
  {10.1093/mnras/sts465}, \href
  {http://adsabs.harvard.edu/abs/2013MNRAS.429.1734V} {429, 1734}

\bibitem[\protect\citeauthoryear{{Walther}, {Hennawi}, {Hiss}, {O{\~n}orbe},
  {Lee}, {Rorai}  \& {O'Meara}}{{Walther} et~al.}{2018}]{walther18}
{Walther} M.,  {Hennawi} J.~F.,  {Hiss} H.,  {O{\~n}orbe} J.,  {Lee} K.-G.,
  {Rorai} A.,   {O'Meara} J.,  2018, \mn@doi [\apj] {10.3847/1538-4357/aa9c81},
  \href {https://ui.adsabs.harvard.edu/abs/2018ApJ...852...22W} {852, 22}

\bibitem[\protect\citeauthoryear{{Walther}, {O{\~n}orbe}, {Hennawi}  \&
  {Luki{\'c}}}{{Walther} et~al.}{2019}]{walther19}
{Walther} M.,  {O{\~n}orbe} J.,  {Hennawi} J.~F.,   {Luki{\'c}} Z.,  2019,
  \mn@doi [\apj] {10.3847/1538-4357/aafad1}, \href
  {https://ui.adsabs.harvard.edu/abs/2019ApJ...872...13W} {872, 13}

\bibitem[\protect\citeauthoryear{{Wu}, {McQuinn}, {Kannan}, {D'Aloisio},
  {Bird}, {Marinacci}, {Dav{\'e}}  \& {Hernquist}}{{Wu} et~al.}{2019}]{wu19}
{Wu} X.,  {McQuinn} M.,  {Kannan} R.,  {D'Aloisio} A.,  {Bird} S.,  {Marinacci}
  F.,  {Dav{\'e}} R.,   {Hernquist} L.,  2019, \mn@doi [\mnras]
  {10.1093/mnras/stz2807}, \href
  {https://ui.adsabs.harvard.edu/abs/2019MNRAS.490.3177W} {490, 3177}

\bibitem[\protect\citeauthoryear{{Y{\`e}che}, {Palanque-Delabrouille}, {Baur}
  \& {du Mas des Bourboux}}{{Y{\`e}che} et~al.}{2017}]{yeche17}
{Y{\`e}che} C.,  {Palanque-Delabrouille} N.,  {Baur} J.,   {du Mas des
  Bourboux} H.,  2017, \mn@doi [\jcap] {10.1088/1475-7516/2017/06/047}, \href
  {https://ui.adsabs.harvard.edu/abs/2017JCAP...06..047Y} {2017, 047}

\bibitem[\protect\citeauthoryear{{Zaldarriaga}, {Seljak}  \&
  {Hui}}{{Zaldarriaga} et~al.}{2001a}]{zaldarriaga01}
{Zaldarriaga} M.,  {Seljak} U.,   {Hui} L.,  2001a, \mn@doi [\apj]
  {10.1086/320066}, \href {http://adsabs.harvard.edu/abs/2001ApJ...551...48Z}
  {551, 48}

\bibitem[\protect\citeauthoryear{{Zaldarriaga}, {Seljak}  \&
  {Hui}}{{Zaldarriaga} et~al.}{2001b}]{zaldarriaga}
{Zaldarriaga} M.,  {Seljak} U.,   {Hui} L.,  2001b, \mn@doi [\apj]
  {10.1086/320066}, \href
  {https://ui.adsabs.harvard.edu/abs/2001ApJ...551...48Z} {551, 48}

\bibitem[\protect\citeauthoryear{{Zaldarriaga}, {Scoccimarro}  \&
  {Hui}}{{Zaldarriaga} et~al.}{2003}]{zaldarriaga03}
{Zaldarriaga} M.,  {Scoccimarro} R.,   {Hui} L.,  2003, \mn@doi [\apj]
  {10.1086/374407}, \href {http://adsabs.harvard.edu/abs/2003ApJ...590....1Z}
  {590, 1}

\bibitem[\protect\citeauthoryear{{du Mas des Bourboux} et~al.,}{{du Mas des
  Bourboux} et~al.}{2017}]{bourboux17}
{du Mas des Bourboux} H.,  et~al., 2017, \mn@doi [\aap]
  {10.1051/0004-6361/201731731}, \href
  {https://ui.adsabs.harvard.edu/abs/2017A&A...608A.130D} {608, A130}

\makeatother
\end{thebibliography}

\appendix 

\clearpage

\section{Spectrograph resolution correction}
\label{ap:resolution}

Studies of the small scale \lya\ flux power spectrum rely on correcting for spectral smoothing that owes to the spectrograph's resolution. While this is a small correction for studies using high-resolution Keck/HIRES or VLT/UVES spectra even at our maximum wavenumber of $k_{\rm max} = 10^{-1.2}$s~km$^{-1}$, the correction amounts to a factor of as much several for the X-Shooter spectrograph.  Previous studies have disagreed on the effective resolution of X-Shooter at the 20\% level, which can lead to large factor of two differences in the reported power at $k_{\rm max}$ \citep{irsic17, yeche17, walther18}. In this section, we describe the implementation of the resolution correction that is used in this paper.

\begin{figure*}
\includegraphics[scale=0.5]{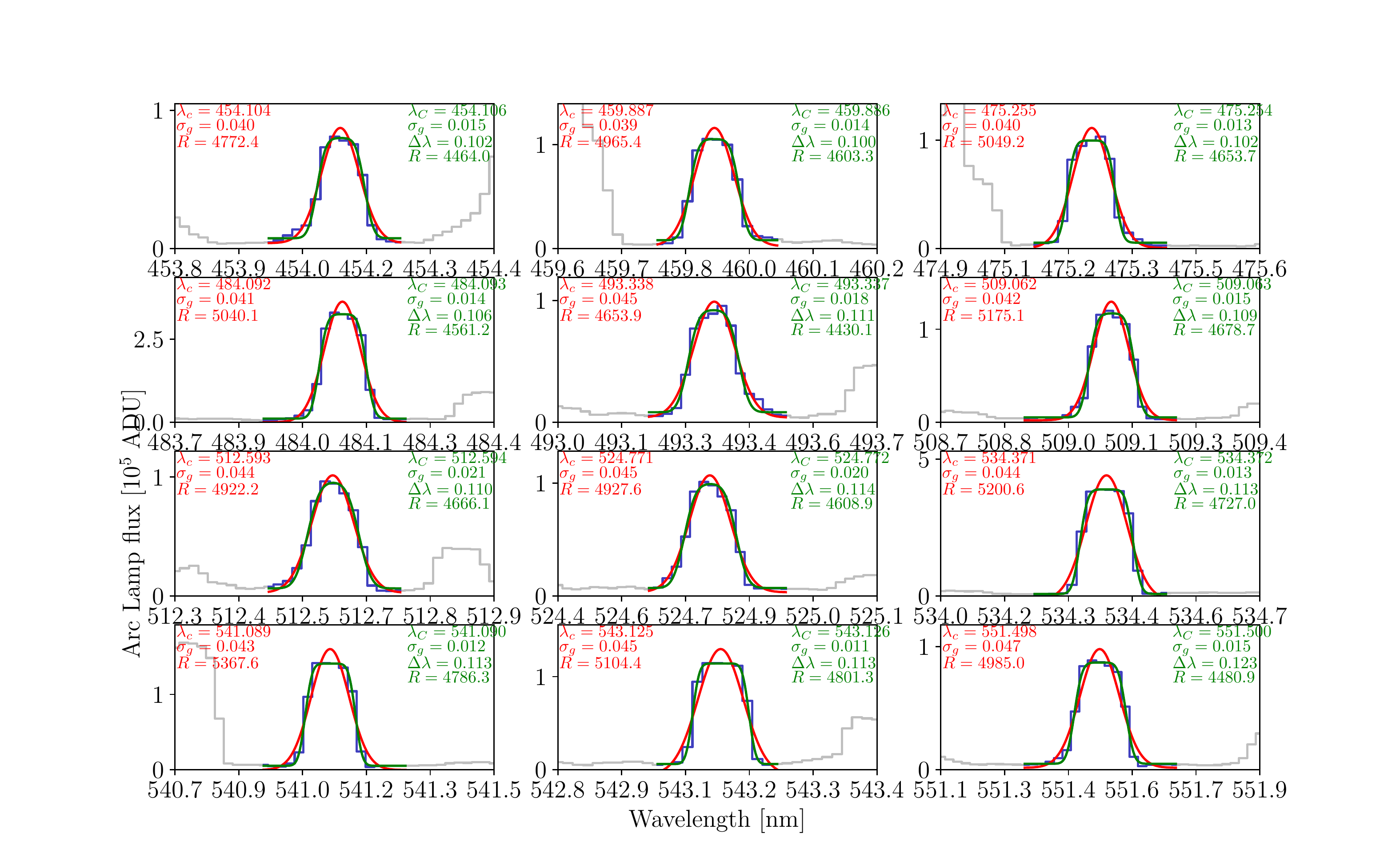}
\caption{The panels show the X-Shooter UV arm ThAr calibration arc lamp spectra of select clean lines that span the wavelength range used here. The arc lamp spectrum in each panel is shown in grey, and highlighted in blue is the region around the line where the fits were performed. In each panel, two fitting models -- Gaussian model (in red) and Box model (in green) -- are overplotted for the best fit parameters, which are quoted in the top left (red) and right (green). \label{fig:uvb_arc}}
\end{figure*}

 First, in order to determine the resolution when the spectrograph's slit is fully illuminated, we use spectra taken of an arc lamp using $1\times2$ binning of spectral pixels.  We use the same slit widths as our data set: $1.0''$ for the UVB arm and $0.9''$ for the VIS arm. The arc-lamp spectra were reduced using the ESO pipeline. The most isolated lines as determined by visual inspection were fitted with a Gaussian.  These fits roughly reproduce the results of a similar analysis in \citet{walther18} that found $R \equiv \lambda/\Delta \lambda \approx 5000$, where $\Delta \lambda$ is the full width half maximum.   See Figure~\ref{fig:uvb_arc}, which shows this exercise for the lines analyzed in the UVB arm.  The blue histograms are the observed line profiles, and the red coloured lines are the best-fit Gaussian models.  The best-fit parameters in each of the panels are also shown in red text.  
  
 However, a Gaussian is not expected to be a good model for the line spread function of a fully illuminated slit, and several of the arc-lamp lines clearly show a more box-like profile.  Thus, we explore a more physically motivated model for the line profile, where the line profile is modeled as a boxcar filter convolved with a Gaussian. This model captures that the slit is a boxcar filter, allowing light to go through spectrograph at different angles. This angular dispersion is larger than the intrinsic dispersion of the spectrograph that owes to diffraction. While this diffractive broadening depends on the details of the spectrograph, such as the number of grooves illuminated, we approximate this lesser contribution with a Gaussian kernel.    This model we refer to as the `Box model'. Unlike our previous one parameter Gaussian model, the Box model is described by two parameters: the full width of the boxcar $R \equiv \lambda/\Delta \lambda$ and the standard deviation of the Gaussian $\sigma_g$. 
   The broader boxcar width $R$ essentially is the FWHM resolution in this model.
 The results of the Box model fits are shown in green in Figure~\ref{fig:uvb_arc}, with green text giving the best-fit parameters.

\begin{figure}
\includegraphics[width=0.50\textwidth]{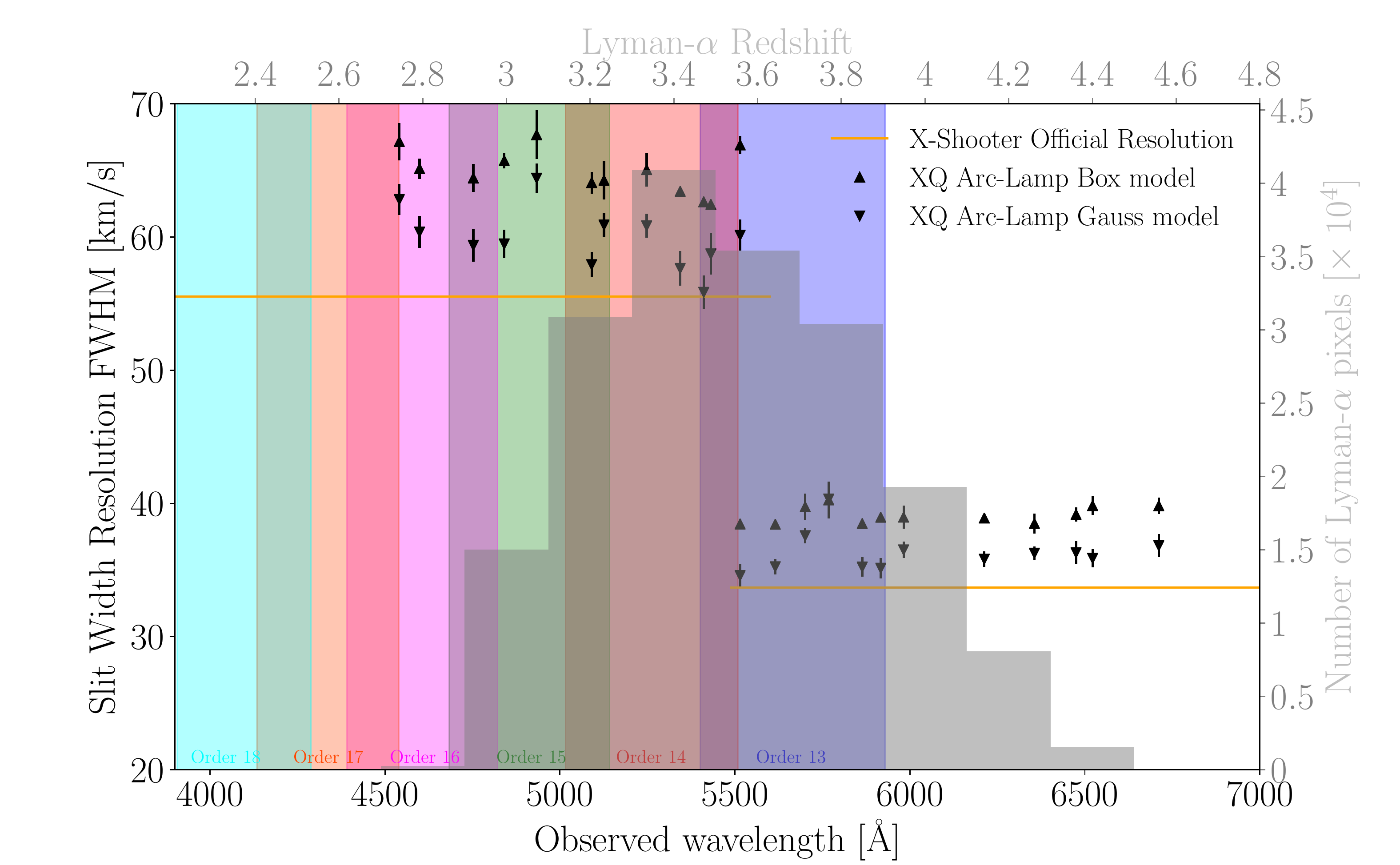}
\caption{Shows the slit width resolution (FWHM in $\mathrm{km\,s^{-1}}$) as estimated from fitting arc lamp lines for the UVB and VIS arms of the XQ-Shooter spectrograph. The two orange lines correspond to the official published X-Shooter resolution for the nominal slit width of the XQ-100 observations, which is: 1.0" for the UVB arm ($R=5400$, shown at shorter wavelenghts); and 0.9" for the VIS arm ($R=8900$, shown at longer wavelengths). The black data points are the measurements of the arc lamp emission lines in both spectral arms for the nominal slit width, using two different methods: the Gaussian model (inverted triangles) and the Box model (triangles).  The Box model is more physically motivated and yields larger FWHMs, possibly explaining some of the discrepant resolutions quoted in previous literature.  The resolution correction applied to the measurements in this paper Fourier transforms the full profile of the Box model.\label{fig:xq_resol}}
\end{figure}

The summary of the fits to the slit-arc spectra are presented in Fig.~\ref{fig:xq_resol}, comparing both models (Gaussian and Box), and comparing the results to the official X-Shooter resolution.   The Box model recovers a $\sim 10\%$ larger FWHM resolution compared to the Gaussian model.  (This difference should be thought of as purely parametric.  These are just different functions!) The VIS arm shows very little wavelength dependent scatter, while the UVB arm shows more structure. This could be partially due to two effects: firstly, the lines in the UVB are more likely to be blends as the lower resolution in this arm makes it harder to find all blends; secondly, the Echelle orders of the spectrograph (visualized by the background colors in Fig.~\ref{fig:xq_resol}) have more overlap in the UVB arm, which might cause resolution variation that owes to the variations in Eschelle orders used. 

We compress the results of our arc-fits to a single resolution number for each of the arms.  Using medians to reduce biases from blended lines, we find a FWHM of $\Delta v =  64.9\;\mathrm{km~s}^{-1}$ for the UVB arm and $\Delta v = 39.1\;\mathrm{km~s}^{-1}$ for the VIS arm.  (These numbers set the width of the boxcar filter.  For the Gaussian that is then convolved with this boxcar, we find a width of $8.76\;\mathrm{km~s}^{-1}$ and $6.64\;\mathrm{km~s}^{-1}$.) 

The second characteristic of the XQ-100 observations that needs to be addressed to properly account for the resolution correction is the atmospheric seeing. The seeing of our XQ-100 observations is generally better than the slit width of the observations \citep{yeche17}, which would result in a lower effective FWHM resolution values when correcting the flux power spectrum measurements of the XQ-100 data (because the angular dispersion through the spectrograph is reduced). In our favored Box car model, there is a simple way to account for this effect.  In a geometric optics picture, each point illuminated the slit contributes to one point in the boxcar profile, with one side of slit contributing to one edge and the other side contributing to the other edge.  Since our model is a Box car convolved with a Gaussian, then in this picture the Gaussian owes to diffraction.  This picture allows us to then weight each part of the slit depending on the seeing, which we model as having a Gaussian point spread function for which the width has been fit observationally during each exposure.  Thus, in contrast to the arclamp spectra in which the slit is uniformly illuminated and the full boxcar populated, each point in the boxcar is weighted by the Gaussian point spread function.  We do this in a manner that weights each point in the slit by the Gaussian seeing profile.  

Thus, this Boxcar model allows us to model the line spread function of the spectrograph for a given seeing.  We are interested in the Fourier transform of the line spread function ($W_{s,X}(k, z)$ that appears in eqn.~\ref{eqn:Pkest}). We find that once we Fourier transform the Boxcar model, the resolution window function can be very well approximated by a Gaussian over the wavenumber range of interest.  We then fit a Gaussian model with standard deviation $\sigma_R$ to our calculations of the seeing dependent window function and use the form
\begin{eqnarray}
  \sigma_R &=&  a  + b (x-0.65)+ c(x-0.65)^2 + d (x-0.65)^3\nonumber \\ &&~~~~~ \text{~if $x<0.8$;} \nonumber\\
    \sigma_R &=&  e  + f (x-1) +g(x-1)^2 +  h(x-1)^3 \nonumber \\ &&~~~~~\text{~otherwise,} \nonumber
\end{eqnarray}
where $x \equiv \theta_{\rm atm}(s)/\theta_{\rm slit}$, $\theta_{\rm atm}(s)$ is the reported seeing for quasar $s$ and $\theta_{\rm slit}$ is the size of the slit. In units of km~s$^{-1}$, we find $(a,b,c,d,e,f,g,h)$ = {\scriptsize $(17.7982, -9.8992, -14.7344, -15.366, 19.9168,-3.4890, -4.6915, -4.6875)$} for the UV arm and $(a,b,c,d,e,f,g,h)$ = {\scriptsize $( 11.4691, -5.2146, -8.0595, -8.7484, 12.5689, -1.7810, -2.4529, -2.4894)$} for the VIS.
Mathematica notebooks with the full calculation pertaining to our resolution correction can be provided upon request. These fits are then used when computing $W_{s,X}(k, z)$ per equation~(\ref{eqn:WSX}).  Our parametrization of $\sigma_R$ is a function of the seeing, which can vary over an observation.  We use the minimum and maximum reported seeing to calculate $W_{s,X}$, and toss out modes where this causes differences beyond our error tolerance (c.f.~eqn.~\ref{eqn:seeing}).  The average of these two $\sigma_R$ is used as our resolution correction.  Finally, we allow for 20\% uncertainty in the resolution parameter, $\sigma_R$, in our main analysis.  Our seeing dependence of the resolution is considerably weaker than the model of \citet{yeche17} where $\sigma_R \propto x$. This $\sigma_R \propto x$ scaling should only hold when $x$ is considerably smaller than $1$ and is not a good approximation over the seeings present in XQ-100.

\section{Metal contamination and Damped \lya\ Absorbers}
\label{ap:metals}

\begin{figure}
\includegraphics[scale=0.40]{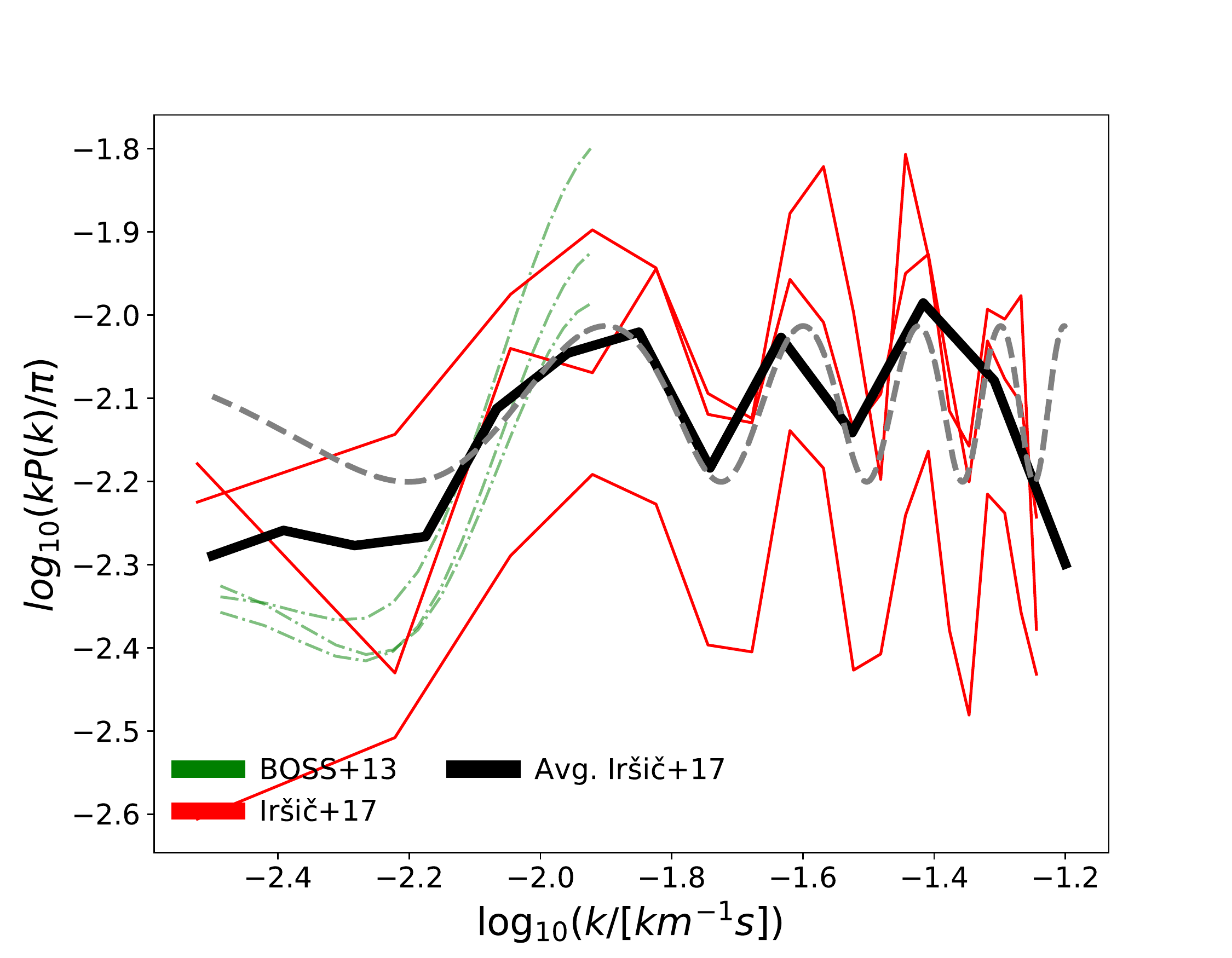}
\caption{Measurements of the red-side metal power spectra described in Appendix~\ref{ap:metals}.  The solid red curves show the XQ-100 measurement of \citet{irsic17} at $z=3.8, 4.0, 4.2$, ordered in decreasing amplitude.  The thick solid black curve averages these three estimates.  The green curves are the SDSS/BOSS measurement of the red-side metal power from \citet{palanque13} for the same redshifts. 
\label{fig:metals}}
\end{figure}

\begin{figure*}
\vspace*{-1cm}
\begin{center}
\includegraphics[scale=.75]{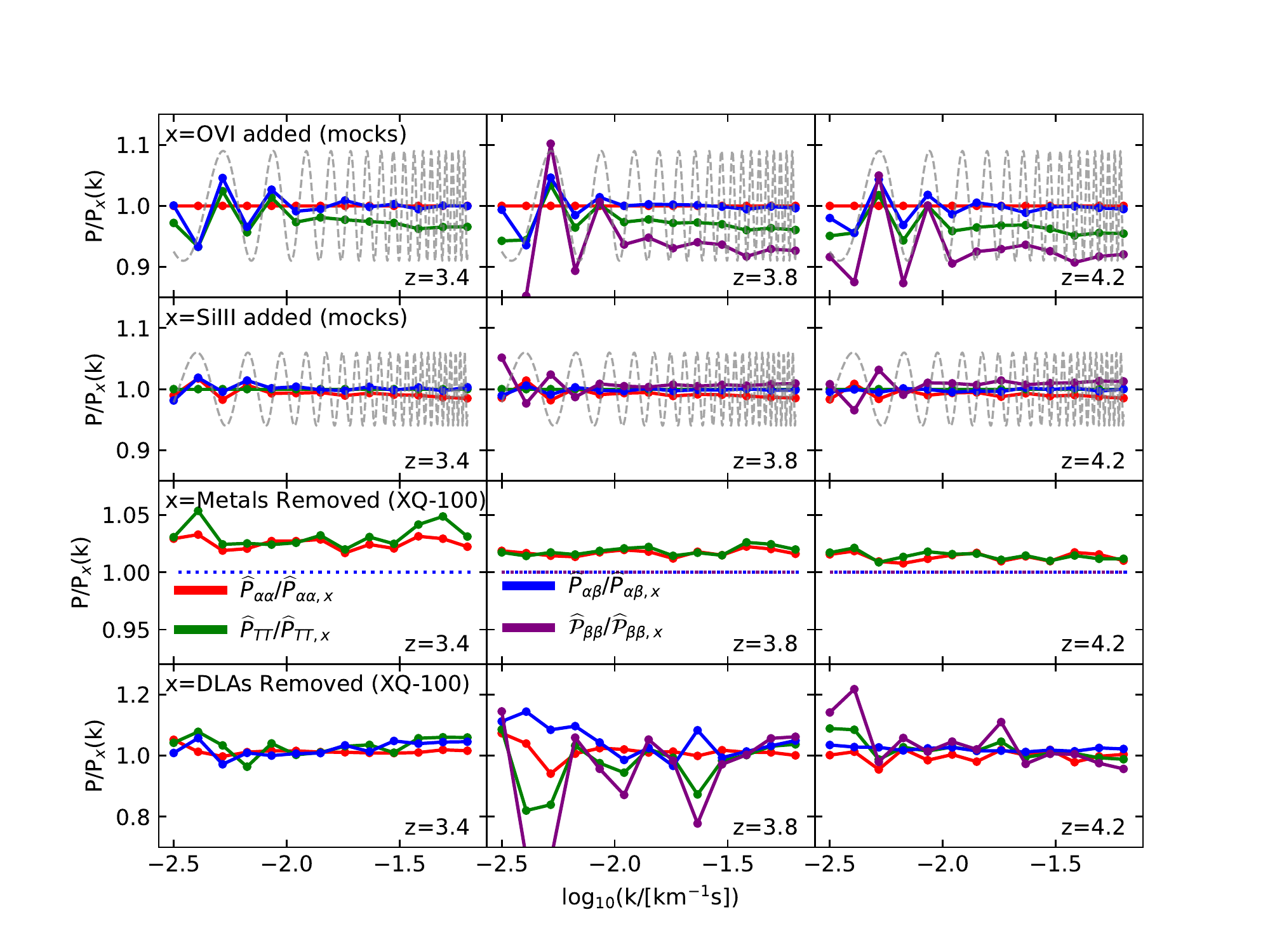}
\vspace{-1.0cm}
\end{center}
\caption{The effect of different systematics for three of the redshift bins we consider.  The first and second rows show respectively predictions for the fractional impact of resonant \OVI\ and \SiIII\ absorption on our power estimates, where this metal contamination is modeled in the manner described in Appendix~\ref{ap:metals}. 
The grey dashed lines show the expected phase from resonances with \OVI\ $\lambda1032$\AA\ and \SiIII\ $\lambda1207$\AA. The third row shows the fractional effect on our estimated power from our non-resonant metal power correction.  Non-resonant metals have no effect on $P_{\alpha \beta}$ and, to the extent they contaminate equally the Ly$\beta$ and foreground Ly$\alpha$ forests (as our procedure assumes), ${\cal P}_{\beta \beta}$.  The fourth row shows the fractional effect of our DLA correction. 
\label{fig:twelve_panels}} 
\end{figure*}

We use estimates for the contaminating metal power spectrum, $\widehat{P}_M(k_i, z_i)$, given in \citet{irsic17} for the same XQ-100 data set. These estimates use the QSO-frame wavelength range just redward of the quasar \lya\ line, $ 1268 < \lambda_{\rm metal} < 1380$\AA, to estimate the forest metal contamination. Isolating to these wavelengths captures the bulk of metal absorption in the \lya\ forest at $z = \lambda_{\rm metal}/\lambda_\alpha(z_{\rm qso}+1)-1$ from lines redward of \lya, and similarly for the \lyb\ forest. Because the redside contains the strongest metal absorption line, \CIV\ $\lambda,\lambda 1548,1550$ \AA, the redside contribution to the power should be larger than the blueside.  The three solid red curves Figure~\ref{fig:metals} shows the XQ-100 estimates of \citet{irsic17} for $z= 3.8, 4.0, 4.2$.  The oscillatory behavior owes to the doublet structure of the \CIV\ line.  The figure also shows the measurement of BOSS in \citet{palanque13}, which used many more spectra (green curves) for the same three redshifts.  The BOSS metal power estimates show less variation in redshift than those of \citet{irsic17}, suggesting that the variation between the three redshift bins in \citet{irsic17} owes largely to cosmic variance and noise.  Thus, to subtract metals we average these three \citet{irsic17} measurements (which yields the solid black curve) and, then, subtract this redshift-average power off of $P_{\alpha\alpha}$ and $P_{TT}$ to obtain our power spectra as given in eqn.~(\ref{eqn:Pkest}). We note that the relative size of the metal correction is less than 3\%, as shown in the third row in Fig.~\ref{fig:twelve_panels}.  The cross power, $P_{\alpha \beta}$, is unaffected by non-resonant metals -- metals that fall far enough from our \HI\ lines that their effect is not resonantly enhanced at the beat wavenumber.  We consider the resonant effect of metals below.  Furthermore, to the extent the metal power does not evolve with redshift as we have assumed, our estimates for ${\cal P}_{\beta \beta}$ are unaffected by metal contamination.

This metal subtraction scheme does not capture the beating effect of metals that have wavelengths close to our transitions \citep{mcdonald06}. The metal lines that are the most problematic for our analysis are \SiIII\ $\lambda$1207\AA\ (2000km~s$^{-1}$ from the \lya\ line), and \OVI\ $\lambda,\lambda$1032, 1038\AA\ (1800,~ 3500km~s$^{-1}$ from \lyb). The beating from \SiIII\ is famously apparent in SDSS analyses of the \lya\ forest power spectra at low wavenumbers \citep{mcdonald06}.  \OVI\ is by far the strongest absorption line at these redshifts for moderately overdense systems \citep{2008ApJ...689..851A} and so a similar beating effect may be important between \lyb\ and \OVI.  

To estimate the potential contribution of resonant metals, we have created mock spectra with \SiIII\ and \OVI\ absorption.  To do this, we used the mean relationship between $\tau_{\rm SiIII}(\lambda = 1207\text{\AA})$ or $\tau_{\rm OVI}(\lambda = 1032\text{\AA})$ and the \HI\ Ly$\alpha$ optical depth $\tau_{\alpha}$ measured in \citet{2004ApJ...602...38A} and \citet{2008ApJ...689..851A} using the pixel optical depth method.  These studies find an average mapping of $\tau_{\rm SiIII} = [10^{-4}-10^{-3}] \tau_{\alpha}$ and $\tau_{\rm OVI} = [10^{-3}- 10^{-2.5}] \tau_{\alpha}$, where these ranges have been chosen to generously bracket the findings of these studies. We also put in the other doublet of the \OVI\ line by using the ratio of oscillator strengths.  The top two rows in Fig.~\ref{fig:twelve_panels} show the fractional effect of these resonances with \OVI\ and \SiIII\ assuming the maximum multiplier in the ranges quoted above of $\tau_{\rm OVI} = 10^{-2.5} \tau_{\alpha}$ and $\tau_{\rm SiIII} = 10^{-3.5} \tau_{\alpha}$.  The dashed curves are to guide the eye show the effective beat frequency of the strongest \OVI\ line, $\cos([1800 {\rm km~s}^{-1}] k)$ for the 1800~km~s$^{-1}$ shift form \OVI\ $\lambda1032$~\AA, and the analogous cosine for \SiIII.  The solid lines show these estimates for fractional amplitude of this resonant metal contamination for the different power spectra we measure.  Most important for our study are $P_{\alpha \alpha}$  and $P_{\alpha \beta}$, where the effects are small, peaking at $\sim 5\%$ at low wavenumbers.  However, this model predicts that for ${\cal P}_{\beta\beta}$, resonant \OVI\ can be a 10\% effect across the board. Since we have taken the maximum of our range of mapping from \HI\ to \OVI, our models may overestimate the effect. However, we are not modeling the stochasticity of the cosmic metal distribution, such that some \HI\ systems will have much stronger lines. This results in a larger stochastic component to the power that could increase the contamination over the level predicted.

We now turn to a second form of contamination, from the damping wings of \HI\ lines from dense hydrogen systems associated with intervening galaxies. These so-called damped \lya\ absorbers (DLAs) are a known source of large-scale power that can bias analyses as damping wings are not included in our mock spectra (nor do our simulations likely capture the properties of DLAs accurately).   DLAs that contaminate the XQ-100 quasar spectra are masked using the DLA sample of the XQ-100 survey team \citep{sanchez16}.  For each DLA, we mask out the spectral region within $1/2$ of the equivalent width from the \lya\ and \lyb\ line centers. In the region outside of the mask, we then divide out DLA absorption profile using the $N_{\rm HI}$ measured in \citet{sanchez16} to remove any remaining effect from the broad damping wings. In the bottom row in Fig.~\ref{fig:twelve_panels}, the ratio between the power spectrum with and without DLAs show that DLAs are a significant contaminant, with a $10-20\%$ effect at $z=3.8$ in $P_{TT}$ and ${\cal P}_{\beta\beta}$. The effects are smaller, $<5\%$ ($\lesssim 10\%$), on the $P_{\alpha \alpha}$ (${P}_{\alpha\beta})$ power spectra we use for our thermal analysis. This shows that DLAs are not a negligible contaminant and so our correction is important.

\section{Wavelength calibration}
\label{ap:calibration}

\begin{figure*}
\includegraphics[scale=0.45]{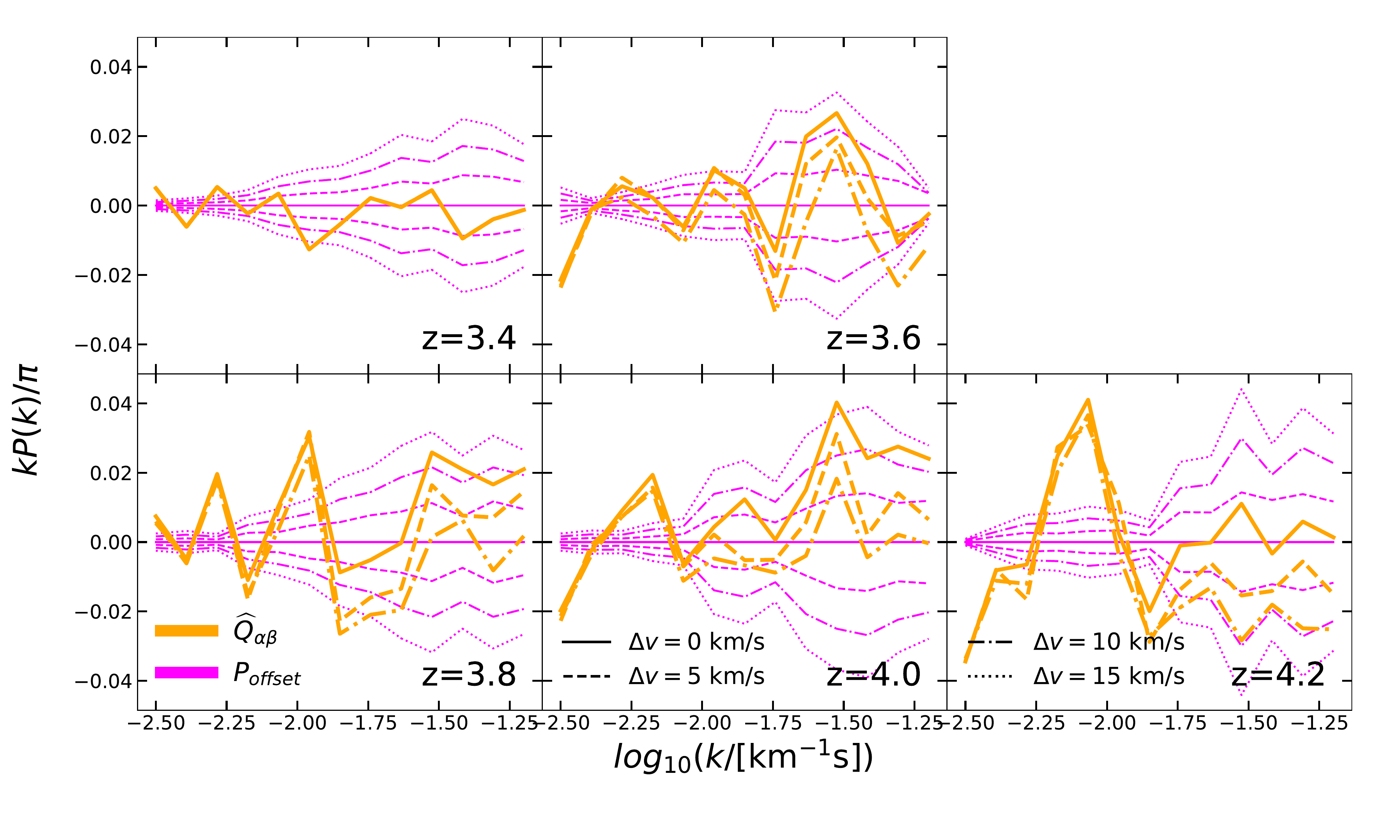}
\vspace{-.8cm}
\caption{The imaginary component of the \lya--\lyb\ cross power spectrum, denoted as $Q_{\alpha \beta}$, as well as the real part of it multiplied by $\tan(k\Delta v)$ for $\Delta v = 0,5,~10, ~15~$km~s$^{-1}$ (pink curves denoted by $P_{\rm offset}$, where $P_{\rm offset} \equiv P_{\alpha \beta} \tan(k\Delta v)$).  Also shown are the resulting $Q_{\alpha \beta}$ if we input a shift into the analysis of the specified amounts.  These pink curves represent the expected imaginary component in the presence of a systematic wavelength miscalibration of size $\Delta v$.  The measured $Q_{\alpha \beta}$ (solid orange) may suggest a slight systematic offset, particularly at $z =3.8$ and $z=4.0$ ($z=3.4$ is only redshift where both measurements use the UV arm and so any offset would be surprising).  To see this, first note that at low $k$ behavior of the orange curves is just noise, and focus on the offsets at the higher $k$. 
\label{fig:calibation}}
\end{figure*}

Another concern is the wavelength calibration of X-Shooter.  Wavelength calibration errors become substantial once they are comparable to the thermal scale of  $\sim10$~km~s$^{-1}$, as quantified below.   Within a single arm of the spectrograph, the wavelength calibration is done by centroiding either arc or skylines, and then fitting a polynomial for how the wavelength depends on pixel number.  This procedure is first applied to a single well-characterized “master” exposure, which is given extra scrutiny.  This master fit is translated onto other exposures using offsets calculated from the positions of the skylines \citep{lopez16}. The expectation is that the errors in interpolation are well below our $10$~km~s$^{-1}$ tolerance (G.~Becker private communication). Another systematic shift owes to the relative positions of the objects in the slits for the UV and VIS arms (as the skylines illuminate the whole slit and so calibrating off of them does not reflect these offsets).  For example, observer centrioiding differences will result in a correlated shift but that does not cancel as the arms have different resolutions. 
Additionally, the slits themselves (as spectrograph arms are observed simultaneously) may not be perfectly aligned mechanically, creating a similar effect.  As the full width half maximum for the UVB arm is $\sim 50$km~s$^{-1}$, a 0.2'' offset in the 1'' slit would result in a $10$km~s$^{-1}$ shift relative to an arm where the source is not offset in its sit. If the centroid is similarly displaced in both slits as per observer miscentering, the effect would be more than halved reflecting the difference in resolutions between the UV and VIS arms.  

For the auto-power spectra, wavelength calibration is likely not a significant systematic as this measurement uses pixels that are relatively close to each other and, furthermore, systematic shifts between arms would have no effect.  However, our cross power spectra typically correlate different spectrograph arms for all redshifts besides $z=3.4$:  All of our \lyb\ measurements use the UVB arm, and these are correlated with \lya, which below $z<3.5$ also uses the UVB arm and $z>3.60$ uses the VIS arm.  

As a shift in velocity of $\Delta v$ results in a mode acquiring an additional phase of $\Delta v k$, such that the imaginary and real parts of the cross power spectrum, $P_{\alpha \beta}$ and $Q_{\alpha \beta}$, are changed from the true purely-real power spectrum  $P^T_{\alpha \beta}$, becoming
\begin{equation}
P_{\alpha \beta} =P^T_{\alpha \beta}(k) \cos[\Delta v k]; ~~~~
Q_{\alpha \beta} =P^T_{\alpha \beta}(k) \sin[\Delta v k].
\end{equation}
where $P^T_{\alpha \beta}(k)$ is the true cross power spectrum. For the wavenumbers we measure and the likely shifts, $\Delta v k$ is smaller than unity and, hence, the effect is much larger in the imaginary term ($Q_{\alpha \beta}$), which we use below to diagnose any offsets.  A shift of $\Delta v=5 \;(10)$km~s$^{-1}$ would bias the cross power measurement studied in the main body of this paper by a factor of $\cos[\Delta v k_{\rm max}] = 0.95 \;(0.81)$, where $k_{\rm max} =10^{-1.2} {\rm s^{-1} km}$ is the center of our highest wavenumber bin used in our analysis. Thus, given our errors, the tipping point where the real component (i.e. our power spectrum measurement) is appreciably biased is offsets of at least $\Delta v \approx 10$km~s$^{-1}$. We make the previous statement for the case where there is no systematic error budget for resolution as it reflects our statistical errors:  Really the tipping point in our thermal analysis in \S~\ref{sec:analysis} should be at larger $\Delta v$ than $10$km~s$^{-1}$ as our 20\% resolution error budget really allows for a factor of $\sim 2$ error at $k_{\rm max}$.

Figure~\ref{fig:calibation} shows this systematic check. The yellow solid lines are the imaginary component of the \lya--\lyb\ cross power spectrum, $\widehat{Q}_{\alpha \beta}$.  The pink curves are the real part of it multiplied by $\tan(k\Delta v)$ for $\Delta v = 0,5,~10, ~15$km~s$^{-1}$ (denoted in Figure as $P_{\rm offset}$, where $P_{\rm offset} \equiv \widehat{P}_{\alpha \beta} \tan(k\Delta v)$).  These pink curves represent the expected imaginary component in the presence of a systematic wavelength miscalibaration of size $\Delta v$.   Also shown are the resulting $\widehat{Q}_{\alpha \beta}$ if we input a shift into the analysis of the specified amounts. The measured $Q_{\alpha \beta}$ (solid orange) may suggest a slight systematic offset, particularly at $z =3.8$ and $z=4.0$ ($z=3.4$ is only redshift where both measurements use the UV arm and so an offset would be surprising).  To see this, first note that at low $k$ the structure is set by noise, and focus on the offsets at the higher $k$.  For example, at $z=4.2$ there is no appreciable offset at high-$k$ without a shift, making it so that when we shift by $5$ and $10$km/s the resulting orange curves follows the pink predictions for the effect of such offsets.  The $\widehat{Q}_{\alpha \beta}$ in the $z=3.8$ and $4.0$ bins suggest at $10$km/s offset, perhaps (again look at the orange solid curve, and the dashed and dot dashed orange curves then show how putting in a possibly corrective offset changes $\widehat{Q}_{\alpha \beta}$).   A $10$km/s shift should not affect our thermal constraints in the paper: we have redone this analysis with a $5$km/s shift, splitting the difference with unoffset and possible offset of $\Delta v = 10$km/s motivated by the systematic error scaling quadradically in $\Delta v$, and find only shifts at the tenths of standard deviations in our $T_0-\gamma$ inferences.

 Unfortunately, for the case where the calibration errors are uncorrelated between observations (and not systematic), as might occur from random observer miscentering on the slit, then the imaginary component no longer constrains the miscalculation. Rather, the cross power spectrum would be purely real with
\begin{equation}
P_{\alpha \beta} =P^T_{\alpha \beta} (k)\; \exp[-\sigma_v^2 k^2/2],
\end{equation}
where we have assumed $\Delta v$ is drawn from a Gaussian with standard deviation $\sigma_v$.  The extra Gaussian damping would bias our measurement.  If $\sigma_v = 5~(10)$km~s$^{-1}$, our rough upper bound from observer miscentering above assuming 20\% offsets, $\exp[-\sigma_v^2 k_{\rm max}^2/2] = 0.95 ~(0.82)$. Thus, similarly to with a systematic offset of $\Delta v$, the bias would be significant for a random offset that are larger than $\sigma_v = 10~$km~s$^{-1}$.

\end{document}